\documentclass{ws-procs9x6}
\begin{document}

\newcommand{\half}{\frac12}
\newcommand{\third}{\frac13}
\newcommand{\eqn}[1]{\label{eq:#1}}
\newcommand{\refeq}[1]{Eq.~(\ref{eq:#1})}
\newcommand{\figgg}[1]{\label{fig:#1}}
\newcommand{\refiggg}[1]{Fig.~\ref{fig:#1}}
\newcommand{\eqsdash}[2]{Eqs.~(\ref{eq:#1}-\ref{eq:#2})}
\newcommand{\eqscomma}[2]{Eqs.~(\ref{eq:#1}, \ref{eq:#2})}
\newcommand{\beq}{\begin{eqnarray}}
\newcommand{\eeq}{\end{eqnarray}}
\newcommand{\nn}{\nonumber}
\def\CO{{\mathcal O}}
\def\CL{{\mathcal L}}
\def\CM{{\mathcal M}}
\def\tr{{\rm\ Tr}}
\def\dim{{\rm dim}\ }
\def\al{\alpha}
\def\bt{\beta}
\def\eps{\epsilon}
\def\mn{{\mu\nu}}
\newcommand{\rep}[1]{{\bf #1}}
\newcommand{\vev}[1]{\langle#1\rangle}
\def\be{\begin{equation}}
\newcommand{\bel}[1]{\be\label{#1}}
\def\ee{\end{equation}}
\newcommand{\eref}[1]{(\ref{#1})}
\newcommand{\Eref}[1]{Eq.~(\ref{#1})}
\newcommand{\rem}[1]{}
\def\tr{{\rm tr}}
\def\half{{1\over 2}}
\def\NN{{\mathcal N}}

\def\nzero{$\NN=0$}
\def\none{$\NN=1$}
\def\ntwo{$\NN=2$}
\def\nntwo{$\NN\approx 2$}
\def\nnfour{$\NN\approx 4$}
\def\nlfour{$\NN\leq 2$}
\def\nfour{$\NN=4$}
\def\neight{$\NN=8$}
\def\Pf{{\rm Pf}\ }
\def\susy{supersymmetry}
\def\susic{supersymmetric}
\def\diag{{\rm diag}}
\def\rarr{\rightarrow}
\def\drarr{\Rightarrow}
\def\betau{\beta_{1/g^2}}
\def\OO{{\mathcal O}}
\def\ZZ{{\bf Z}}
\def\SS{{\bf S}}
\def\TT{{\bf T}}
\def\RR{{\bf R}}
\def\DD{{\tilde D}}
\def\LLam{{\overline{\Lambda}}}
\newcommand{\VV}[1]{{\bf V}_#1}
\newcommand{\II}{{\bf I}}
\def\yy{{\bf \hat y}}
\def\PP{{\mathcal P}}
\def\QQ{{\mathcal Q}}
\def\bQQ{\bar{\mathcal Q}}
\def\ZN{\ZZ_N}
\def\OM{Olive-Montonen}
\def\NO{Nielsen-Olesen}
\def\Lgr{{\mathcal L}}
\def\slz{$SL(2,\ZZ)$}
\def\tim{\tilde m^2}
\newcommand{\dbyd}[2]{{\partial #1\over\partial #2}}
\newcommand{\ddbyd}[2]{{\partial^2 #1\over\partial #2^2}}
\newcommand{\ddbydd}[3]{{\partial^2 #1\over\partial #2\ \partial #3}}
\def\ta{\theta}
\def\bit{\begin{itemize}}
\def\eit{\end{itemize}}
\def\iddx{\int\ d^dx\ }
\def\EXE{{{\underline{\it Exercise:} }}}
\newcommand{\EX}[1]{\vskip 0.2 in {\noindent\underline{\bf Exercise:} #1 \vskip 0.2 in }}
\def\dslash{\partial\ \!\!\!\!\!\slash}
\def\Dslash{D\!\!\!\!\slash\ }


\title{ AN UNORTHODOX INTRODUCTION TO SUPERSYMMETRIC GAUGE THEORY}
\author{Matthew J. Strassler}
\address{Department of Physics \\ 
         University of Washington\\
         Box 351560 \\ 
         Seattle, WA 98195 \\
         E-mail: strassler@phys.washington.edu}
\maketitle

\abstracts{         
Numerous topics in three and four dimensional supersymmetric
gauge theories are covered.  The organizing principle in this
presentation is scaling (Wilsonian renormalization group flow.)
A brief introduction to scaling and to
supersymmetric field theory, with
examples, is followed by discussions of nonrenormalization
theorems and beta functions.  Abelian gauge theories are
discussed in some detail, with special focus on 
three-dimensional versions of supersymmetric QED, which
exhibit solitons, dimensional antitransmutation,
duality, and other interesting phenomena.  Many of
the same features are seen in four-dimensional non-abelian
gauge theories, which are discussed in the final sections.
These notes are based on lectures given at TASI 2001.
}

\section{Introduction}

These lectures are the briefest possible introduction to some
important physical ideas in supersymmetric field theory.

I have specifically avoided restating what the textbooks already
contain, and instead have sought to provide a unique view of the
subject.  The usual presentation on superfields is sidestepped; no
introduction to the supersymmetric algebra is given; superconformal
invariance is used but not explained carefully.  Instead the focus is
on the renormalization group, and the special and not-so-special
qualitative features that it displays in supersymmetric theories.
This is done by discussing the often-ignored classical renormalization
group (the best way to introduce beta functions, in my opinion) which
is then generalized to include quantum corrections.  This approach is
most effective using theories in both three and four dimensions.
Initially, models with only scalars and fermions are studied, and the
classic nonrenormalization theorem is presented.  Then I turn to
abelian gauge theories, and finally non-abelian gauge theories.  Fixed
points, unitarity theorems, duality, exactly marginal operators, and a
few other amusing concepts surface along the way.  Enormously
important subjects are left out, meriting only a sad mention in my
conclusions or a brief discussion in the appendix.

Clearly these lectures are an introduction to many things and a
proper summary of none.  I have tried to avoid being too cryptic, and
in some sections I feel I have failed. I hope that the reader can still
make something useful of the offending passages.  I have also completely
failed to compile a decent bibliography.  My apologies.

My final advice before beginning: {\it Go with the flow.}  If you
aren't sure why, read the lectures.

\section{ Classical theory}
\label{sec:classicaltheory}

\subsection{Free massless fields}
\label{subsec:freefields}

Consider classical free scalar and spinor fields in $d$
 space-time dimensions.
\bel{freescalar}
S_\phi = \iddx  \partial_\mu\phi^\dagger \partial^\mu\phi
\ee
\bel{freefermion}
S_\psi = \iddx i\bar\psi \dslash \psi
\ee

By simple dimension counting, since space-time coordinates have 
mass dimension $-1$ and space-time derivatives have mass dimension $+1$,
the dimensions of these free scalar and spinor fields are
$$
\dim \phi = {d-2\over 2} \ \ ; \ \ \dim\psi = {d-1\over 2}
$$
so in $d=3$ scalars (fermions) have dimension $1/2$ (1) while in
$d=4$ they have dimension 1 ($\frac32$).
These free theories are scale-invariant --- for any $s>0$, 
$$
x\to sx \ ; \phi \to s^{-(d-2)/2}\phi \ ;  
\psi \to s^{-(d-1)/2}\psi \ \Rightarrow
S_\phi \to S_\phi \ ; \ S_\psi \to S_\psi \ .
$$
Less obviously, they are conformally invariant.  (If you don't already
know anything about conformal invariance, don't worry; for the
purposes of these lectures you can just think about scale invariance,
and can separately study conformal symmetry at another time.)

\subsection{Free massive fields}
\label{subsec:freemassive}

Now let's consider adding some mass terms.
\bel{massivescalar}
S = \iddx  \left[
\partial_\mu\phi^\dagger \partial^\mu\phi - m^2\phi^\dagger \phi
\right]
\ee
Scalar masses always have dimension $\dim m = 1$ for all $d$.
The propagator
$$
\vev{\phi(k)\phi(-k)} = {i\over k^2 - m^2} 
$$
is a power law $1/|x-y|^{d-2}$ (the Fourier transform of the
propagator $i/k^2$) for $|x-y|\ll m^{-1}$, and acts like a delta
function $\delta^4(x)$ (the Fourier transform of the propagator
$-i/m^2$) at $|x-y|\gg m^{-1}$.  The scale-dependent propagator
interpolates between these two scale-invariant limits.

\EX{Compute the propagator in position space and show that it does
indeed interpolate between two scale-invariant limits.}

What is the renormalization group flow associated with this theory?
It can't be completely trivial even though the theory is free and
therefore soluble. The number of degrees of freedom is two (one
complex scalar equals two real scalars) in the ultraviolet and zero in
the infrared, so obviously the theory is scale dependent.  In
particular, Green's functions are not pure power laws, as we just saw
for the propagator.  But how are we to properly discuss this given
that the one coupling constant in the theory, $m$, does not itself
change with scale?  The approach we will use is to define a {\it
dimensionless} coupling $\nu^2\equiv m^2/\mu^2$ where $\mu$ is the
renormalization-group scale -- the scale at which we observe the
theory. We can think of this theory as transitioning, as in
\refiggg{nuflow}, between two even simpler theories: the
scale-invariant theory at $\mu\to\infty$ where the mass of $\phi$ is
negligible and $\nu\to 0$, and the empty though scale-invariant
$\nu\to\infty$ theory in the infrared, where the scalar does not
propagate.

\begin{figure}[th]
\begin{center}
 \centerline{\psfig{figure=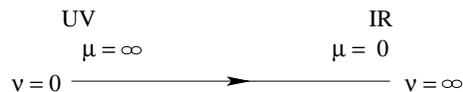,width=6cm,clip=}}
 \caption{\small The effect of a mass term grows in the infrared. }
\figgg{nuflow}
\end{center}
\end{figure}

Now consider a massive fermion.
\bel{massivefermion}
S = \iddx \left[ i\bar\psi \dslash \psi -m\psi\psi\right]
\ee
Again, fermion masses always have dimension $\dim m = 1$ for all
$d$.\footnote{Here and throughout I am using the notation of Wess and
Bagger.  In particular, $\psi$ is a {\it two-component} complex
left-handed Weyl fermion, with a spinor index $a=1,2$; $\bar \psi$,
its conjugate, a right-handed Weyl anti-fermion, has a conjugate
spinor index $\dot a=1,2$ which cannot be contracted with $a$!  A
kinetic term $\bar\psi^{\dot a}\sigma_{\dot a a}^\mu \partial_\mu
\psi^a$ ($\sigma^0 = i\delta_{\dot a a}$, $\sigma^{1,2,3}$ are Pauli
matrices) simply translates a left-handed Weyl fermion; it conserves a
fermion number symmetry (the symmetry $\psi\to \psi e^{i\alpha}$).  A
so-called ``Majorana'' mass term $\psi^a\epsilon_{ab}\psi^b$
($\epsilon_{11}=\epsilon_{22}=0$, $\epsilon_{12}=1=-\epsilon_{21}$)
breaks fermion number and mixes the particle with its antiparticle.
It is very possible that left-handed neutrinos have such mass terms.
Note Majorana mass terms can only 
be written for particles which are gauge-neutral; for instance,
electrons cannot have them as long as QED is an unbroken
gauge symmetry.  Instead, an electron mass connects the
two-component charge 1 left-handed electron $\psi$ to the conjugate of
the two-component charge -1
left-handed positron ${\chi}$, via a coupling $\epsilon_{ab}\chi^a\psi^b$.}
Better yet, consider a theory with two free fermions.
\bel{twomassivefermions}
S = \iddx \sum_{n=1}^2 \left[ i\bar\psi_n \dslash \psi_n 
-m_n\psi_n\psi_n\right]
\ee
Now we have a dimensionless coupling constant $\rho=m_1/m_2$.

Consider the classical scaling behavior (renormalization group flow)
shown in \refiggg{twomasses} below.  Note there are four
scale-invariant field theories in this picture: one with two massless
fermions, two with one massless and one infinitely massive fermion,
and one with no matter content (indeed, no content at all!)  We will
call such scale-invariant theories ``conformal fixed points'', or
simply ``fixed points'', to indicate that the dimensionless couplings
of the theory, if placed exactly at such a point, do not change with
scale.

\begin{figure}[th]
\begin{center}
 \centerline{\psfig{figure=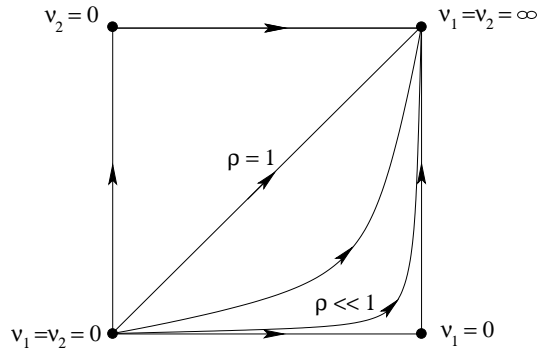,width=7cm,clip=}}
 \caption{\small The scaling flow 
of two masses, with $\rho = m_1/m_2=\nu_1/\nu_2$. }
\figgg{twomasses}
\end{center}
\end{figure}

Whatever are $m_1$ and $m_2$, or for our purposes the dimensionless
couplings $\nu_1$ and $\nu_2$, scale transformations take the theory
from the $\nu_1=\nu_2=0$ conformal fixed point to one of the other
fixed points.  The parameter $\rho\equiv \nu_1/\nu_2$, which is
scale-invariant, parameterizes the ``flows'' shown in the graph.  The
arrows indicate the change in the theory as one considers it at larger
and larger length scales.  In addition to flows which actually end at
$\nu_1=0, \nu_2=\infty$, note there are also interesting flows from
$\nu_1=\nu_2=0$ to $\nu_1=\nu_2=\infty$ which pass {\it arbitrarily}
close to the fixed point $\nu_1=0, \nu_2=\infty$.  These can remain
close to the intermediate fixed point for an {\it arbitrarily} large range of
energy, (namely between the scales $\mu=m_2$ and $\mu=m_1 = \rho m_2$)
although this is not obvious from the graph of the flow; it is
something one must keep separately in mind.

A mass term is known as a ``relevant'' operator, where the relevance
in question is  {\it at long distances} (low energies.)  Although the mass
term has no effect in the ultraviolet --- at short distance --- it
dominates the infrared (in this case by removing degrees of freedom.)
We can see this from the fact that the dimensionless coupling $\nu$
grows as we scale from the ultraviolet toward the infrared.  In fact
we can define a beta function for $\nu=m/\mu$ as follows:
\bel{betar} \beta_\nu
\equiv \mu \dbyd{\nu}{\mu} = -\nu \ee 
That $\nu$ grows in the
infrared is indicated by the negative beta function.  More
specifically, the fact that the coefficient is $-1$ indicates that $\nu$ scales
like $1/\mu$.  This tells us that the mass $m$ has dimension 1.  We
will see examples of irrelevant operators shortly.

\subsection{Supersymmetry! The Wess-Zumino model}
\label{subsec:WZmodel}

Now let's add some ``interactions'' 
(more precisely, let's consider nonquadratic theories.)
\bel{Yukawa}
S_{{\rm Yukawa}} = \iddx  \left[\partial_\mu\phi^\dagger \partial^\mu\phi
+i\bar\psi \dslash \psi - y \phi\psi\psi-y^* \phi^\dagger\bar\psi\bar\psi 
- \lambda|\phi^\dagger\phi|^2 \right]
\ee
The coefficients $\lambda$ and $y$ are dimensionful for $d<4$ but
dimensionless for $d=4$.  Appropriate dimensionless
coefficients are $\lambda\mu^{d-4}$ and $y\mu^{(d-4)/2}$, with
classical beta functions 
$$
\beta_{\lambda\mu^{d-4}} = (d-4)\lambda\mu^{d-4} \ ; \
\beta_{y\mu^{(d-4)/2}} = \half (d-4)y\mu^{(d-4)/2}$$
showing the interactions are classically relevant for $d<4$ (and
irrelevant for $d>4$!)  but are ``marginal'' for $d=4$ --- they do not
scale.  The theories in question are thus classically scale invariant
for $d=4$ (and in fact conformally invariant.)

This Yukawa-type theory can be conveniently written in the following form
\begin{eqnarray}
\label{WessZumino}
S_{{\rm Yukawa}} &=& S_{kin} + S_{int} + S_{int}^\dagger
\cr \  & & \ \cr
S_{kin} &=& \iddx  \left[\partial_\mu\phi^\dagger \partial^\mu\phi
+i\bar\psi \dslash \psi + F^\dagger F\right] \cr \  & & \  \cr
S_{int} &=&   \iddx \left[- y \phi\psi\psi+ h \phi^2 F \right]
\end{eqnarray}
Here $F$ is just another complex scalar field, except for one thing -- it has
no ordinary kinetic term, just a wrong-sign mass term.  This looks
sick at first, but it isn't.  $F$ is what is known as an ``auxiliary
field'', introduced simply to induce the $|\phi^\dagger\phi|^2$
interaction in \eref{Yukawa}.  The classical equation of motion for
$F$ is simply $F^* \propto \phi^2$.

We can break the conformal invariance of the $d=4$ theory by adding 
mass terms and cubic scalar terms:
\bel{massiveYukawa}
S_{{\rm Yukawa}} \to S_{{\rm Yukawa}} - \iddx  \left[M^2 \phi^\dagger\phi
+\half m\psi\psi+h^*\phi|\phi^2| + h\phi^\dagger |\phi|^2  \right]
 \ .
\ee
If $M$, $h$ and $\lambda$ are related so that the scalar potential is a 
perfect square, then we may write this in the form \eref{WessZumino}, 
\bel{massiveWessZumino}
S_{int} = \iddx  \left[ - \half m\psi\psi-y \phi\psi\psi+ 
(M\phi + s \phi^2) F \right]
\ee
where $s$ is a constant, and
with $S_{kin}$ as before.

When $y$ and $s$ are equal, and $M$ and $m$ are equal, the theory has
supersymmetry.  Given any spinor $\zeta$, the transformations
\bel{susytransf}
\delta \phi = \sqrt{2}\zeta\psi \ ; 
\delta\psi_a = i\sqrt{2}\sigma^\mu_{a\dot a}\bar\zeta^{\dot a}\partial_\mu \phi
+ \sqrt{2}\zeta_a F ; \delta F = i\sqrt{2}\bar\zeta^{\dot a} 
\dslash_{\dot a a}\psi^a
\ee
change the Lagrangian only by a total derivative, which integrates to
nothing in the action.  I leave it to you to check this.  This
supersymmetric theory is called the ``Wess-Zumino model''; it dates to
1974.

At this point there is a huge amount of supersymmetric
superfield and superspace
formalism one can introduce.  There are lots of books on this subject
and many good review articles.  You don't need me to write another
(and many of you have already read one or more of them) so I will not
cover this at all.  Instead I will state without proof how one may
construct supersymmetric theories in a simple way.  Those of you who
haven't seen this before can take this on faith and learn it later.
Those of you who have seen it will recognize it quickly.

Let us define a chiral multiplet, which we will represent using something
we will call a chiral superfield $\Phi$.  $\Phi$ contains three
``component'' fields: $\phi, \psi, F$.  
Note that $\Phi$ is complex: $\phi$ is
a complex scalar (with 
two real components), $\psi$ is a Weyl fermion (two complex
components, reduced to one by the Dirac equation) and $F$ is a complex
auxiliary field (no propagating components.)  Thus before accounting
for the equations of motion there are four real bosonic and four
real fermionic degrees of freedom; after using the equations of motion,
there are two real
bosonic propagating degrees of freedom and two fermionic ones.  That
the numbers of bosonic and fermionic degrees of freedom are equal is a
requirement (outside two dimensions) for any supersymmetric theory.

Since $\Phi$ is complex we can distinguish holomorphic functions
$W(\Phi)$, and antiholomorphic functions, from general functions
$K(\Phi,\Phi^\dagger)$.  As we will see, this fact
is the essential feature which explains why we know so much more about
supersymmetric theories than nonsupersymmetric ones --- the difference
is the power of complex analysis compared with real analysis.

Consider any holomorphic function (for now let it be polynomial) $W(\Phi)$.  A
well-behaved supersymmetric classical field theory can be written in
the form of Eqs.~\eref{Yukawa}-\eref{massiveWessZumino} with
\bel{generalWessZumino}
S_{int} = \iddx  \left[
 -  \half\ddbyd{W(\phi)}{\phi}\psi\psi +\dbyd{W(\phi)}{\phi}F \right]
\ee
(In this expression the function $W(\Phi)$ is evaluated at
$\Phi=\phi$.)  The function $W$ is called the ``superpotential.''  (Note
that $\dim W = d-1$.)  

\EX{Check that we recover the previous case by
taking $W = \half m\Phi^2 + {1\over 3} y\Phi^3$.}

The potential for the scalar field $\phi$ is always
\bel{Vphi}
V(\phi) = |F|^2 = \left|\dbyd{W(\phi)}{\phi}\right|^2 
\ee
Notice this is positive or zero for all values of $\phi$.

It is a theorem that supersymmetry is broken if $\vev F\neq 0$; you
can see this from the transformation laws \eref{susytransf}, in which
$F$ appears explicitly.  (By contrast, $\vev{\phi}$ does not appear in
the transformation laws, so it may be nonzero without breaking
supersymmetry automatically.)  From \eref{Vphi} we see that
$V(\phi)=0$ in a supersymmetry-preserving vacuum.  This also follows
from the fact that the square of one of the supersymmetry generators
is the Hamiltonian; if the Hamiltonian, acting on the vacuum, does not
vanish, than the supersymmetry transformation will not leave the
vacuum invariant, and the vacuum thus will not preserve the
supersymmetry generator.

To find the vacua of a supersymmetric theory is therefore much
easier\footnote{Usually!  we will return to a subtlety later.} than in
a nonsupersymmetric theory.  In the nonsupersymmetric case we must
find all $\phi$ for which $\dbyd{V}{\phi}=0$, and check which of these
extrema is a local or global minimum.  In the supersymmetric case, we
need only find those $\Phi$ for which $\dbyd{W}{\Phi}=0$; we are guaranteed
that any solution of these equations has $V=0$ and is therefore a
supersymmetric, global minimum (although it may be one of many).

For example, if we take the theory $W = \third y\Phi^3$, then $V(\phi)
= |\phi^2|^2$; we can see it has only one supersymmetric minimum, at
$\phi=0$.

\EX{ Find the scalar potential and the  
supersymmetric minima for $W = \third y \Phi^3 + \half
m\Phi^2$, $W = \third y \Phi^3 + \xi \Phi$, $W = \third y\Phi^3 +
c$, where $\xi$ and $c$ are constants.  
In each case, note and carefully interpret what happens as $y$ goes
to zero.  }

\subsection{The XYZ model}
\label{subsec:XYZmodel}

Let us consider a theory with three chiral superfields $X,Y,Z$ with
scalars $x,y,z$, and a superpotential $W=hXYZ$.  Then the potential
$V(x,y,z)=|h^2|(|xy|^2 + |xz|^2 + |yz|^2)$ has minima whenever two of
the three fields are zero.  In other words, there are three complex
planes worth of vacua (any $x$ if $y=z=0$, any $y$ if $z=x=0$, or any
$z$ if $x=y=0$) which intersect at the point $x=y=z=0$. This rather
elaborate space of degenerate vacua --- noncompact, continuous, and
consisting of three intersecting branches --- is called a ``moduli
space''.  The massless complex fields whose expectation values
parameterize the vacua in question ($x$ on the X-branch, {\it etc.})
are called ``moduli''.  It is useful to represent the three complex
planes as three intersecting cones, as in \refiggg{XYZ}.  The point
where the cones intersect, and all fields have vanishing expectation
values, is called the ``origin of moduli space.''

\begin{figure}[th]
\begin{center}
 \centerline{\psfig{figure=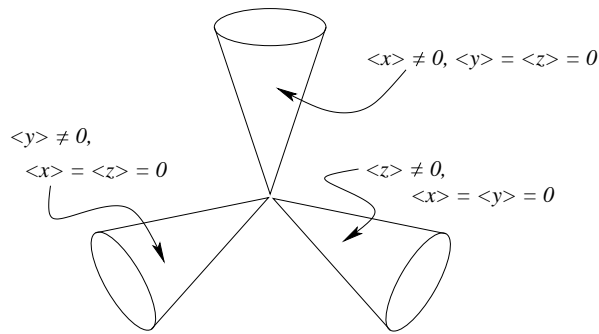,width=8cm,clip=}}
 \caption{\small The moduli space of the XYZ model. }
\figgg{XYZ}
\end{center}
\end{figure}
 
\EX{ Prove that these vacua are not all physically equivalent (easy
--- compute the masses of the various fields.)  Note that there are
extra massless fields at the singular point $x=y=z=0$ where the three
branches intersect.  Then look at the symmetries --- discrete and
continuous --- of the theory and determine which of the vacua are
isomorphic to one another.}

\subsection{XYZ with a mass}
\label{subsec:XYZmass}

Let's quickly consider what happens in the theory $W= hXYZ + \half m
X^2$. Then the $Y$ and $Z$ branches remain while the $X$ branch
is removed, as in \refiggg{XYZplusXX}.  
It is interesting  to consider the classical
renormalization group flow.  Let's think of this in $d=4$; then the
theory with $m=0$ is classically conformal, with $h$ a marginal
coupling, and $m$ a relevant one. 

\begin{figure}[th]
\begin{center}
 \centerline{\psfig{figure=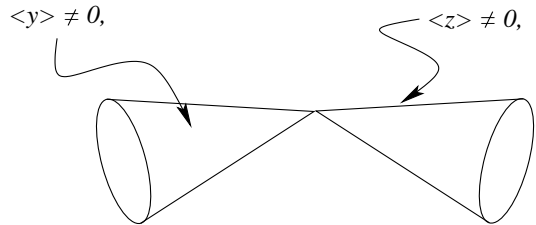,width=7cm,clip=}}
 \caption{\small The moduli space once a mass for $X$ is added. }
\figgg{XYZplusXX}
\end{center}
\end{figure}

What is happening in the far infrared?  The theory satisfies the equation
$$\nabla^2 X^\dagger = m^\dagger(mX + hYZ)$$  
which for momenta much lower than $m$ (where $\nabla^2 X \ll |m^2|X$) 
simply beomes $mX=-hYZ$.  In this limit the kinetic term for $X$ plays
no role, and $X$ acts like an auxiliary field.  We may therefore
substitute its equation of motion back into the Lagrangian, obtaining
a low-energy effective theory for $Y$ and $Z$ with superpotential
$$
W_L(Y,Z) = \kappa Y^2Z^2 \ \ ; \ \ \kappa= -h^2/2m
$$
At the origin of moduli space, where
$\vev{y}=0=\vev{z}$, $Y$ and $Z$ are massless.

The superpotential $Y^2Z^2$ leads to interactions such as
$\kappa^2|y^2z|^2$ and $\kappa y^2\psi_z\psi_z$ which have dimension
higher than 4; equivalently, $\kappa$ has negative mass dimension
$-1$.  Defining a dimensionless quantity $k=\kappa\mu = -h^2/2\nu$, we
see it has a {\it positive} beta function ($\beta_k =
2\beta_h-\beta_\nu=+k$) so it becomes unimportant in the infrared.  We
therefore call $\kappa$, or $k$, an ``irrelevant coupling'', and
$Y^2Z^2$ an ``irrelevant operator.''  More precisely, at the origin of
moduli space there is a fixed point in the infrared at which the
massless chiral superfields $Y,Z$ have $W(Y,Z)=0$ (since the physical
coupling $k\to 0$ in the infrared.)  We say that $\kappa$ (or $k$) is
an irrelevant coupling, and $Y^2Z^2$ is an irrelevant
operator\footnote{Still more precisely, the irrelevant operators are
those which appear in the Lagrangian; it is not $W$ but $|\partial
W/\partial Y|^2$ which has dimension higher than 4.  This shorthand is
a convenient but sometimes confusing abuse of language.  The assertion
that $\kappa$ is an irrelevant coupling is not subject to this
ambiguity.}  {\it with respect to this infrared fixed point}. The
renormalization group flow for $\vev{y}=\vev{z}=0$ is shown in
\refiggg{RGinXYZmass}.

\begin{figure}[th]
\begin{center}
 \centerline{\psfig{figure=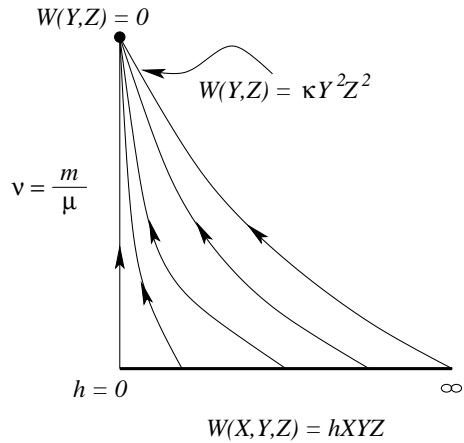,width=6cm,clip=}}
 \caption{\small The flow at the origin of moduli space; the effect of
the $Y^2Z^2$ superpotential renormalizes away in the far infrared.}
\figgg{RGinXYZmass}
\end{center}
\end{figure}

Thus, we may think of this field theory as a flow from a (classically)
conformal fixed point with $W(X,Y,Z)=hXYZ$ --- one of a continuous
class of fixed points with coupling $h$ --- to the isolated conformal
field theory with $W(Y,Z)=0$.  In this flow, the mass term $X^2$ acts
as a relevant operator on the ultraviolet fixed point, causing the flow
to begin, and the flow into the infrared fixed point occurs along the
direction given by the irrelevant operator $(YZ)^2$.

However, this description is only correct if
$\vev{y}=\vev{z}=0$, that is, exactly at the origin of moduli space.  Suppose
$\vev{y}$ is not zero; then $Z$ is massive at the scale $hy^2$ and we
must re-analyze the flow below this scale.  The far-low-energy theory in
this case would have only $Y$ in it.  Here lies a key subtlety.
The existence of the fixed point with {\it vanishing} superpotential 
$W(Y,Z)$ at the origin of
moduli space might seem, naively, to 
imply that somehow $Z$ would remain massless
even when $\vev{y}\neq 0$.  But this is inconsistent
with the original classical analysis, using either the $XYZ+\half mX^2$
superpotential or the $kY^2Z^2$ superpotential.  
How is this contradiction resolved?
To understand this better, consider  more carefully 
the effect of the nonvanishing
superpotential for finite $\vev{y}$.
At momenta $\mu\gg \kappa\vev{y^2}$ the theory has
light fields $Y,Z$, as does the conformal point, but at momenta 
$\mu\ll \kappa\vev{y^2}$
only $Y$ is light; see \refiggg{RGinXYZvev}. {\it Thus the limits $\mu\to
0$ and $\vev{y}\to 0$ do not commute!}   This is one way to see that the
infrared conformal theory at the origin of moduli space is 
effectively disconnected from the infrared theory at $\vev{y}\neq 0$.

\begin{figure}[th]
\begin{center}
 \centerline{\psfig{figure=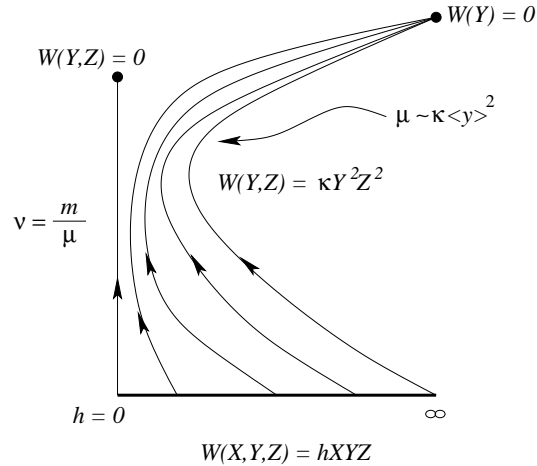,width=7cm,clip=}}
 \caption{\small Away from the origin of moduli space, the combination of
$W$ and $\vev{y}$ drive the flow to a different endpoint. }
\figgg{RGinXYZvev}
\end{center}
\end{figure}

\EX{ Show that the theory with $W(X,Y,Z) = hXY^2 + mYZ +\xi X$ has no
supersymmetric vacuum.  Supersymmetry is spontaneously broken.  Find
the supersymmetry-breaking minimum and check that there is a massless
fermion in the spectrum (the so-called Goldstino.)  }
 
\subsection{Solitons in XYZ}
\label{subsec:XYZSols}

What happens to the classical theory if we take the superpotential $W
= hXYZ + \xi X$?  The equations $\dbyd{W}{X} = hYZ+\xi X = 0$,
$\dbyd{W}{Y} = hXZ = 0$, $\dbyd{W}{Z} = hXY = 0$, have solutions
$X=0$, $YZ = \xi/h$.  (Henceforth we will usally not distinguish the
chiral superfield's expectation value from that of its component
scalar field; thus $X=0$ means $\vev{x}=0$.)  Now the equation $YZ = c$, $c$
a constant, is a hyperbola.  To see this consider $c$ real (without
loss of generality) and now note that for $Y$ real the equation $YZ=c$
gives a real hyperbola; rotating the phase of $Y$ gives a complex
hyperbola.  What has happened to the three branches is this: the
X-branch has been removed, while the two cones of the Y and Z-branches
have been joined together as in the figure below. The moduli space
$YZ=c$ (\refiggg{XYZplusX})
can be parameterized by a single modulus $Y\over \sqrt c$ which
lives on the complex plane with the point at zero removed.

\begin{figure}[th]
\begin{center}
 \centerline{\psfig{figure=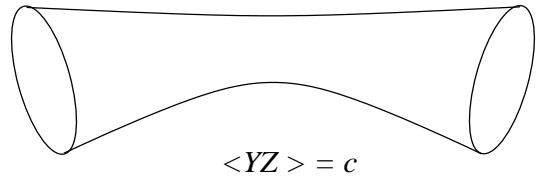,width=7cm,clip=}}
 \caption{\small The moduli space when $W=XYZ+\xi X$. }
\figgg{XYZplusX}
\end{center}
\end{figure}

This seemingly innocent theory hides something highly nontrivial.  Let
us take the case of $d=3$, and use polar coordinates $r,\theta$ on
the two spatial directions.  Now suppose that we can find a
time-independent circularly-symmetric solution to the classical
equations, of characteristic size $r_0$, in which
$$
Y(r,\theta) = \sqrt{c}f(r)e^{i\theta} \ , \ Z(r,\theta) 
= \sqrt{c}f(r)e^{-i\theta}
$$
 where $f(r\to 0) = 0$ (to avoid multivalued fields at the origin) and
$f(r\gg r_0) =1$ (so that $YZ=c$ at large $r$, meaning the potential
energy density in this solution is locally zero for $r\gg r_0$.)  This
object would be a candidate for a ``vortex'' soliton, a composite
particle-like object, in which $Y$ and $Z$ wind (in opposite
directions, maintaining $YZ=c$) around the circle at spatial infinity,
and in which the energy density is large only in a ``core'' inside
$r=r_0$ (\refiggg{vortex}).

\begin{figure}[th]
\begin{center}
 \centerline{\psfig{figure=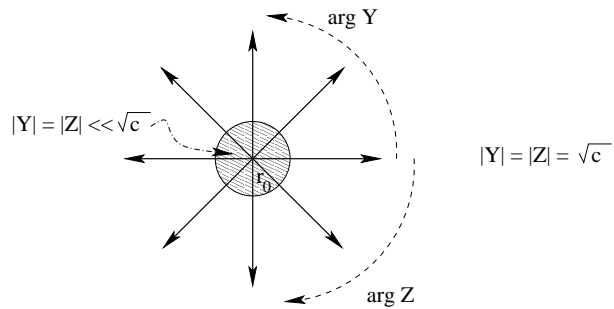,width=8cm,clip=}}
 \caption{\small The vortex, which has logarithmically divergent energy. }
\figgg{vortex}
\end{center}
\end{figure}

However, although the energy density falls to zero rapidly for $r\gg
r_0$, the total energy of this vortex diverges.  The kinetic energy
from the winding of $Y$ and $Z$ is logarithmically divergent --- so
any soliton of this type would have infinite energy.  Why even
consider it?  Well, a vortex and antivortex {\it pair}, a distance
$\Delta$ apart, will have finite energy, and this even if they are
quite far apart, with $\Delta\gg r_0$, as in \refiggg{vtxantivtx}.  For
such a pair, $Y$ and $Z$ would wind locally but not at spatial
infinity.  Consequently there would be no logarithmic divergence in
the energy; instead the energy would go as twice the core energy plus
a term proportional to $\log \Delta$.  Since we cannot pull the
solitons infinitely far apart, we should think of these objects as
logarithmically confined solitons.

\begin{figure}[th]
\begin{center}
 \centerline{\psfig{figure=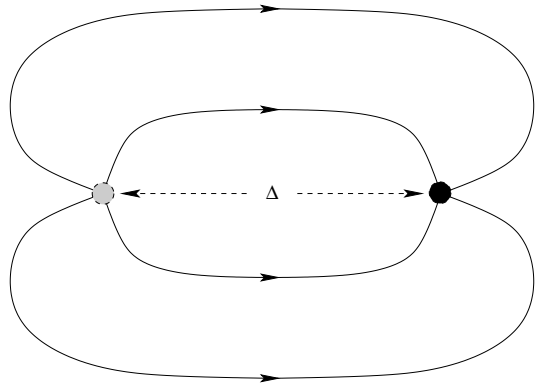,width=7cm,clip=}}
 \caption{\small A finite-energy vortex/antivortex configuration; there is
 no winding of $\arg Y$ at infinity. }
\figgg{vtxantivtx}
\end{center}
\end{figure}

This is not as strange as it sounds.  {\it In three dimensions,
massive electrons exchanging photons are logarithmically confined!}
In other words, these vortices are no worse than electrons!  and
indeed they behave as though they are exchanging some massless
particle.  We'll come back to this idea later.

In four dimensions, we may repeat the same analysis.  Instead of
particle-like vortices, we would find
logarithmically-confined vortex {\it strings}, extended in the third spatial
dimension.  Again, these really can be pair-produced in the theory
(or rather, as in \refiggg{stringloop},
a closed loop of this string can be created at
finite energy cost.)

\begin{figure}[th]
\begin{center}
 \centerline{\psfig{figure=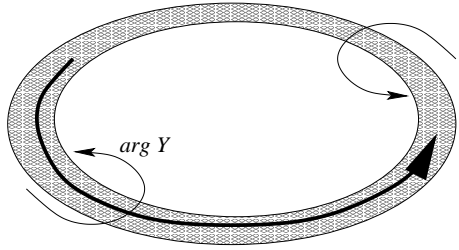,width=6cm,clip=}}
 \caption{\small In $d=4$, a closed vortex plays the role of the
 vortex/antivortex configuration of $d=3$. }
\figgg{stringloop}
\end{center}
\end{figure}

\subsection{A more general form}
\label{subsec:generalform}

Finally, I should point out that I have by no means written the most general
supersymmetric theory.  Taking any real function 
$K(\Phi^i,\Phi^{i\dagger})$ of 
chiral superfields
$\Phi^i$ and their conjugates, 
called the ``K\"ahler'' potential, and a superpotential $W(\Phi)$,
one can construct a supersymmetric theory by writing
\begin{eqnarray}\label{generalKahler}
S_{{\rm general}} &=& S_{kin} + S_{int} + S_{int}^\dagger
\cr \  & & \ \cr
S_{kin} &=& \iddx  \Bigg[\ddbydd{K(\phi,\phi^\dagger)}{\phi^{j\dagger}}{\phi^i}
\left(\partial_\mu\phi^{j\dagger} \partial^\mu\phi^i
+i\bar\psi^j\dslash \psi^i + F^{j\dagger} F^i \right) 
\cr
& &
\ \ \ \ \ \ + {\rm \ higher \ order \ terms}  \Bigg]\cr & &
\cr 
S_{int} &=&   \iddx  \left[-\half\ddbydd{W(\phi)}{\phi^i}{\phi^j}\psi^i\psi^j 
+ \dbyd{W(\phi)}{\phi^i}F^i\right] \cr
\end{eqnarray}
where repeated indices are to be summed over.  (The higher
order terms in $S_{kin}$ are given in Wess and Bagger; we will not need
them but they are quite interesting.)
More compactly, through the definition of the ``K\"ahler metric'' 
$$
K_{i\bar j}\equiv \ddbydd{K(\phi,\phi^\dagger)}{\phi^{\bar j\dagger}}{\phi^i}
$$
and
$$
W_i \equiv \dbyd{W(\phi)}{\phi^i} \ ; \ 
W_{ij} \equiv  \ddbydd{W(\phi)}{\phi^i}{\phi^j} \ ,
$$
we may write
\begin{eqnarray}\label{compactgeneral}
S_{kin} &=& \iddx  \left[K_{i\bar j}
\left(\partial_\mu\phi^{\bar j\dagger} \partial^\mu\phi^i
+i\bar\psi^{\bar j}\dslash \psi^i + F^{\bar j\dagger} F^i\right) 
+ \cdots \right]\cr & &
\cr 
S_{int} &=&   \iddx  \left[-\half W_{ij}\psi^i\psi^j 
+ W_i F^i\right] \ .
\end{eqnarray}

Note that the scalar potential of the theory is now modified!  It is now
$$
V(\phi_i) = W^\dagger_{\bar j}\left(K^{-1}\right)^{i\bar j} W_i
$$ 
(Previously we considered only a ``canonical'' K\"ahler potential $K
= \sum_i \Phi_i^\dagger \Phi_i$, with a metric $K_{i\bar j} =
\delta_{i\bar j}$.)  The above potential is still positive or zero,
however, so the condition for a supersymmetric vacuum, which was
$$ W_i= 0 \ \ {\rm for \ all \ }i \ ,
$$
remains true {\it unless} the metric $K_{i\bar j}$ is singular!  Such
singularities do occur (and have definite physical origins) so one
must not neglect this possibility.

\EX{ Take the theory of a single chiral superfield $\Phi$, with $W =
y\Phi^3$, and rewrite it by defining a new chiral superfield $\Sigma
\equiv \Phi^3$.  The superpotential is now $W = y\Sigma$, for which
$dW/d\Sigma\neq 0$, even for $\Sigma=0$.  Compute the K\"ahler
potential and the K\"ahler metric, and show that the theory does have
a supersymmetic vacuum at $\Sigma=0$.  The moral: one cannot determine
from the superpotential alone whether a theory breaks supersymmetry!
At a minimum, additional qualitative information about the K\"ahler
metric is required.  }

Even this set of supersymmetric theories is a small subset of the
whole.  For example, we have considered only theories with
two-derivative terms.  However, there is no reason to restrict
ourselves in this way.  For example, in the theory $W(X,Y,Z)= hXYZ
+\half m X^2$ and a canonical K\"ahler potential, the classical
effective theory at scales $\mu\ll m$ for the fields $Y$ and $Z$ most
certainly has terms in its Lagrangian with four or more
derivatives of $y$
and/or $z$, suppressed by inverse powers of $m$.  For these more
general cases, the Lagrangian is not fully specified by the K\"ahler
and superpotential alone.

\EX{ Check this last claim by substituting the equation of motion for
$X$ into the action of the original theory and expanding in
$1/|m^2|$.}

\section{Perturbation Theory}

\subsection{The quantum Wess-Zumino model}\label{subsec:QWZ}

Now let's return to the theory with a single field $\Phi$ and a
superpotential $W=\third y \Phi^3 + \half m \Phi^2$, with Lagrangian
obtained using \eref{generalWessZumino}.  Classically $m$ is a
relevant coupling; when $m$ is zero, $y$ is scale-invariant and the
theory is conformal.  What happens quantum mechanically?

Since most such theories are divergent, we must regulate them.  We can
do this by putting in a cutoff at a scale $\Lambda_{UV}$ (though this
is difficult in supersymmetric theories since most cutoffs violate
supersymmetry) or by introducing ghost fields (called Pauli-Villars
regulators) of mass $\Lambda_{UV}$ which cancel the degrees of freedom
at very high momentum while leaving those at low momentum.

Having done this, we can guess the form of perturbative corrections to
any coupling constant using dimensional analysis.  If the theory has
one or more dimensionless coupling constants, then we expect any
coupling of dimension $p$ to get a correction of order
$(\Lambda_{UV})^p$. (For $p=0$ we expect a $\log \Lambda_{UV}$
correction.) In particular, in four dimensions a $\lambda\phi^4$ 
interaction gives
a quadratic divergence ($\Lambda_{UV}^2$ times a function
of $\lambda$) to $M^2|\phi|^2$.  This is
not so for $\phi^4$ theory in $d=3$; its coupling $\lambda$ has
dimension 1, and this reduces the degree of divergence
possible in any diagram.  In fact the divergence in $M^2$ is now
proportional to $\lambda\Lambda_{UV}$.  Furthermore, while $\lambda$
itself gets a logarithmic divergence in $d=4$, in $d=3$ it gets only
finite corrections.

\EX{ Check these statements about $\phi^4$ theory. }

However, even correct dimensional analysis can overestimate the degree
of divergence if there are symmetries around.  A
coupling constant which {\it breaks} a symmetry cannot get an additive
divergence, but only a multiplicative one.  For example, in the Yukawa
theory in \Eref{Yukawa}, the fermion mass term might be expected to
get a linear divergence of order $|y^2|\Lambda_{UV}$.  However, the
theory \eref{Yukawa} has an explicitly broken chiral symmetry
$\psi\to\psi e^{i\alpha}$; it is broken by both $m$ and the coupling
$y$.  It also has a symmetry $\phi\to\phi e^{i\beta}$ broken by $y$
and $h$.  But what can you do with broken symmetries?  Just ask our
teachers, who understood the chiral Lagrangian of QCD!  Replace the
broken symmetries with ``spurious'' ones, under which the
symmetry-breaking couplings $m, y,h$ --- thought of as though they
were background scalar fields, called ``spurions'' --- transform with
definite charges.  Here's a table of dimensions and of spurious
charges under two spurious symmetries:
\bel{chargesA}
\begin{tabular}[h]{|l||c|c|c|c|c|c|c|}
\hline
 & $\phi$ & $\psi$ & $M^2$ & $h$ & $\lambda$ & $m$ & $y$\\ 
\hline 
$d=4$ {\rm\ dimension}& $1$& ${\frac32}$  &$ 2 $&$ 1 $&$ 0 $&$ 1 $&$ 0$\\ 
\hline 
$d=3$ {\rm\ dimension}&$\half $&$1$&$ 2 $&$ \half $&$ 1 $&$ 1 $&$ \half$\\
\hline 
\hline 
$U(1)_\alpha $&$  0 $&$ 1 $&$ 0 $&$ 0 $&$ 0 $&$ -2 $&$ -2$\\
\hline 
$U(1)_\beta $&$ 1$&$0$&$0$&$1$&$0$&$0$&$-1$ \\
\hline
\end{tabular}
\ee
For simplicity suppose for the moment that $h=0$; now let us see
why $m$ cannot have a linear divergence in four dimensions and is
finite in three.  We want to know $\Delta m$, the quantum mechanical
corrections to the fermion mass.  These must be in the form of
polynomials in the coupling constants times a possible power of
$\Lambda_{UV}$.  But by the above spurious symmetries, $\Delta m$ {\it
must} be proportional to $m$; otherwise $m$ and $\Delta m$ could not
possibly have the same charge under the spurious symmetries.
(Equivalently, this is required by the fact that when $m=0$ the
spurious symmetry $U(1)_{2\alpha-\beta}$ becomes a real one, and this
true symmetry then forbids a non-zero $\Delta m$.)  Therefore
the linear dimension of $\Delta m$ has already been soaked up by the
factor of $m$, and we can therefore have only a logarithm of
$\Lambda_{UV}$ appearing in $\Delta m$.  In $d=3$ in this theory, it
is even better; $\Delta m$ comes from interactions, but all of the
couplings have positive mass dimension, making even a logarithmic
divergence impossible.

\EX{ Show that in this theory $M^2$ has a divergence
$\Lambda_{UV}^{d-2}$ while {\it all} other couplings, as well as the
wave-function renormalization factors for $\phi$ and $\psi$, are log
divergent in $d=4$ and {\it finite} in $d=3$ (once $M^2$ has been
renormalized.)}

Once you've gone through this exercise, you're ready to see why
supersymmetry is so powerful.  Supersymmetry requires $M^2 = |m^2|$;
but this must hold even quantum mechanically (assuming supersymmetry
is preserved) so the divergences in $M^2$ must be reduced down to
those for $m$!  Thus in $d=4$ the above Wess-Zumino model has at most
log divergences; and in $d=3$ it is completely finite!

How can this happen?  Let's look at the diagrams in \refiggg{WZlogs}.

\begin{figure}[th]
\begin{center}
 \centerline{\psfig{figure=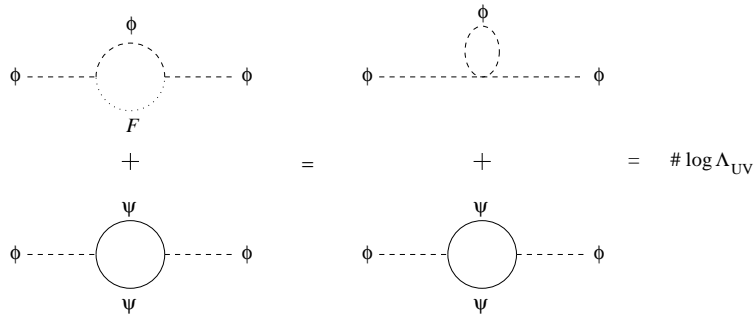,width=10cm,clip=}}
 \caption{\small The quadratic divergences to the scalar mass cancel. }
\figgg{WZlogs}
\end{center}
\end{figure}

Now, spurious symmetries can do more; they can constrain finite as
well as infinite quantum corrections.  For example, if $h=0$ and $y=0$
then the $U(1)_\beta$ symmetry is genuine; therefore the effective
potential for $\phi$ can only be a function of $\phi^\dagger \phi$,
and a quintic $\phi^5$ is clearly forbidden.  Once $h\neq 0$ (still
with $y=0$ for simplicity) then that symmetry is lost.  But it is easy
to see that the coefficient of $\phi |\phi^2|^2$ {\it must} be
proportional to $h^*$ (times a polynomial in $\lambda$) since
otherwise there is no way for the quantum effective Lagrangian to
respect the spurious symmetries!

Now comes the astounding part.  The spurious symmetries of
supersymmetric theories profoundly constrain the finite as well as
infinite corrections to supersymmetric theories.  In particular, {\it
the superpotential cannot be renormalized at any order in perturbation
theory!}  All quantum corrections must appear in the K\"ahler
potential or in higher-derivative operators.\footnote{The original
proofs of this, made in the 1980s, did not use spurions, so this
language is somewhat ahistorical.  It was Seiberg's great insights in
1993 which led him to this proof.}

Let us prove this, using modern methods, in the model with $W=\third
y\Phi^3 + \half m \Phi^2$; the generalization is straightforward
though tedious.  The key is that the renormalized effective
superpotential (which must be carefully defined, in a way that I am
avoiding getting into here) is itself a homomorphic function of the
chiral fields $\Phi$ (and not their conjugates) {\it and} of the
coupling constants $y$ and $m$ (and not {\it their} conjugates).  In
fact, one should think of $y$ and $m$ as additional ``background''
chiral fields!  That is, they act as though they are the expectation
values of scalar components of other, nonpropagating, chiral fields.
The mathematics of supersymmetry, and in particular the Feynman graph
expansion, automatically treats them this way.

We start with two special cases.  First, suppose $y=0$ and $m\neq 0$.
In this case there is obviously no renormalization since there are no
quantum effects.  Next, suppose $m=0$ and $y\neq 0$.  This is more
interesting.  The theory has one real symmetry and one spurious
symmetry.  The real symmetry is especially curious, in that it takes
$$
\phi \to \phi e^{2i\alpha/3} \ ; \ \psi \to \psi e^{-i\alpha/3} \ ; 
\ F\to Fe^{-4i\alpha/3} \ ; \
y \to y \ ; W \to e^{2i\alpha}  W
$$ 
which means that different parts of the $\Phi$ superfield transform
differently.  In fact the charge of $\phi$ is one unit greater than
that of $\psi$ and two units greater than that of $F$.  Such a
symmetry is called an R-symmetry, and it plays a special role since it
does not commute with supersymmetry transformations.  Note that the
superpotential, thought of as a function of $\phi$, transforms with
charge 2; this is a requirement of any R-symmetry, as can be seen from
the action \eref{generalKahler}.  The spurious symmetry is more
ordinary, except that $y$ transforms:
$$
\phi \to \phi e^{i\beta} \ ; \ \psi \to \psi e^{i\beta} \ ; \ F\to Fe^{i\beta}
; \ y\to ye^{-3i\beta} \ ; \ W \to W
$$
leaves the action invariant.  Now, what terms can we write in the
effective superpotential?  We can only write objects which carry
charge 2 under the R-symmetry and are neutral under the spurious
symmetry.  In perturbation theory, every term in the superpotential
must be of the form $y^p\Phi^q$, $p,q$ integers; since $W$ is
holomorphic we cannot write any powers of $y^*$ or $\Phi^\dagger$.
This is very important.  We cannot have {\it any} functions of
$|y|^2$, in contrast to what would have occurred in a
nonsupersymmetric theory where holomorphy is not an issue.  Clearly
$y\Phi^3$ is the unique choice, and therefore the superpotential
remains, even quantum mechanically, $W=\third y\Phi^3$.  This is
remarkable; no mass term can be generated in the effective
superpotential.  Moreover, no $\Phi^4$ term can be generated either.
This can be understood by thinking about the component fields; the
resulting term $\phi^2\psi^2$ in the Lagrangian is simply forbidden by
the chiral symmetries.  Clearly this is also true for any $\Phi^k$,
$k>3$.

Now finally let us consider the more complicated case $W = \third
y\Phi^3 + \half m\Phi^2$.  In this case there are no real symmetries.
However there are two spurious symmetry, one of them an R-symmetry.
Under the ordinary symmetry, $\Phi, y, m$ have charge $1,-3,-2$
respectively.  Invariance under this symmetry requires the
superpotential depend only on $m\Phi^2$ and $u\equiv y\Phi/m$.  The
choice of R-symmetry is a bit arbitrary (since we may take linear
combinations of any R-symmetry and the spurious but ordinary symmetry
to get a new R-symmetry) but a simple choice is to assign, as before,
charges $2/3,0,2/3$ to $\Phi,y,m$.  This means $\phi,\psi, F$ have
charge $2/3,-1/3,-4/3$ as before.
\bel{chargesB}
\begin{tabular}[h]{|l||c|c|c|c|c|c|c|}
\hline
& $\phi (\Phi) $&$ \psi $&$ F $&$ W  $&$ m $&$ y $&$ u=y\Phi/m$
\\ \hline \hline
$d=4$ {\rm\ dimension}&$ 1 $&$ {\frac32} $&$ 2 $&$ 3$&$ 1 $&$ 0 $&$ 0$ 
\\ \hline
$d=3$ {\rm\ dimension}&$\half $&$1$&$ {\frac32}$&$ 2 $&$ 1 $&$ \half $&$ 0$
\\ \hline \hline
$U(1)_R $&$ {\frac23} $&$ -{\frac13}
 $&$ -{\frac43} $&$ 2 $&$ {\frac23} $&$ 0 $&$ 0$
\\ \hline
$ U(1)_\Phi $&$ 1$&$1$&$1$&$0$&$-2$&$-3 $&$ 0$
\\ \hline
\end{tabular}
\ee

Notice that $u$ is invariant under this as well, while $m\phi^2$
has R-charge 2, so the superpotential must take the form
$$
W_{eff} = \half m\Phi^2 f\left({y\Phi\over m}\right)
$$ 
where $f$ is a {\it holomorphic} function of its argument.  This
fact is crucial.  We know that when $y=0$, $W_{eff}=\half m\Phi^2$, so
$f(0)=1$.  Therefore the mass term in the superpotential (we will
soon see how important it is to say ``in the superpotential'')
cannot be renormalized even when it is nonzero; any attempt
to correct $m$ with a factor of $y$ is always accompanied by
a field $\Phi$, which means this correction does not give a
contribution to $\Phi^2$.

We also determined just a moment ago that when $m\to 0$, $W= \third
y\Phi^3$, so $f(|u|\to\infty) = 2u/3$.  This in turn guarantees there
can be no correction to the coefficient of $\Phi^3$.

What about $\Phi^4$?  Its coefficient is $y^2/m$ times a number, which
might be zero.  Let's consider first what graphs might lead to a
corresponding $\phi^2\psi^2$ term in the Lagrangian.  The mass term
means that we can draw a non-vanishing tree graph, the first in
\refiggg{nonrenorm}, which is proportional to $m$, with two
$\phi\psi\psi$ vertices and one chirality-flipping mass insertion,
with a factor of $1/m^2$ coming from the propagator of the virtual
$\psi$.  However, we are interested in quantum effective actions, to
which tree graphs do not contribute.  To get a quantum contribution to
a $\phi^2\psi^2$ term, we see that we need {\it more} than two
$\phi\psi\psi$ vertices.  But there's the rub; this means that the
coefficient of this term in the quantum superpotential is proportional
at least to $y^4$, or more precisely (if you look at the second
diagram in \refiggg{nonrenorm} carefully) $y^2|y|^2$.  This is in
contradiction to the general form of the superpotential; therefore
this term must vanish.  And so on, for all perturbative contributions
to terms in the superpotential.  To all orders in perturbation theory,
$f(u) = f_{classical}(u) = 1+{2\over 3}u$, and $W_{eff} =
W_{classical}$.

\begin{figure}[th]
\begin{center}
 \centerline{\psfig{figure=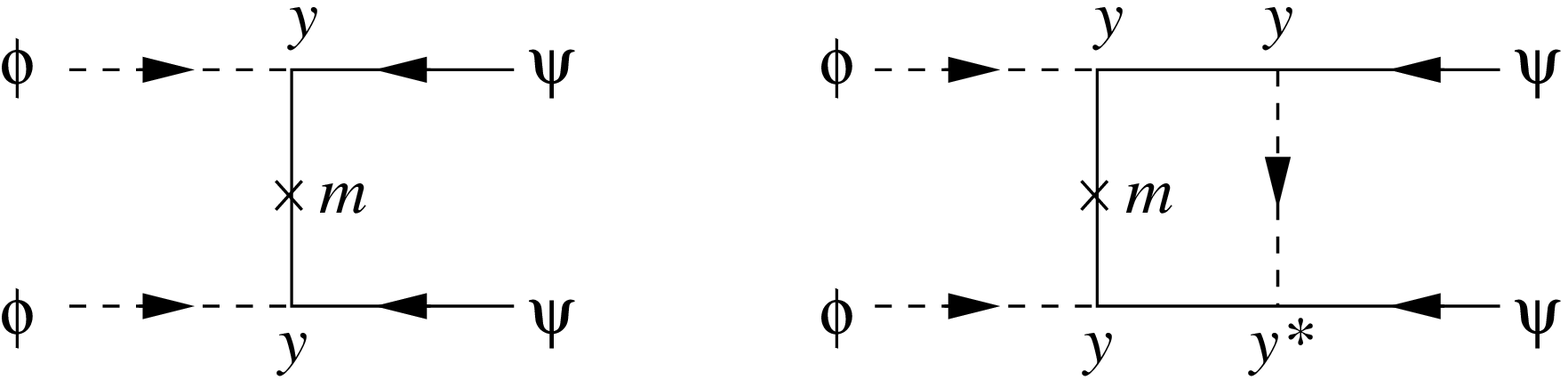,width=7cm,clip=}}
 \caption{\small The first diagram has the right form, but is classical;
the second diagram contributes to the quantum effective action, but has
the wrong form. }
\figgg{nonrenorm}
\end{center}
\end{figure}

But how far can we carry this argument?  What about non-perturbative
corrections?  We know these corrections must be very small in the limit that
$y$ is very small and $m$ is finite; therefore $f(u)\approx 1+\frac23 u$,
to all orders in $u$,
even nonperturbatively near $u=0$.  And we still know that $f(u)\to
\frac23 u$ as $u\to\infty$, because our arguments using the real
symmetry of the $m=0$ case left no room, in that case, for an unknown
function even non-perturbatively.  Now, the claim is that there are no
holomorphic functions {\it except} $f_{classical}$ which have these
properties --- and therefore $f=f_{classical}$ {\it exactly}.

Note holomorphy is essential here, as is the fact that we know the
superpotential when $|u|$ is large but has arbitrary phase.  For instance,
holomorphy 
rules out functions such as
$$
e^{-1/|u|^2} + 
{\frac23} u
$$
whereas our constraint on $u\to\infty$ rules out functions such as
$$
e^{-1/u^2} + 
{\frac23} u
$$
which has the wrong behavior for small imaginary $u$.  

Fantastic.  The superpotential for this theory gets no quantum
corrections; the coupling constants appearing there are unaltered.  It
would seem, then, naively, that the coupling constants of this theory
do not run.  But this sounds wrong. We have already argued that all
the coupling constants of the theory should have logarithmic
divergences in $d=4$; has supersymmetry eliminated them?  And should
there be no finite renormalizations whatsoever?  Indeed, this is far
too facile.  The effective superpotential is well under control, but
the effective K\"ahler potential is not.  The latter potential is {\it
real}, so it can contain real functions of $y^*y$ and $m^*m$ appearing
all over the place.  Consequently we cannot make any strong statements
about its renormalization.  But how does this affect the coupling
constants?

The resolution of this puzzle is that the coupling $y$ appearing in
$W(\Phi)$ is not a physical quantity.  Let us rename it $\hat y$.  By
construction it is a holomorphic quantity.  Note that we can change it
by redefining our fields by $\Phi\to a\Phi$, where $a$ is any complex
constant.  This changes $\hat y$, and it also changes the K\"ahler
potential, making the kinetic terms noncanonical.  Physical quantities
(such as the running coupling constants, as measured, say, in scattering
amplitudes at particular scales) must be independent of such
field redefinitions.  To define physical quantities, we should be more
careful.  Let us take the K\"ahler potential to have the form
$$
K(\Phi^\dagger,\Phi) = Z\Phi^\dagger \Phi \ .
$$
$Z$ gives the normalization of the wave-function of $\Phi$.  Note that
a propagator representing an incoming or outgoing particle state
should be $i/(k^2-m^2)$.     Thus the presence of the factor
$Z$, giving $i/Z(k^2-m^2)$ for the propagator, implies that 
we will have to take care in normalizing Green functions, a point
we will return to shortly.  

The rescaling of $\Phi$ by a factor $a$ changes $\hat y$ by $a^{-3}$
and $Z$ by $|a|^{-2}$.  A natural definition of an invariant coupling
is then the quantity ${|y|}^2 \equiv \hat y^\dagger Z^{-3} \hat y$,
which is clearly invariant under field redefinitions of the sort we
were just considering.  The coupling $|y|$ is physical but
non-holomorphic, in contrast to the unphysical but holomorphic $\hat
y$.

But now we see how renormalization of the {\it wave function} ---
divergences or even finite renormalizations which affect the kinetic
terms in the K\"ahler potential --- can in turn renormalize {\it
physical} coupling constants.  While $\hat y$ cannot be renormalized
and become a scale-dependent function, $Z$ can indeed become a
scale-dependent function of $\mu$.  In fact, we may expect that the
graph in \refiggg{oneloopZ}
\begin{figure}[th]
\begin{center}
 \centerline{\psfig{figure=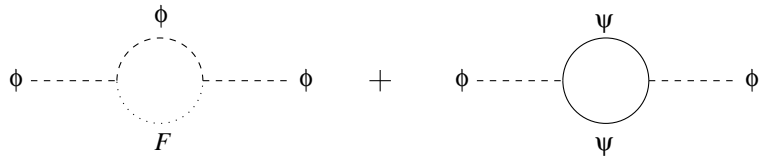,width=10cm,clip=}}
 \caption{\small One-loop diagrams contributing to $Z(\mu)$. }
\figgg{oneloopZ} 
\end{center}
\end{figure}
will renormalize the wave function of $\Phi$ by
\bel{oneloopZeq}
Z(\mu) = 1 + \tilde c_0 {|y|^2\over 16\pi^2} \log(\mu/\Lambda_{UV})
= Z(\mu_0) + \tilde c_0 {|y|^2\over 16\pi^2} \log(\mu/\mu_0)
\ee
where $\tilde c_0$ is a constant of order 1 and $\mu_0$ is an
arbitrary scale.  Shortly we will determine
the sign of $\tilde c_0$.

We may now determine the scaling behavior of the physical coupling $y$
(again, to be distinguished from the holomorphic coupling $\hat y$
appearing in the superpotential.)
$$
\beta_{{|y|}^2} = { y^*}\beta_y + y\beta_{y^*} 
= -3 { |y|}^2 \dbyd{\ln Z}{\ln \mu}.
$$

Let us define the {\it anomalous mass dimension} $\gamma(y)$
of the field $\Phi$ by
$$
\gamma = -\dbyd{\ln Z}{\ln \mu} \  
$$
which tells us how $Z$ renormalizes with energy scale.

\EX{ Why should we think of this as an ``anomalous dimension''?  What
is the relation between $Z(\mu)$ and the dimension of a field?  Show
$\dim \Phi = 1 + \half \gamma$.}

Then we have an {\it exact} relation
\bel{betay}
\beta_y = {3\over 2} y\gamma(y)
\ee
We can use \Eref{oneloopZeq}
at small $y$ to find the approximate result
$$
\gamma = -\tilde c_0 {|y|^2\over 16\pi^2} + {\rm order \ }(|y|^4) \
\Rightarrow
\beta_y = - {\frac32} 
\tilde c_0 y{|y|^2\over 16\pi^2} + {\rm order \ }(|y|^4) \ ,
$$ but when $y$ is larger we have no hope of computing $\gamma(y)$,
and therefore none of computing $\beta_y$.  Fortunately, even though
$\gamma(y)$ itself is an unknown function, relations such as
\eref{betay} can be extremely powerful in and of themselves, as
we will see shortly.

Let us understand where this relation \eref{betay} came from by looking
at diagrams.  \refiggg{FullZ} shows the full propagator,
which is proportional to $Z^{-1}$; therefore
we must normalize the external fields 
in any physical process by a factor of $1/\sqrt{Z}$.  
\begin{figure}[th]
\begin{center}
 \centerline{\psfig{figure=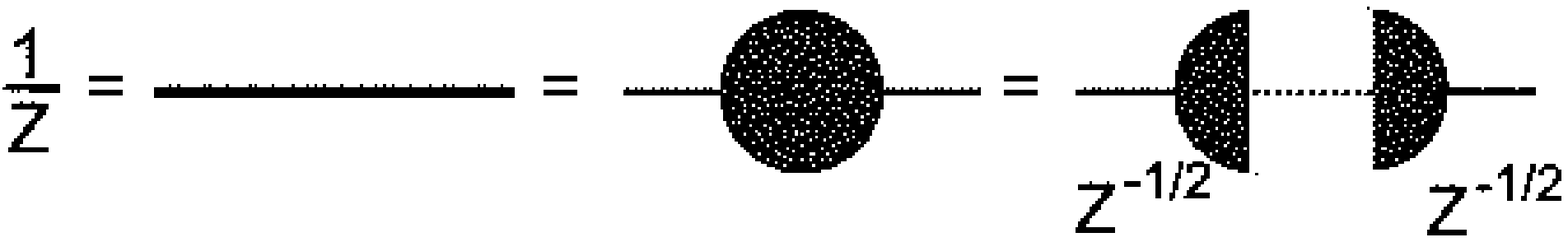,width=10cm,clip=}}
 \caption{\small The full propagator is proportional to $Z^{-1}$;
external fields must then be normalized with a factor of $Z^{-1/2}$. }
\figgg{FullZ}
\end{center}
\end{figure}
The graphs contributing
to the physical $|y|^2$ take the form of \refiggg{yphysical}; 
but supersymmetry eliminates all
corrections to the holomorphic $\Phi^3$ vertex, making the graphs
much simpler --- and (remembering that we must normalize
the fields) proportional to $Z^{-3}$.
\begin{figure}[th]
\begin{center}
 \centerline{\psfig{figure=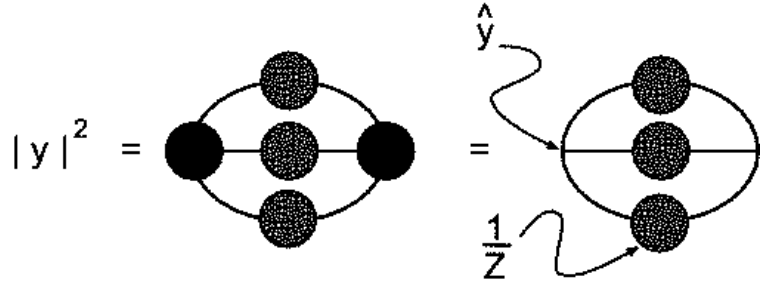,width=10cm,clip=}}
 \caption{\small The physical coupling $|y|^2$ gets quantum corrections only
 from $Z$ (grey circles); the vertex factors (black circles) get no
 quantum contributions, since the holomorphic coupling $\hat y$ is
 unrenormalized.}
\figgg{yphysical}
\end{center}
\end{figure}

As an important aside, let me note that chiral superfields have a very
special property, namely that products of chiral superfields have {\it
no short-distance singularities!}  In contrast to
expectations from non-supersymmetric field theories, composite
operators built from chiral fields (but no antichiral or real
fields)
have the property that they may defined without a short-distance
subtraction.  The dimension of such a composite is the sum of the
dimensions of its components.

\EX{ Show that $\beta_y = y[\dim(\Phi^3) -\dim W]
= y[3\ \dim(\Phi) -\dim W].$}

Now, there is a very important theorem which we may  put to use.
Near any conformal fixed point (including a free field theory) all
gauge invariant operators $\OO$ must have dimension greater than or
equal to 1 (or more generally, $(d-2)/2$).  If its dimension is 1 (or
more generally, $(d-2)/2$), then $\nabla^2\OO = 0$ (i.e., the operator
satisfies the Klein-Gordon equation.)  This is true without any appeal
to supersymmetry!

The theorem applies to the scalar field $\phi$ which is the lowest
component of $\Phi$.  {\it Therefore, $\gamma\geq 0$; and $\gamma=0$
if and only if $y=0$.}  From this we may conclude that
$$
\gamma(y) = c_0 {|y|^2\over 16\pi^2} + \ {\rm order\ }( |y|^4)
$$
where $c_0$ is a {\it positive} 
constant.  This in turn implies $\beta_y>0$, and so
$y$ flows to zero in the far infrared.  

\EX{ Calculate $c_0$.}

Thus, rather than being a conformal field theory, as it was
classically, with an {\it exactly} marginal coupling $y$, the quantum
$W=y\Phi^3$ theory flows logarithmically to a free conformal field
theory with $y=0$.  We refer to $y$ as a {\it marginally irrelevant}
operator; it is marginal to zeroth order in $y$, but when $y$ is
nonzero then $\beta_y>0$.  The quantum
renormalization group flow of the theory with nonzero $y$ and $m$ is
shown in \refiggg{ClQuPhicubed}. 
\begin{figure}[th]
\begin{center}
 \centerline{\psfig{figure=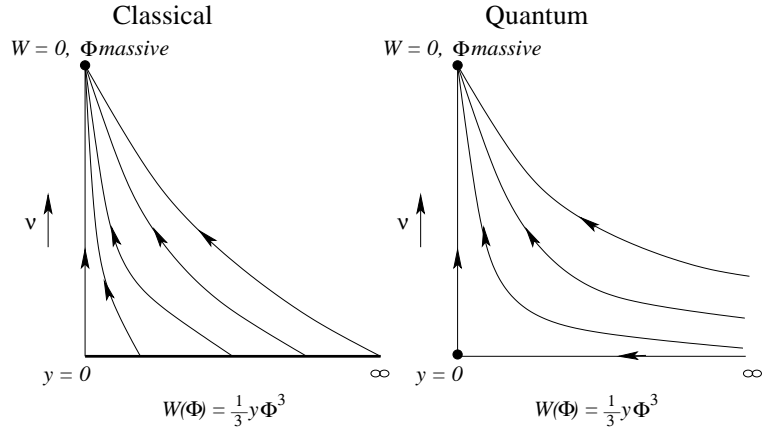,width=10cm,clip=}}
 \caption{\small The scale-invariance of $y$ is lost through quantum
effects; even if $m=0$ the coupling $y$ flows. }
\figgg{ClQuPhicubed}
\end{center}
\end{figure}

\subsection{Wess-Zumino model in $d=3$}
Now, what do we expect to happen in three dimensions?  Here the
formula \eref{betay} is not really appropriate, because it leaves out
the {\it classical} dimension of $y$.  Perturbation theory can only be
done in dimensionless quantities, so we should study not $y$ but
$\omega = y/\sqrt{\mu}$.  Already we notice a problem.  The coupling
$\omega$ is large, classically, when $\mu\ll 1/y^2$, so perturbation
theory can't possibly work in the infrared!  At long distances this
theory will automatically be strongly coupled, unless large quantum
effects change the scaling of $\omega$ drastically.  But quantum
effects will generally be small unless $\omega$ is large --- so this
can't happen self-consistently.

Let's be more explicit.  The beta function for $\omega$ is
$$
\beta_\omega = \omega \left[-\half + {\frac32}\gamma(\omega)\right]
$$ 
Again $\gamma$ must be positive (by the above theorem) and a
perturbative power series in $\omega$, beginning at order
$\omega^2/16\pi^2$.  It has a large negative beta function (meaning it
grows toward the infrared) and will only stop growing if
$\gamma(\omega) = \third$.  However, this can only occur if
$\omega/4\pi\sim 1$, so a one-loop analysis will be insufficient by
the time this occurs. Consequently, the most important behavior of the
theory will occur in regimes where the perturbative expansion is 
breaking down, and nonperturbative effects might be
important.  We cannot expect perturbation theory to tell us everything
about this theory, and specifically we cannot reliably calculate
$\gamma(\omega)$.

However, suppose that there is {\it some} coupling $\omega_*$ for
which $\gamma(\omega_*) = \third$.  This need not be the case; it
could be that $\gamma<\third$ for all values of $\omega$.  But if it
is the case, then at $\omega_*$ the beta function of the dimensionless
coupling $\omega$ vanishes, and the theory becomes truly scale
invariant.  (Notice that the beta function for $y$ is nonzero there;
but scale invariance requires that {\it dimensionless} couplings not
run.)  In fact, since $\gamma<\third$ for $\omega<\omega_*$, and since
(barring a special cancellation) we may therefore expect
$\gamma>\third$ for $\omega>\omega_*$, the beta function for $\omega$
is negative below $\omega_*$ and positive above it.  The
renormalization group flow for $\omega$ then is illustrated in
\refiggg{omegaflow}.  The point $\omega=\omega_*$ is a stable infrared
fixed point; if at some scale $\mu$ the physical coupling $\omega$
takes a value near $\omega_*$, then, at smaller $\mu$, $\omega$ will
approach $\omega_*$.\footnote{We know a scale can be generated in a
classically scale-invariant theory; this is called ``dimensional
transmutation'' and is familiar from QCD.  Here we have a theory with
a classical scale $\hat y^2$; but this scale is {\it lost} quantum
mechanically.  This equally common phenomenon is often called
``dimensional antitransmutation.''}

\begin{figure}[th]
\begin{center}
 \centerline{\psfig{figure=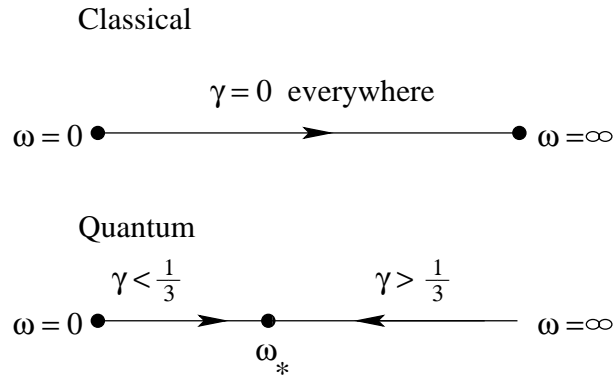,width=8cm,clip=}}
 \caption{\small The scale-dependence of
the coupling $\omega = y/\sqrt{\mu}$ is lost, due
to large quantum corrections, at the value $\omega_*$.  }
\figgg{omegaflow}
\end{center}
\end{figure}

\EX{ What is the behavior of the wave function  $Z(\mu)$
once this fixed point is reached?}

At this conformal fixed point, the field $\Phi$ will have
dimension $\frac23$.  Thus the operator $\Phi^4$, when appearing in
the superpotential, represents not a marginal operator of dimension 2
(as in the free theory) but an irrelevant operator of dimension
${\frac83}>2$.

\EX{ For the mass term $m\Phi^2$, what is the dimension of $m$ at
the fixed point?}

Should we skeptical of the existence of this fixed point?  No;
massless nonsupersymmetric $\lambda\phi^4$ theory has a fixed point
for $d= 4-\epsilon$ dimensions, with $\lambda\sim\epsilon$.  This
``Wilson-Fisher'' fixed point is easy to find in perturbation theory.
It has been verified that it continues all the way to $d=3$
dimensions.  A similar analysis can be done for a complex scalar and
two-component complex fermion coupled as in \eref{Yukawa}; this too
shows a similar fixed point.

As another example, a theory (the $O(N)$ model) with $N$ massless
scalar fields $\phi_i$ and a potential $V = \lambda (\sum_i
\phi_i^2)^2$ can be shown, directly in $d=3$ and at leading order
in a $1/N$ expansion, to have a nontrivial fixed point $\lambda\sim
1/N$.  A corresponding analysis can be done for a supersymmetric
theory with $N$
chiral fields $\Phi_i$ and a single chiral field $X$, with
superpotential $hX(\sum_i \Phi_i^2)$.

\EX{ Verify the claims of the previous paragraph for the $O(N)$ model.
Then try the supersymmetric theory with $W(X,\Phi_i) = hX(\sum_i \Phi_i^2)$;
show it has a fixed point.  Compute
the anomalous dimensions of $X$ and $\Phi$ at the fixed point
(hint --- do the easiest one,
then use conformal invariance to determine the other.)  Work
only to leading nonvanishing order in $1/N$.
You may want to see Coleman's Erice lectures on the $O(N)$ model.}

Thus the putative fixed point at $\omega=\omega_*$ is very plausible,
and we will assume henceforth that it exists.  Note that at this fixed
point the effective theory is highly nonlocal.  It has no particle
states (the propagator $\vev{\phi^\dagger(x)\phi(0)}=x^{-4/3}$ does
not look like a propagating particle or set of particles, which would
have to have integer or half-integer dimension.)  In this and all
similar cases the superpotential and K\"ahler potential together are
insufficient.  Indeed it is very difficult to imagine writing down an
explicit Lagrangian for this fixed point theory.  Even in two
dimensions it is rarely known how to write a Lagrangian for a
nontrivial fixed point, although in two dimensions there are direct
constructive techniques for determining the properties of many
conformal field theories in detail.  In $d=3$ there are no such
techniques known; we have only a few pieces of information, including
the dimension of $\Phi$.

\subsection{Dimensions and R-charge}
In fact there is another way (which at first will appear trivial) to
determine the dimension of $\Phi$ at the fixed point.  This requires
looking a bit more closely at the symmetries of the Lagrangian
\eref{Yukawa}.  In particular, for $W=\frac13\hat y\Phi^3$, we saw
earlier that there is an R-symmetry
$$
\phi\to\phi e^{2i\alpha/3} \ ; \ \psi\to\psi e^{-i\alpha/3}\ ; 
\ F\to F e^{-4i\alpha/3}
$$
which is the unique R-symmetry of the theory.  Under this symmety $\Phi$
has charge $2/3$ and the superpotential has charge $2$.

At a conformal fixed point, there is a close relation between the
dimensions of many chiral operators and the R-charges that they carry.
The energy-momentum tensor (of which the scale-changing operator, the
``dilation'' or ``dilatation generator'', is a moment) and the current
of the R-charge are actually part of a single supermultiplet of
curents.  At a conformal fixed point, the dilation current and the R
current are both conserved quantities, and the superconformal algebra
can then be used to show that the dimension of a chiral operator is
simply $(d-1)/2$ times its R charge.  This implies that at any fixed
point with $W\propto\Phi^3$, the dimension of $\Phi$ is $1$ in $d=4$,
$2/3$ in $d=3$, and $1/3$ in $d=2$.  (In all cases, $\dim \Phi = \dim
\phi = \dim\psi-\half = \dim F-1$.) Since the dimension of $\Phi$ must
be strictly greater than $1$ $(1/2)$ $[0]$ at any nontrivial fixed
point in $d=4(3)[2]$, it follows that there can be nontrivial fixed
points in two or three dimensions, but not for $d=4$.

\subsection{The quantum XYZ model in $d=3$}
\label{subsec:QXYZ}
Now let us turn to some other fixed points.  If we have three fields
$X,Y,Z$ and superpotential $W=\hat hXYZ$, then by symmetry
$\gamma_X=\gamma_Y= \gamma_Z\equiv\gamma_0$.  In four dimensions there
can again be no nontrivial fixed point; the coupling $h$ is marginally
irrelevant.  But in three dimensions, it is relevant and interesting
dynamics can be expected.  In particular, the coupling $\eta =
h/\sqrt\mu$ has
$$
\beta_\eta =
\half\eta[-1+\gamma_X(\eta)+\gamma_Y(\eta)+\gamma_Z(\eta)] =
\half\eta[-1+3\gamma_0(\eta)] \ .
$$
As in the previous theory we see that
if there exists some $\eta_*$ for which
$\gamma_0(\eta_*)=\third$, then the theory is conformal there.  Let us
assume $\eta_*$ exists and the flow is as given in \refiggg{etaflow}.  

\begin{figure}[th]
\begin{center}
 \centerline{\psfig{figure=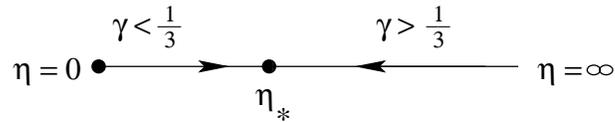,width=8cm,clip=}}
 \caption{\small A fixed point for $\eta$. }
\figgg{etaflow}
\end{center}
\end{figure}

\EX{ List the marginal and relevant operators at this fixed point.
Note that some apparently relevant operators are actually just
field redefinitions and are not actually present; for example,
the addition of $mXY$ to the superpotential $W=hXYZ$
has no effect, since we may simply redefine $Z \to Z-m$ and eliminate
the coupling altogether.  Such eliminable operators are called
``redundant.''}

Now we can consider the effect of adding other
operators to the theory.  For example what happens if we add
$\omega_x\sqrt{\mu}X^3$ to the theory?

\EX{ Prove that $\omega_x$ is a marginally irrelevant coupling.  To do
this use the facts that (1) at $\eta=0$, $\omega_x\neq 0$, we know
$\gamma_X>0 =\gamma_Y=\gamma_Z$, and (2) $\gamma_X-\gamma_Y$, a
continuous real function of the couplings, is known to be zero when
$\omega_x = 0$.}

Since $\omega_x$ (and similarly $\omega_y$ and $\omega_z$, when we add
them in as well) are
marginally irrelevant couplings, we may wonder if this fixed point, 
located at $(\eta,\omega_x,\omega_y,\omega_z) = (\eta_*,0,0,0)$, is
an isolated point
in the space of the four coupling constants.  In fact, the answer is no.
Examine the four beta functions
\begin{eqnarray}
\beta_{\eta} &=& \half \eta(-1+\gamma_X+\gamma_Y+\gamma_Z)  \cr
\cr 
\beta_{\omega_x} &=& \half\omega_x (-1 + 3\gamma_X) \cr
\cr
\beta_{\omega_y} &=& \half\omega_y (-1 + 3\gamma_Y) \cr
\cr
\beta_{\omega_z} &=& \half\omega_z (-1 + 3\gamma_Z) 
\end{eqnarray}
where $\gamma_X,\gamma_Y,\gamma_Z$ are functions of the four
couplings.  We see that a condition for all four beta functions to
vanish simultaneously puts only {\it three} conditions on the
anomalous dimensions $\gamma_X,\gamma_Y,\gamma_Z$.  Specifically, the
conditions are $\gamma_X(\eta,\omega_x,\omega_y,\omega_z)=\third$ and
similarly for $\gamma_Y$ and $\gamma_Z$.  Three conditions on four
couplings imply that any solutions occur generally on one-dimensional
subspaces (and since these couplings are complex, the subspace is
one-complex-dimensional in extent.)  Since the three anomalous
dimensions must be equal on this subspace, the symmetry permuting the
three fields is presumably unbroken on it.  Let us therefore take
$\omega_x=\omega_y=\omega_z=\omega_0$ and examine the anomalous
dimension $\gamma_0(\eta,\omega_0)$.  \refiggg{etaomegaflow} indicates
the renormalization group flow of the couplings.  Notice that there is
a {\it line} of conformal field theories ending at $\eta=\eta_*,
\omega_0=0$ and extending into the $\eta,\omega_0$ plane.  The line
ends at $\eta=0, \omega_0=\omega_*$, clearly the same $\omega_*$ as in
the $W=\third\hat{y}\Phi^3$ model (since for $\eta=0$ we have three
noninteracting copies of the latter model)  The precise location of
this line is totally unknown, since we do not know
$\gamma_0(\eta,\omega_0)$; but if $\omega_*$ and/or $\eta_*$ exists,
then the line must exist also. We can define a new coupling
$\rho(\eta,\omega_0)$ which tells us where we are along this line.
This coupling is called an ``exactly marginal coupling,'' and the
operator to which it couples is called an ``exactly marginal
operator.''

\begin{figure}[th]
\begin{center}
 \centerline{\psfig{figure=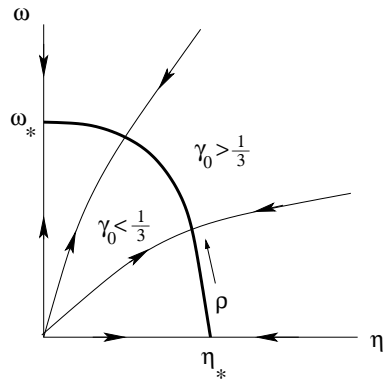,width=5cm,clip=}}
 \caption{\small The (complex) line of fixed points lies at
 $\gamma_0(\omega,\eta)=\frac13$ and may be parameterized by a single
 (complex) variable $\rho$.}
\figgg{etaomegaflow}
\end{center}
\end{figure}

\EX{ Argue there are no nontrivial fixed points in $d=4$ in  
supersymmetric theories with only chiral superfields (and 
no gauge interactions.)  Use the fact that since all chiral superfields
are gauge invariant operators, their dimensions are greater than one.
Then 
consider all possible
interactions which are relevant at the free fixed point and might drive
the theory to a nontrivial fixed point.  Note loopholes in your proof.}

\section{Abelian gauge theories}

\subsection{The classical theory}

Time to turn to gauge theories.  Gauge bosons are contained in vector
supermultiplets, given by superfields $V$, which are real and contain,
in $d=4$, a real vector potential $A_\mu$, a Majorana fermion
$\lambda_a$ called a ``gaugino'', and a real auxiliary field $D$.  In
three dimensions, the only change is that the vector potential has one
less component, which is made up by the presence of a single real
scalar field $\varphi$. All of these are in the adjoint representation
of the gauge group.  In the case of $U(1)$, which we now turn to, they
are all neutral.\footnote{ Instead of $V$, it is often convenient to
use $W_\alpha$, a superfield containing the gaugino $\lambda$, the
field strength $F^{\mu\nu}$, the auxiliary field $D$, and (in $d=3$)
$\varphi$.  This object transforms homogeneously under gauge
transformations (and is gauge invariant in the abelian case.)  There
is yet another useful superfield in $d=3$ but I'll skip that here.}

The kinetic terms of the pure $U(1)$ theory are
$$
S_{gauge} = {1\over e^2} \iddx \left[ -{1\over 4}F_{\mu\nu}^2 - 
i\bar\lambda\dslash\lambda
+\half D^2\left\{+ \half (\partial_\mu\varphi)^2\right\}\right]
$$
The last term is absent in four dimensions.  Notice I have normalized
all of the fields with a $1/e^2$ out front; this is convenient for
many purposes.  However, since $e^2$ has mass dimension $4-d$, this
means that I have made the dimension of the gauge field somewhat
unusual.  In four dimensions, it has dimension 1, like any free
bosonic field, but in $d=3$, the gauge field and scalar {\it also}
have dimension 1, in contrast to the scalars in the chiral multiplet
which were normalized with dimension $\half$.  This choice is
arbitrary; but we will see soon why this is physically convenient.

It is instructive to count degrees of freedom.  Accounting for gauge
invariance but not the equations of motion, the gauge boson has $d-1$
degrees of freedom, the Majorana fermion $4$ real degrees of freedom
(we will write them as 2 complex), the auxiliary field has 1 and the
scalar field has $4-d$; thus there are four bosonic and four fermionic
degrees of freedom.  After the equations of motion, the gauge boson
has $d-2$, the fermion $2$, and the scalar $4-d$, so there are two
bosonic and two fermionic {\it propagating} degrees of freedom.

What we have defined above is the vector supermultiplet of a theory
with four supersymmetry generators.  The chiral multiplet also
comes from such a theory.  The fact that the Majorana fermion
has four real degrees of freedom is related to the number of
supersymmetry generators. Confusingly, this much supersymmetry is
known as \none\ in $d=4$ and \ntwo\ in $d=3$.  This is because in four
dimensions there is {\it one} gaugino, while in $d=3$ the gaugino
defined above is actually a reducible spinor, representing {\it two}
copies of the smallest possible spinor.

\subsection{Extended supersymmetry}
There are other supersymmetries, each with their own
vector and matter multiplets.  A
theory with {\it eight} supersymmetry generators has two Majorana
spinors in $d=4$ and four of the smallest spinors in $d=3$; it is
therefore called \ntwo\ in $d=4$ and \nfour\ in $d=3$. Its vector multiplet
contains one vector multiplet plus one chiral multiplet (both in the adjoint)
from the case of four generators. Altogether it
contains a gauge boson $A_\mu$, {\it two} Majorana fermions
$\lambda,\psi$, a complex scalar $\Phi$, and three real auxiliary
fields $D,{\rm Re}\ F, {\rm Im}\ F$, as well as (in three dimensions
only) a real scalar $\varphi$.  (With this much supersymmetry there is
another multiplet, called a hypermultiplet, consisting of two chiral
multiplets of opposite charge under the gauge symmetry; more on this
below.)  A theory with 16 supersymmetry generators --- the maximum
allowed without introducing gravity --- has only vector multiplets,
each of which contains one vector
multiplet and three chiral multiplets of the 4-generator case ({\it
i.e.}, one vector multiplet and one hypermultiplet of the 8-generator
case,) all in the adjoint representation.  It is called \nfour\ in
$d=4$ and \neight\ in $d=3$.  Finally, in $d=3$ there are some cases
with 2, 6 and 12 supersymmetry generators, which we will not have time to
discuss.

In these lectures we will use the language of $d=4$ \none\ (which is
almost the same as $d=3$ \ntwo) even to describe the other cases.
This is common practise, since there is little convenient
superfield notation for more than four supersymmetry generators.

\begin{equation}
\label{properties}
\begin{tabular}[h]{|l||c|c|c|c|c|}
\hline
{Number of SUSY generators}  & $d=3$ & $d=4$           \\
\hline
 4         &   \ntwo\  &    \none\   \\
\hline
 8         &    \nfour\  &  \ntwo\   \\
\hline
 16         &    \neight\  &    \nfour\                      \\
\hline
\end{tabular}
\end{equation}

\subsection{The gauge kinetic function}
Before proceeding further it is important to mention the $\theta$
angle in $d=4$ gauge theories.  In the $d=4$ action, we should also
include a term\footnote{We will not discuss the dimensional reduction
of this object to three dimensions.}
$$
\int\ d^4x\ {\theta\over 32 \pi^2} F^{\mu\nu}\tilde F_{\mu\nu} = 
\int\ d^4x\ {\theta\over 32 \pi^2} 
F^{\mu\nu} F^{\rho\sigma} \epsilon_{\mu\nu\rho\sigma}
$$
a term to which electrically charged objects are insensitive but which
strongly affects magnetically charged objects.  In fact we should define a
generalized holomorphic gauge coupling
\bel{taudef}
\tau \equiv {1\over 2\pi} \left[\theta+ i{8\pi^2\over e^2}\right] \ .
\ee
Then we may write the action as 
$$
S_{gauge} =
{i\tau\over 8\pi}\int\ d^4x\ \left[{1\over 4}(F^2+iF\tilde F) + 
i\bar\lambda\dslash\lambda - {1\over 2}D^2\right]  \  
+  \  {\rm hermitean
\ conjugate.}
$$

Even this is not sufficiently general.  Consider, for example, adding
a neutral chiral multiplet $\Phi$ to a theory with a $U(1)$ vector
multiplet $V$.  The complex scalar $\phi$ can have an expectation
value.  In principle, just as the low-energy QED coupling in nature
depends on the Higgs expectation value through radiative effects, the
gauge coupling for $V$ could depend functionally on $\vev\phi$.  In
other words, we could write a theory of the form
\bel{gaugekinetic}
{i\over 8\pi}\int\ d^4x\ \tau(\phi)\left[{1\over 4}(F^2+iF\tilde F) + 
i\bar\lambda\dslash\lambda - {1\over 2}D^2\right]  
\  +  \  {\rm hermitean
\ conjugate.} + \cdots
\ee
where the dots indicate the presence of many other terms required
by supersymmetry, which I will neglect here.  Since $\tau$ is a
holomorphic quantity, it  must be a holomorphic function
of the chiral superfield $\Phi$.  We will refer to this new 
holomorphic function as the
``gauge kinetic function.''
Thus, to define our gauge theory, we need to specify at least a
superpotential, a gauge kinetic function, and a K\"ahler potential;
the first two are holomorphic, and the latter is real.

We can modify the theory of $V$ and $\Phi$ to have $d=4$ \ntwo\
invariance.  To do this, we must make sure that the K\"ahler potential
$K(\Phi,\Phi^\dagger)$ and the gauge kinetic function $\tau(\Phi)$ are
related such that the two gauginos of the \ntwo\ vector multiplet (one
of which, in the above notation, is in the \none\ vector multiplet,
while the other is in the multiplet $\Phi$) have the same kinetic
term.  The simplest theory with \ntwo\ has $\tau$ a constant and $K =
(1/g^2) \Phi^\dagger \Phi$.   There is
no superpotential in this theory; the moduli space is simply the
complex $\phi$ plane.

\EX{Derive the above-mentioned condition!}

The \nfour\ $U(1)$ gauge theory in four dimensions has three
complex scalars $\phi_i$, $i=1,2,3$, from its three \none\ chiral
multiplets, and no superpotential.  This means it has a moduli space
which is simply six dimensional unconstrained flat space, with an
$SO(6)$ symmetry rotating the six real scalars into each other.  This
is an R-symmetry, since the four fermions of \nfour\ are spinors of
$SO(6)$, while the vector boson is a singlet and the scalars are in
the ${\bf 6}$ representation.

What about in three dimensions?  The \ntwo\ vector multiplet has a single
real scalar, so the classical moduli space is simply the real line.
Similarly, the \nfour\ vector multiplet has three real scalars, and
the \neight\ vector multiplet has seven.  But quantum mechanically
this will not be the whole story.  To see why, we must discuss
duality.

\subsection{Dualities in three and four dimensions}
The pure Maxwell theory in $d=4$ has a famous symmetry between its
electric and magnetic fields.  One may phrase this as follows: given
physical electric and magnetic fields $E$ and $B$, one may find a
gauge potential $A_\mu$ with $F_{\mu\nu} = \partial_\mu A_\nu -
\partial_\nu A_\mu$, and one may {\it also} find a potential $C_\mu$
with $F_{\mu\nu} = \epsilon_{\mu\nu}^{\ \ \rho\sigma} (\partial_\rho
C_\sigma - \partial_\sigma C_\rho)$.  The electric fields of one
potential are the magnetic fields of the other.  Notice {\it both}
gauge potentials have separate $U(1)$ gauge invariances, $A_\mu\to
A_\mu+\partial_\mu\varphi$ and $C_\mu\to C_\mu + \partial_\mu\chi$;
both invariances are unphysical, since the physical fields $E$ and $B$
are unaffected by them.  Always remember that gauge symmetries are
{\it not} physical symmetries; they are {\it redundancies} introduced
only when we simply our calculations by replacing the physical $E$ and
$B$ by the partly unphysical potential $A_\mu$ (or $C_\mu$).  The two
gauge invariances simply remove the unphysical longitudinal parts of
$A$ and $C$.  ``Electric-magnetic'' duality exchanges the electric
currents (to which $A$ couples simply) with magnetic currents (to
which $C$ couples simply) and as such exchanges electrically charged
particles with magnetic monopoles.  We know from Dirac that the charge
of a monopole is $2\pi/e$, so this transformation must exchange $e^2$
with $4\pi^2/e^2$ --- weak coupling with strong coupling --- and more
generally $\tau \to -{1\over \tau}$.  (In modern parlance, this kind
of duality, whose quantum version is discussed in more detail in my
Trieste 2001 lectures, is often called an S-duality.)

In $d=3$ there is also an electric-magnetic duality, but it does not
exchange a gauge potential with another gauge potential.  $E$ and $B$
are both three-vectors in $d=4$, but in $d=3$ the electric field is a
two-vector and $B$ is a spatial scalar.  We may exchange the
two-vector with the gradient of a scalar $\sigma$, and $B$ with its
time derivative; thus $F_{\mu\nu} = \partial_\mu A_\nu - \partial_\nu
A_\mu = \epsilon_{\mu\nu\rho}\partial^\rho\sigma$ in three dimensions.
Now $\sigma\to\sigma+c$ is a global symmetry. But the expectation
value of the scalar $\sigma$ represents a new degree of of freedom,
one which spontaneously breaks this symmetry.  The Goldstone boson of
this breaking, $\partial \sigma$, is just the photon that we started
with.  It can be shown that $\sigma$ is periodic and takes values only
between $0$ and $2\pi e^2$.  Thus, when determining the moduli space
of the pure $d=3$ \ntwo\ $U(1)$ gauge theory, we need to include not
only the scalar $\varphi$ but also $\sigma$, and in particular these
two combine as $\varphi+i\sigma$ into a complex field $\Sigma$.  Since
$\sigma$ is periodic, the moduli space of the theory --- the allowed
values for $\vev{\Sigma}$ --- is a {\it cylinder}, shown in
\refiggg{Sigmaspace}.  (Similarly, the \nfour\ theory in $d=3$ actually
has a four-dimensional moduli space, while the \neight\ theory has an
eight-dimensional moduli space.)  Note the cylinder becomes the entire
complex plane in the limit $e\to\infty$, which in $d=3$ (since $e^2$
has mass dimension 1) is equivalent to the far infrared limit; thus
the theory acquires an {\it accidental} $SO(2)$ symmetry, as in figure
\refiggg{Sigmaspace}.  Similarly, the \nfour\ and \neight\ cases have
$SO(3)$ enhanced to $SO(4)$ and $SO(7)$ enhanced to $SO(8)$ in the
infrared.\footnote{There is one more duality transformation in $d=3$
that {\it does} exchange one gauge potential with another. Identify
the gauge field strength as the dual of another gauge field
$F_{\mu\nu} = \partial_\mu A_\nu - \partial_\nu A_\mu =
\epsilon_{\mu\nu\rho} V^\rho$ (which looks gauge non-invariant, but
read on.)  The equation of motion $\partial_\mu F^{\mu\nu} = J_e^\nu$,
where $J_e$ is the conserved electric current, tells us that $
J_e^\rho = \epsilon^{\mu\nu\rho}F^V_{\mu\nu} $ (and conservation of
$J_e$ is the Bianchi identity for $F^V$.)  An electric charge (nonzero
$J_e^0$) corresponds to localized nonzero magnetic field $F^V_{12}$
for $V$.  An object carrying nonzero magnetic field is a magnetic
vortex.  Thus unlike electric-magnetic duality, which exchanges
electrically charged particles and magnetic monopoles, which are
particles in $3+1$ dimensions, this duality transformation exchanges
electrically charged particles and magetic vortices, which are
particles in $2+1$ dimensions.  For this reason we will call it
``particle-vortex'' duality; see Intriligator and Seiberg (1996).}

\begin{figure}[th]
\begin{center}
 \centerline{\psfig{figure=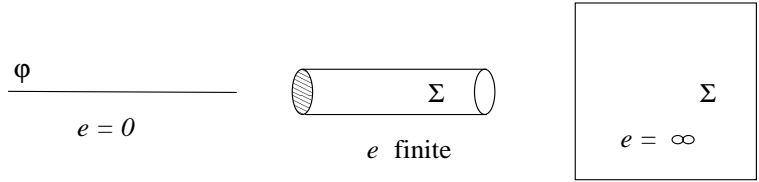,width=10cm,clip=}}
 \caption{\small The moduli space is a cylinder of radius $\propto g$.  }
\figgg{Sigmaspace}
\end{center}
\end{figure}

I have discussed these duality transformations in the context of
pure $U(1)$ gauge theories.  However, it is easy to extend them
to the supersymmetric case, since all the superpartners in
an abelian vector multiplet are gauge-neutral.  I will do so
as we need them.

\subsection{Classical $d=4$ \none\ SQED}
Now we are ready to add matter to the theory.  Let us add $N_f$ chiral
multiplets $Q_r$ of charge $k_r$ and $N_f$ $\tilde Q^s$ of charge
$-\tilde k_s$.  The superpotential $W(Q_r,\tilde Q^s)$ and the gauge
kinetic function $\tau(Q_r,\tilde Q^s$) must be holomorphic functions
of gauge-invariant combinations of the chiral multiplets.  For now,
classically, we will take $W=0$ and $\tau = 4\pi i/e^2$.  The kinetic
terms for the charged fields are modified by the gauge interactions.
Taking a canonical K\"ahler potential for simplicity, we have
\begin{eqnarray}
S_{kin} &= \sum_r\ \int\ d^dx\  \Big[D_\mu q_r^\dagger D^\mu q_r
+i\bar\psi_r \Dslash \psi_r
 + F_r^\dagger F_r + \nonumber \\
&k_r(\lambda \psi_r q_r^\dagger 
 + \bar \lambda \bar \psi_r q_r + q_r^\dagger D q_r) \Big]
\nonumber \\ \nonumber \\
&+\sum_s\ \iddx 
 \Big[
D_\mu \tilde q^{s\dagger} D^\mu \tilde q^s
+i\overline{\tilde \psi^s} \Dslash \tilde \psi^s + \tilde F^{s\dagger} 
\tilde F^s \nonumber \\
&-\tilde k_s
(\overline\lambda\overline{ \tilde\psi^s} \tilde q^s 
+  \lambda \tilde\psi^s \tilde q^{s\dagger} 
+\tilde q^{s\dagger}
D \tilde q^s)
\Big] \cr
\end{eqnarray}
Here $Q_r$ contains the superfields $q_r, \psi_r, F_r$ and similarly
for $\tilde Q^s$; the covariant derivative is $D_\mu = \partial_\mu + i k
A_\mu$ acting on a particle of charge $k$ (remember that the coupling
$e$ appears not here but in the kinetic term of the gauge boson); and
$D$ with no index is the auxiliary field in the vector multiplet.

In the special case where all the $k_r=k_s=1$, then the $Q$'s are
rotated by an $U(N_f)$ global symmetry and the $\tilde Q$'s are
rotated by a different $U(N_f)$ symmetry; let us call then $U(N_f)_L$
and $U(N_f)_R$ in analogy with terminology in QCD.  The diagonal
$U(1)_V$ which rotates $Q$ and $\tilde Q$ oppositely is the symmetry
which is gauged (and so is a redundancy, not a true symmetry.)  The
gauge-invariant chiral operators take the form $M_r^s \equiv Q_r\tilde
Q^s$; all other gauge-invariant combinations of chiral superfields
reduce to products of the $M_r^s$ fields.

Since there is a term $D^2$ in the vector multiplet kinetic terms
\eref{gaugekinetic}, we see that there will be a new contribution to
the potential energy of the theory.  In particular, for a canonical
K\"ahler potential and $\tau$ a constant the potential will be
$$
V(q_r, \tilde q_s) = {1\over 2g^2} D^2 + \sum_r |F_r|^2 + \sum_s
|\tilde F_s|^2
$$
where the equation of motion for $D$ reads 
$$
D = \sum_r k_r |q_r|^2 - \sum_s k_s |\tilde q^s|^2
$$
Again, a supersymmetric vacuum must have $V=0$, and therefore all the
auxiliary fields must {\it separately vanish.}

The condition $D=0$ is very special.  Let us consider first the
simplest possible case, namely $N_f=1$, with $Q$ and $\tilde Q$ having
charge 1 and -1.  Although there are two complex fields, with four
degrees of freedom, the gauge symmetry removes one of these, since we
may use it to give $q$ and $\tilde q$ the same phase.  The real
condition $D = |q|^2 - |\tilde q|^2 =0$ removes one more degree of
freedom and ensures that both $q$ and $\tilde q$ have the same {\it
magnitude}.  In fact, it acts as though the gauge invariance of the
theory were complexified! at least as far as the moduli space of the
theory is concerned.  The moduli space is then given by one complex
parameter $v = \vev{q}= \vev{\tilde q}$, which we may also write in
gauge invariant form as $v^2 = \vev{M} = \vev{Q\tilde Q}$.  In short,
the moduli space is simply the complex $M$ plane.  We began with two
chiral multiplets; only one is needed to describe the moduli
space.\footnote{We have assumed so far that the superpotential is
zero.  A superpotential
$W=mQ\tilde Q$ simply gives the chiral multiplets masses, leaving only
the massless vector multiplet and its unique vacuum at $q=\tilde q=0$.
If we add a superpotential $W = y(Q\tilde Q)^2$,
the resulting potential again has a vacuum only at $q=\tilde q=0$,
but $Q$ and $\tilde Q$ are massless there. }

Why did one chiral multiplet of freedom have to disappear?  Well, when
$M$ is nonzero, the gauge group is broken, and as we know very well,
the photon is massive.  But a massive gauge boson has to absorb a
scalar field to generate its third polarization state, as we know from
the electroweak Higgs mechanism.  However, in supersymmetry it must be
that a vector multiplet must absorb an entire chiral multiplet;
otherwise there would be partial multiplets left over, which would
violate supersymmetry.  One massive photon means that one of the two
massless chiral multiplets has paired up with the massless vector
multiplet; the remaining fields form the massless and neutral chiral
multiplet $M$.

How about $N_f=2$, with $Q_1,Q_2$ of charge 1 and $\tilde Q_1,\tilde
Q_2$ of charge $-1$, and no superpotential?  In this case the
condition $D = |q_1|^2 + |q_2|^2- |\tilde q_1|^2 - |\tilde q_2|^2 =0$
(combined with gauge invariance) leaves three massless chiral
multiplets.  It turns out that the solution to this equation is
$M_1^1M_2^2 = M_2^1M_1^2$.  Thus the {\it four} gauge invariant
operators $M_r^s$, subject to the constraint $\det M=0$, give us the
three-complex-dimensional moduli space.

\EX{ Verify that $\det M=0$ is the solution to the above equation.
Hint; use the $SU(2)\times SU(2)$ flavor symmetry to rotate the vevs
into a convenient form.}

\EX{ Verify that for $N_f>2$  the D-term
constraints imply that the gauge-invariant operators $M_r^s$,
subject to the constraint that $M$ be a matrix of rank zero or one,
parameterize the moduli space.}

\subsection{\ntwo\ $d=4$ SQED}
\label{subsec:ddddNN}

Now let us slightly complicate the story by considering the \ntwo\
$d=4$ gauge theory.  The \ntwo\ vector multiplet has an extra chiral
multiplet $\Phi$.  The fields $Q_r$ and $\tilde Q^s$ can be organized
into $N_f$ hypermultiplets in which the indices $r$ and $s$ should now
be identified.  The global $U(N_f) \times U(N_f)$ symmetry will now be
reduced to a single $U(N_f)$, because of the superpotential 
$$
W(\Phi, Q_r,\tilde Q^r) = \sqrt 2\Phi\sum_r Q_r\tilde Q^r
$$
required by the \ntwo\ invariance.
In normalizing the superpotential this way, I have assumed that the
kinetic terms for $\Phi$ are normalized
$$
K = {1\over e^2} \Phi^\dagger \Phi
$$
to agree with the normalization of the kinetic terms of the \none\
vector multiplet $V$.  Sometimes it is more convenient to normalize
$\Phi$ canonically; then a factor of the gauge coupling $e$ appears
in front of the superpotential.

Let us begin with the case $N_f=1$.  Now we have several
conditions\footnote{Henceforth we will not distinguish between
superfields and their scalar components, since it is generally clear
from context which is relevant; also we will generally write $\Phi$ to
represent $\vev{\Phi}$.}
$$
D = |Q|^2-|\tilde Q|^2 = 0 \ ; \ F^\dagger_\Phi= Q\tilde Q = 0 \ ; \ 
F^\dagger = \Phi Q =0; \ \tilde F^\dagger = \Phi\tilde Q=0 \ ,
$$ which clearly have no solution with nonzero $Q$ and/or $\tilde Q$.
In fact, in the language of the operator $M=Q\tilde Q$, which we know
satisfies the D-term conditions, the $F_\Phi$ equation explicitly says
$M=0$, while $Q\tilde F^\dagger$ gives $\Phi M = 0$. This allows
any nonzero $\Phi$.
When only $\Phi$ is nonzero, the
gauge group is unbroken, so the photon is massless
and the electric potential of a point charge
is $1/r$ at large $r$.  For this reason, this
branch of moduli space is called the ``Coulomb branch.''

Does this structure make sense?  Suppose $Q$ and $\tilde Q$ were
nonzero, so that the Higgs mechanism were operative; would this be
consistent?  As before, were the vector multiplet to become massive it
would have to absorb an entire charged multiplet, which in this case
would have to be the entire hypermultiplet.  This would leave no
massless fields to serve as moduli.  Therefore this theory cannot have
a branch of moduli space on which the photon is massive.  By
constrast, the vector multiplet scalars $\Phi$ can have expectation
values without breaking the gauge symmetry at all; instead $\Phi$
simply makes $Q$ and $\tilde Q$ massive, while itself remaining
massless.  Thus the only branch of moduli space in this theory is the
Coulomb branch, in the form of the complex $\Phi$ plane, with a
special point at the origin where the hypermultiplet is
massless. 

Now consider $N_f=2$.  We can expect that there will again 
be no obstruction
to having $\vev\Phi\neq 0$; such an expectation value will
make the hypermultiplets massive, preventing them from having
expectation values.  We can also expect that if the charged scalars do
have expectation values, they will make the vector multiplet massive,
preventing $\Phi$ from having an expectation value; and since only one
hypermultiplet will be eaten by the vector multiplet, there should be
an entire hypermultiplet --- two chiral superfields --- describing the
moduli space.  Is this true?  As in the case
of \none\ $N_f=2$ SQED, the D-term conditions are
satisfied by using the operators $M_r^s$ with $\det M =0$.  The
conditions
$$
F^\dagger_\Phi= Q_1\tilde Q_1 + Q_2\tilde Q_2 = 0 \ ; \ 
F^\dagger_r = \Phi Q_r=0 ; \ \tilde F^\dagger_r = \Phi\tilde Q_r=0
$$
can be rewritten as 
$$
F^\dagger_\Phi = \tr \ M =0 \ ; Q_s\tilde F^\dagger_r = \Phi M_r^s = 0
$$
which indeed imply that either $\Phi$ or $M$ can be nonzero, but not
both simultaneously, and that there are two complex degrees of freedom
in $M$ which can be nonzero, making a single neutral hypermultiplet.

\begin{figure}[th]
\begin{center}
 \centerline{\psfig{figure=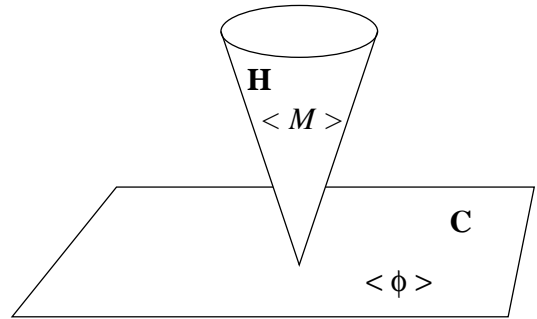,width=7cm,clip=}}
 \caption{\small The classical two-complex-dimensional Higgs branch ({\bf H}) 
and the one-complex-dimensional Coulomb branch ({\bf C}) of 
$d=4$ \ntwo\ $U(1)$ with two charged hypermultiplets. }
\figgg{ddNNFF}
\end{center}
\end{figure}

This last example illustrates the branch structure of these theories,
indicated schematically in \refiggg{ddNNFF}.  Either $\Phi$ is
nonzero, with the hypermultiplets massive and the gauge group unbroken
(the Coulomb phase), or $M$ is nonzero, with the gauge group broken
(the Higgs phase) and only some massless neutral hypermultiplets
remaining.  In this case the Coulomb branch {\bf C} has complex
dimension 1 while the ``Higgs branch'' {\bf H} has complex dimension 2
(in fact quaternionic dimension 1.)  The two branches meet at the
point where all the fields are massless.

\EX{ Show that for $N_f>2$ \ntwo\ $U(1)$ gauge theory, this Higgs and
Coulomb branch structure continues to be found, with the quaternionic
dimension of the Higgs branch being $N_f-1$.}

\subsection{Classical $d=3$ \ntwo$,4$ SQED}
\label{subsec:dddNN}

In three dimensions, the physics is slightly more elaborate, because
even the \ntwo\ multiplet has a real scalar field $\varphi$.  This
means that even for \ntwo\ $N_f=1$ $U(1)$ gauge theory with no
superpotential, there is a Coulomb branch with $\vev{\varphi}$
nonzero, in addition to the Higgs branch with nonzero
$\vev{M}=\vev{Q\tilde Q}$.  This is in contrast to the $d=4$ \none\
$N_f=1$ $U(1)$ gauge theory, which has only a Higgs branch.  The
Coulomb branch of the $d=3$ \ntwo\ theory is similar, classically, to
the one we encountered in Sec.~\ref{subsec:ddddNN} 
when we considered the pure
$d=4$ \ntwo\ abelian gauge theory.  

The details of the moduli space are controlled
partly by a new term in the Lagrangian
\bel{newinteraction}
\varphi^2(|Q^2|+|\tilde Q|^2) \ .
\ee
The existence of this term can be inferred from the $d=4$ \none\ gauge
theory as follows.  When we go from four dimension to three, the
component of the photon $A_3$ becomes the scalar $\varphi$.  All
derivatives $\partial_3$ are to be discarded in this dimensional
reduction, but {\it covariant} derivatives $D_3 = \partial_3 + i A_3$
become $i\varphi$.  The interaction \eref{newinteraction} is simply
the dimensional reduction of the $d=4$ kinetic term $|D_3Q|^2
+|D_3\tilde Q|^2$.  For \eref{newinteraction} to be zero, it must
always be that either $\varphi=0$ or $Q=\tilde Q=0$; this gives two
branches.  Note that the charged chiral multiplets are massive on the
Coulomb branch, as usual, because of this quartic potential.  The
classical moduli space is shown in \refiggg{dNN}.

\begin{figure}[th]
\begin{center}
 \centerline{\psfig{figure=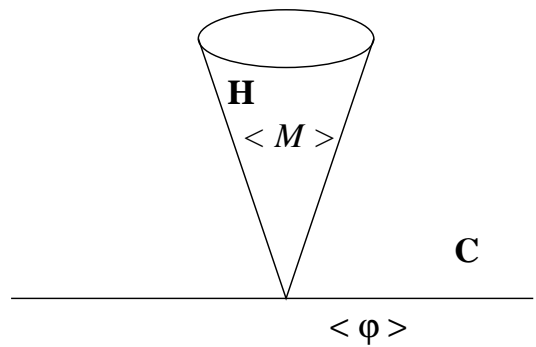,width=7cm,clip=}}
 \caption{\small The classical one-complex-dimensional Higgs branch 
and the one-real-dimensional Coulomb branch  of 
$d=3$ \ntwo\ $U(1)$ with $N_f=1$.  }
\figgg{dNN}
\end{center}
\end{figure}

What about the case of \nfour\ supersymmetry?  Here the $N_f=1$ case
is not so confusing, because there is no Higgs branch; in $d=4$ \ntwo\
there was a one-complex-dimensional Coulomb branch, whereas here there
should be (classically!) a three-real-dimensional Coulomb branch, made from
the scalars $\varphi,{\rm Re}\ \phi, {\rm Im}\ \phi$.  There is an
$SO(3)$ symmetry acting on the scalars (which, as usual for theories
with more than four supercharges, is an extended R-symmetry.)

\subsection{Quantum SQED in $d=4$}
To go further, we have to do some quantum mechanics.  Let's begin with
perturbation theory.

In four dimensions, perturbation theory is familiar.  Just as
electrons generate a positive logarithmic running for the
electromagnetic coupling, via the one-loop graph above, so do scalar
charged particles; and the combination of $N_f$ chiral multiplets
$Q_r$ of charge 1 and $N_f$ $\tilde Q_s$ of charge $-1$ gives a one-loop
beta function
$$
\beta_e = {e^3\over 16\pi^2} N_f 
$$
It is often convenient to write formulas not for $e$ but for $1/e^2$:
$$
\beta_{{1\over e^2}} = -{N_f\over 8\pi^2} \ \Rightarrow
{1\over e^2(\mu) } = 
{1\over e^2(\mu_0) } + {N_f\over 8\pi^2} \ln\left({\mu_0\over\mu}\right)
$$
so $e$ shrinks as we head toward the infrared.  This means
that $e$ becomes large in the ultraviolet, which means that perturbation
theory breaks down there, making it difficult to define the theory.  We can
avoid this problem by defining the theory with some additional Pauli-Villars
regulator fields, $N_f$ ghost chiral superfields of charge 1 and $N_f$ of
charge -1, all of mass $M$.  In this case there are no charged fields above
the scale $M$, so $\beta_{e}=0$; thus
for $\mu>M$ the gauge coupling is a constant $e_0$, and
$$
{1\over e^2(\mu) } = 
{1\over e^2_0} + {N_f\over 8\pi^2} \ln\left({M\over\mu}\right)
$$
for $\mu<M$.
(Remember this is only accurate at one-loop, so it only makes sense if 
$e_0\ll 1$)
If the fields $Q$ and $\tilde Q$
have masses $m$ then (as in real-world QED) the gauge coupling will stop
running at the scale $m$, as illustrated in \refiggg{eruns}.

\begin{figure}[th]
\begin{center}
 \centerline{\psfig{figure=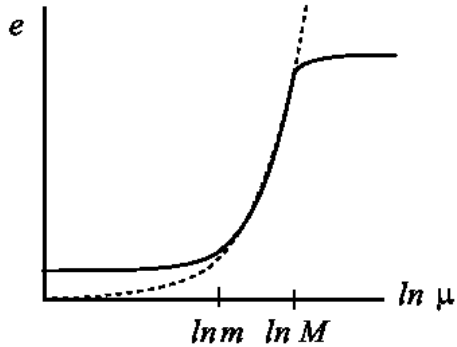,width=6cm,clip=}}
 \caption{\small The coupling runs (approximately) logarithmically
below $M$ and above $m$.}
\figgg{eruns}
\end{center}
\end{figure}

Now, the scale $M$ was just put in to regulate the theory, while
$e^2(\mu)$ is physical for low $\mu$ and should not depend on $M$.  We
therefore should define a physical scale $\Lambda$ by taking it to be
the value of $M$ where the one-loop coupling $e_0$ is formally
infinite (though remember it will differ from the real coupling at
that scale due to higher-loop effects)
$$
{1\over e^2(\mu) } =   {N_f\over 8\pi^2} \ln\left({\Lambda\over\mu}\right)
$$
where
$$
\Lambda^{N_f} = \mu^{N_f} e^{+8\pi^2/e^2(\mu)}
$$

I've been careless here: $m$ and $M$ are complex parameters (while
$\mu$ is real\footnote{You can make it complex, actually --- this is
itself interesting but beyond what I can cover here.}) so $\Lambda$
should be complex also; but $e^2$ is real.  How can we make the above
expressions sensible?  Clearly, we should introduce the $\theta$
angle, and rewrite the previous equation in its final form as
$$
-2\pi i \tau(\mu) \equiv {8\pi^2\over e^2(\mu)}  
- i \theta = N_f  \ln\left({\Lambda\over\mu}\right)
$$
which is a one-loop formula that is only sensible for $\mu\ll\Lambda.$

Note that $\tau$ is a holomorphic coupling constant.  Let us verify
that in perturbation theory this one-loop formula is exact!  In
perturbation theory the expression for $\tau$ must be a perturbative
series in $e^2\propto 1 / {\rm Im}\ \tau$ and cannot contain
$\theta\propto {\rm Re}\ \tau$; but that's impossible if it is to be a
holomorphic expression.  The only term which can appear in quantum
corrections to $\tau$ must then be $\tau$-independent, namely the one
we see above.

But as before, the fact that we have found a simple 
formula for a holomorphic quantity by no means indicates that the
physical coupling is so simple.  The physical coupling gets corrections
from higher loop effects (caution: as we will see, these are cancelled for
theories with eight or more supersymmetry generators).  As before,
these can only occur in the nonholomorphic part of the theory: the K\"ahler
potential.  As we did for the coupling $y$ in Sec.~\ref{subsec:QWZ}, 
we should find a definition
of the coupling constant which is independent of field redefinitions.
We'll come back to this point soon.

Let's first examine $d=4$ \ntwo\ $U(1)$ gauge theory with $N_f$
massless hypermultiplets.  We've already discussed its branch
structure in Sec.~\ref{subsec:ddddNN}; 
there is a Higgs branch on which some of the
gauge-invariant operators $M_r^s = Q_r\tilde Q^s$ act as massless
fields, with the others massive.  There is also a Coulomb branch on
which $\Phi$ has an expectation value and the term $\sqrt 2\Phi
Q_r\tilde Q_s$ in the superpotential gives the charged fields masses.
The branch structure for $N_f>1$ massless hypermultiplets was shown in
\refiggg{ddNNFF}.  On the Coulomb branch, we can ask an interesting
physical question: how does the infrared limit
\be
\tau_L \equiv
\lim_{\mu\to 0}\tau(\mu)
\ee
of the coupling constant $\tau$ depend on $\Phi$?  Since (1) the
theory has a $\Phi$-independent value of $\Lambda$, and (2) for any
value of $\Phi$, the coupling constant stops running at the scale
$\Phi$ at which the charged fields are massive,
$$
-2\pi i\tau_L = N_f  \ln\left({\Lambda\over\Phi}\right) \ \ \
[\Phi \ll \Lambda] \ .
$$
This is singular only at $\Phi=0$, where the Higgs and Coulomb branches
meet and the charged fields are massless.  (Recall that charged
massless fields always drive the electric coupling $e$ to zero, and thus
$\tau\to i\infty$, in the infrared.)  Note we cannot take $|\Phi|$ to
be larger than $|\Lambda|$, since our description of the theory is not
reliable there.  The behavior of $\tau_L$ on the moduli space is
sketched in \refiggg{ddNNFFq}.

\begin{figure}[th]
\begin{center}
 \centerline{\psfig{figure=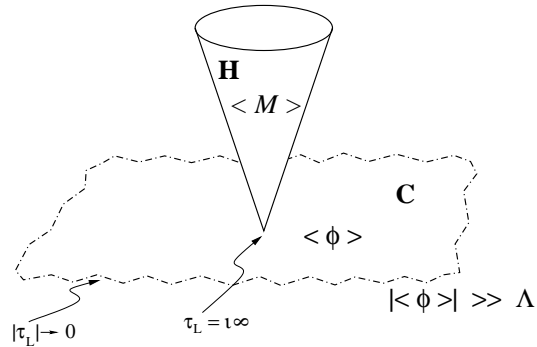,width=7cm,clip=}}
 \caption{\small The quantum version of \refiggg{ddNNFF}; the low-energy
 coupling $e(\mu\to 0)$ grows from $0$ at the origin to $\infty$ at
 the dashed line, where the description of the theory breaks down.}
\figgg{ddNNFFq}
\end{center}
\end{figure}

\EX{ For $N_f=4$,  if two hypermultiplets have mass $m$ and two have
mass $m'$, show that
$$
-2\pi i\tau_L = 2 \ln\left({\Lambda\over\Phi+m}\right)
+2 \ln\left({\Lambda\over\Phi+m'}\right)
$$
so that there are two singular points; at each singular point there are two
massless hypermultiplets, which have a Higgs branch intersecting the Coulomb
branch at that point.  If each hypermultiplet has its own mass, then show
that there are four singular points but {\it no} Higgs branches anywhere.}

What does this function $\tau_L$ really tell us?  Away from the
singular points, for any value of $\Phi$, the charged fields are all
massive, and there is simply a pure $U(1)$ \ntwo\ gauge theory in the
infrared.  Its effective action is of the form \eref{gaugekinetic}
with a nontrivial gauge kinetic function $\tau_L(\Phi)$.  But
like any pure abelian gauge theory, it has duality symmetries.  In
particular, there is the electric-magnetic transformation
$\tau\to-{1\over\tau}$.  There is also the obvious symmetry
$\tau\to\tau+1$, which represents a shift by $2\pi$ of the $\theta$
angle.  These two symmetry transformations generate a group of duality
transformations of the form $SL(2,\ZZ)$, the symmetry group of a
torus.

\begin{figure}[th]
\begin{center}
 \centerline{\psfig{figure=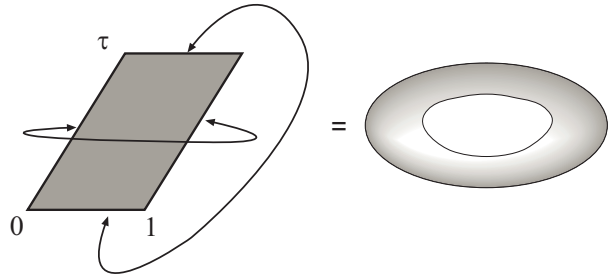,width=8cm,clip=}}
 \caption{\small A torus is a parallelogram with opposite sides identified.
 }
\figgg{torus}
\end{center}
\end{figure}

A torus can be defined by taking a parallelogram and identifying
opposite sides, as in \refiggg{torus}.  Ignore the size of the
parallelogram by taking one side to have length 1; then the other size
has a length and angle with respect to the first that can be specified
by a parameter $\tau$ that lives in the upper-half of the complex
plane.  (For the gauge theory, the gauge coupling must be positive, so
${\rm Im}\ \tau >0$.)  However, exchanging the two sides obviously
leaves the torus unchanged ($\tau\to -{1\over\tau}$) as does shifting
one side by a unit of the other side ($\tau\to\tau+1$) and any
combination of these transformations.

One can therefore take the point of view that the low-energy \ntwo\
$U(1)$ gauge theory should not be specified by $\tau$.  In fact, we
can see this by taking $\Phi\to\Phi e^{2\pi i}$; that is, let us
circle the singular point at $\Phi=0$.  The theory obviously must come
back to itself, since the physics depends only on $\Phi$; but $\tau_L$
shifts by $N_f$ as we make the circle.  Thus $\tau_L$ does not
properly characterize the theory; all values of $\tau_L$ related by
$SL(2,\ZZ)$ transformations are actually giving the same theory.  We
should therefore characterize the low-energy theory by specifying a
torus!  For each value of $\Phi$, there should be a torus with
parameter $\tau_L(\Phi)$ which tells us the properties of the
low-energy theory.  More precisely, this is a fiber bundle, with a
torus fibered over the complex $\Phi$ plane, as expressed in
\refiggg{ddNNC}.  This torus is invariant under $\Phi\to\Phi e^{2\pi
i}$, and becomes singular at $\Phi=0$.

\begin{figure}[th]
\begin{center}
 \centerline{\psfig{figure=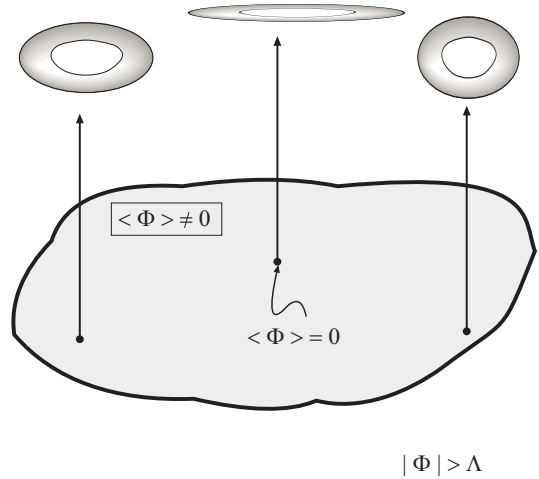,width=7cm,clip=}}
 \caption{\small The low-energy gauge coupling and its dualities are best
understood using a torus fibered over the Coulomb branch; at
the origin, where $\tau\to i \infty$, the torus degenerates.}
\figgg{ddNNC}
\end{center}
\end{figure}

\subsection{Quantum SQED in $d=3$}

Now let's move back to three dimensions.  The gauge coupling is now
dimensionful, so classically it has a negative beta function.  Due to
the wonders of gauge symmetry, the diagram in \refiggg{finitefig}
\begin{figure}[th]
\begin{center}
 \centerline{\psfig{figure=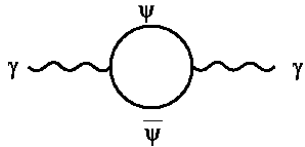,width=4cm,clip=}}
 \caption{\small This one-loop graph is finite in three dimensions.  }
\figgg{finitefig}
\end{center}
\end{figure}
is ultraviolet finite!  But not trivial.  In fact, if the fermion
is masssless, and the momentum flowing through the photon
line is $p^\mu$, this graph is
proportional to ${1\over \sqrt{p^2}}$!

\EX{ Calculate the one-loop correction to ordinary nonsupersymmetric
QED in three dimensions for $N_f$ massless electrons.}

This means that the one-loop gauge coupling in three dimensions has the form
$$
{1\over e^2(\mu)} = {1\over e^2_0} + c_3{N_f\over \mu}
$$ 
where $c_3$ is a positive constant, of order one, which depends
on the specific theory.
Notice that there is no divergence in $e$ as $\mu\to\infty$; $e$
goes to a constant $e_0$ in the ultraviolet, so in $d=3$
supersymmetric QED is well-defined in the ultraviolet.

As always, to define a beta function we should employ a dimensionless
coupling
$$
\zeta\equiv {e^2(\mu)\over \mu} = \left({\mu\over e^2_0} + c_3N_f\right)^{-1}
$$ 
which is infinite for large $\mu$ but --- interestingly --- goes
(at one loop) to a {\it constant} at small $\mu$.  In other words,
$$
\beta_\zeta = {\mu\over e_0}^2\left({\mu\over e^2_0} + c_3N_f\right)^{-2}
 = \zeta(1-c_3N_f\zeta)
$$
so there is a fixed point in the one-loop formula at $\zeta_*=1/c_3N_f$.
This is illustrated in \refiggg{dddxi} and \refiggg{xiflow}.  

\begin{figure}[th]
\begin{center}
 \centerline{\psfig{figure=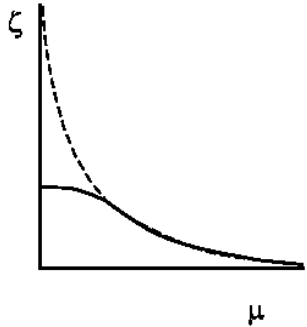,width=4cm,clip=}}
 \caption{\small $\zeta$ as a function of scale $\mu$; the dashed line
shows its classical flow. }
\figgg{dddxi}
\end{center}
\end{figure}

\begin{figure}[th]
\begin{center}
 \centerline{\psfig{figure=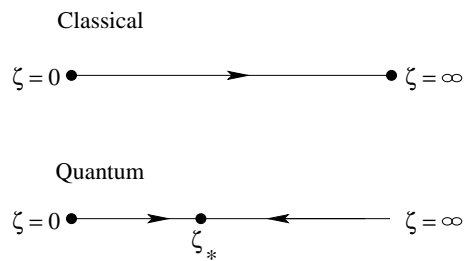,width=6cm,clip=}}
 \caption{\small The coupling $\xi$ has a quantum fixed point. }
\figgg{xiflow}
\end{center}
\end{figure}

This is very interesting. Remembering that perturbation theory is an
expansion in the parameter $e^2/\mu = \zeta$, we see that the one-loop
formula has a fixed point {\it at weak coupling} if $N_f$ is large.
If this is true, then for large $N_f$ two-loop effects such as those
in \refiggg{diags} are always
suppressed by factors of $1/N_f$ and can be neglected.  Thus the
large-$N_f$ behavior of the theory is indeed given by the one-loop
formula, with a fixed point at small $\zeta$ and all diagrams
calculable.  The theory is soluble!

\begin{figure}[th]
\begin{center}
 \centerline{\psfig{figure=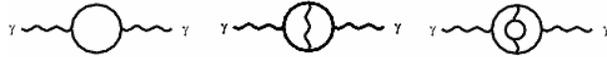,width=8cm,clip=}}
 \caption{\small In $d=3$ $N_f\gg 1$ (S)QED
the one-loop correction $\propto e^2 N_f$
dominates the propagator; the higher-loop corrections are
suppressed by extra powers of $e^2\propto 1/N_f$, and can be dropped. }
\figgg{diags}
\end{center}
\end{figure}

\EX{ Show nonsupersymmetric QED in $d=3$ is soluble
and has a conformal fixed point at large $N_f$.}

In nonsupersymmetric QED, it is believed that there is a value of $N_f$
below which the fixed point disappears and other nonperturbative
phenomena take place.  In \ntwo\  and \nfour\  QED, however,
the fixed point visible at one loop survives for all $N_f$. 
In fact, in the latter case, there are no higher-loop corrections to the
gauge coupling, so the above beta function is exact and the
fixed point at small $N_f$ is completely reliable.

 Let us examine the $N_f=1$ case in both \ntwo\ and \nfour\
supersymmetry.  Both of these theories are truly remarkable.  As we
noted, for $d=3$ a vector boson has a electric-magnetic dual
pseudoscalar $\sigma$.  In \ntwo, this scalar combines with the scalar
$\varphi$ to make a complex field $\Sigma=\varphi+i\sigma$.  The
scalar $\sigma$ is compact, with radius $2\pi e^2$, so classically the
field $\Sigma$ takes values on a cylinder of radius $e^2$.  As $e^2\to
0$ the field $\sigma$ disappears and the cylinder becomes the
$\varphi$ line; when $e^2\to\infty$ the cylinder expands to be become
the entire plane, as we saw in \refiggg{Sigmaspace}.

But in the presence of charged matter, $e^2$ is not so simple.
In particular, the terms
$$
\varphi^2(|Q^2|+|\tilde Q|^2) + \varphi(\bar \psi\psi 
-\overline{\tilde\psi}\tilde\psi)
$$
imply that the charged matter has mass $\varphi$.  Since
$e^2(\mu)$ stops running below this scale, the low energy
value $e^2_L$ of the gauge coupling is
$$
{1\over e^2_L} = {1\over e^2_0} + {1\over \varphi}
$$
For very large $|\varphi|$ we have $e_L\approx e_0$, so
the radius of the cylinder on which $\Sigma$ lives is of order $e_0^2$
far from $\varphi = 0$.  However, for very small $\varphi$ the
$1/e_0^2$ term can be neglected and $e^2_L\sim \varphi$; thus the
cylinder shrinks in size.  At $\varphi = 0$ --- where the Higgs branch
meets the Coulomb branch --- the cylinder shrinks to zero radius.  Thus,
the moduli space has the form of \refiggg{dNNq}.
\begin{figure}[th]
\begin{center}
 \centerline{\psfig{figure=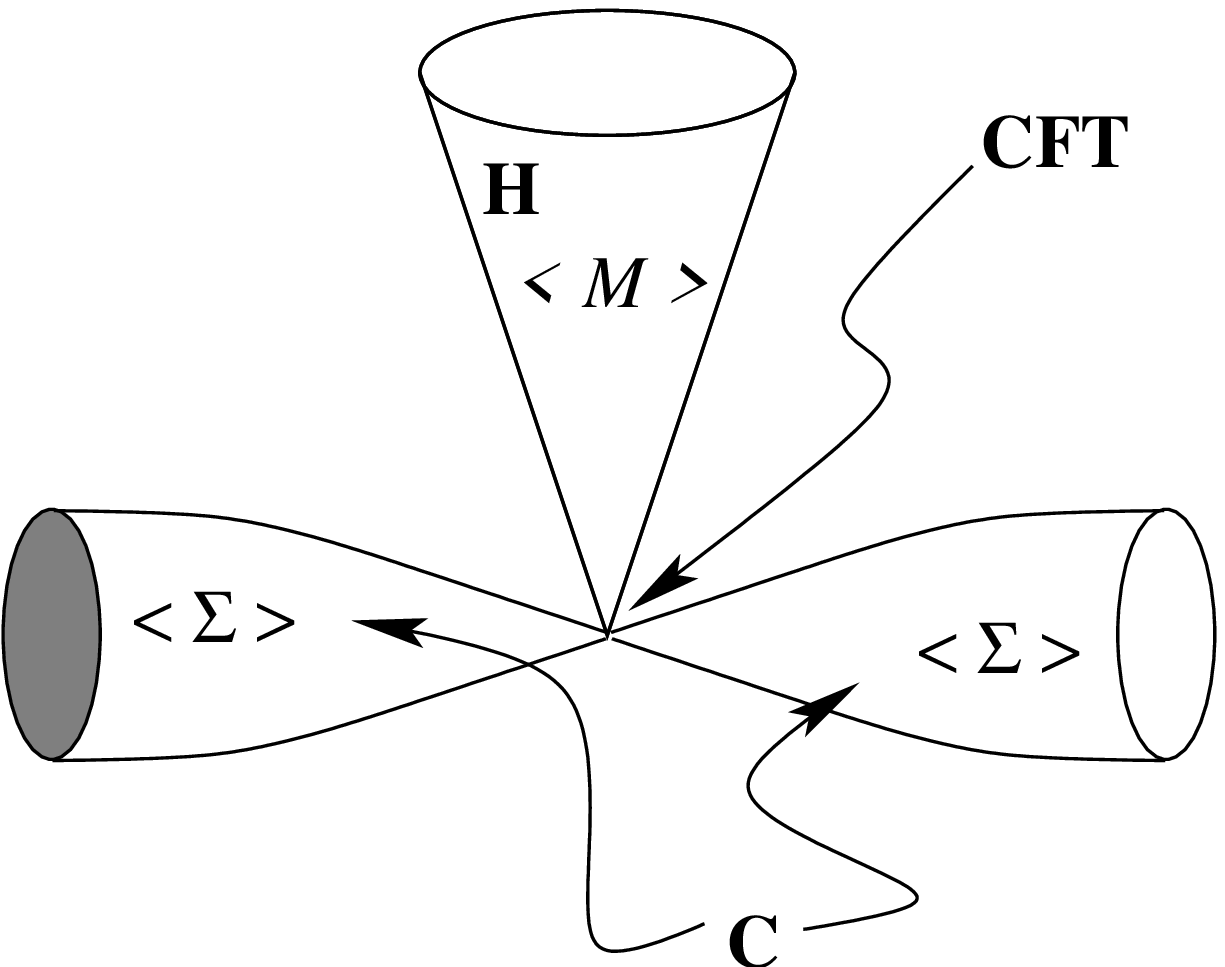,width=6cm,clip=}}
 \caption{\small The quantum version of \refiggg{dNN}, which combines
it with \refiggg{Sigmaspace}. }
\figgg{dNNq}
\end{center}
\end{figure}

This picture should look familiar.  At the meeting point of the three
branches, there is a conformal field theory which looks remarkably
like the $W=hXYZ$ conformal fixed point that 
we considered earlier (\refiggg{XYZ} and Sec.~\ref{subsec:QXYZ}).  And in
fact, it is the same!  In the XYZ model, the three branches had
nonzero expectation values for $X$, $Y$ and $Z$ respectively.  Here,
the branches are the three complex planes labeled by the expectation
values for $M=Q\tilde Q$, $e^\Sigma$, and $e^{-\Sigma}$.  Thus we have
another example of ``duality''; a single conformal fixed point is the
infrared physics of two different field theories, one the $W=hXYZ$
model, the other \ntwo\ super-QED.  The theories are different in the
ultraviolet but gradually approach each other, becoming identical in
the infrared, as shown in \refiggg{SQEDXYZ}.  This is called an
``infrared'' duality.  Notice the $Z_3$ symmetry between the branches
is exact in the XYZ model but is an ``quantum accidental'' symmetry (a
property only of the infrared physics) in super-QED.

\begin{figure}[th]
\begin{center}
 \centerline{\psfig{figure=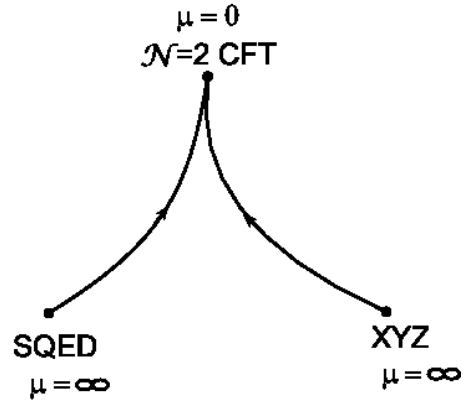,width=6cm,clip=}}
 \caption{\small \ntwo\ SQED and the XYZ model flow to the same
fixed point. }
\figgg{SQEDXYZ}
\end{center}
\end{figure}

\EX{ Calculate the   anomalous dimension of $Q$. Note the sign!  Why is
it allowed here?}

This duality is a particle-vortex duality.  Along the Higgs branch,
where $\vev M\neq 0$, there are vortex solitons of {\it finite mass};
these are similar to the vortices discussed in
Sec.~\ref{subsec:XYZSols}.  The phases of the fields $Q$ and $\tilde
Q$ wind once around the circle at infinity; however, the presence of
the gauge field cuts off the logarithmically divergent energy that was
a feature we dwelt on in Sec.~\ref{subsec:XYZSols}.  (You can read
about how this works in Nielsen and Olesen (1973).)  These solitons
correspond to the fields $Y$ and $Z$, which have (finite) mass when
$\vev{X}\neq 0$.  Thus the vortices of the one theory correspond to
the particles of the other.

\EX{ Since $X$ and $M=Q\tilde Q$ are to be identified, a mass term
$W=mQ\tilde Q$ should correspond to changing the dual theory to
$W=hXYZ + mX$.  The massive fields $Q, \tilde Q$ are logarithmically
confined (as always for weakly-coupled electrically-charged particles
in $d=3$) by the light photon which remains massless.  Look back at
section 1.6, where we showed there are vortex solitons in the dual
$W=hXYZ + mX$ theory which are logarithmically confined, and argue
that it is consistent to identify $Q$ and $\tilde Q$ with these
solitons.  Using the relation beween the gauge field and $\sigma$, try
to show that the electric field surrounding the electrons $Q, \tilde
Q$ corresponds to a variation in $\Sigma$ which agrees with the
properties of $Y$ and $Z$ near these solitons.  }

Now let us examine the theory with \nfour\ supersymmetry, which is
even more amazing.  We can obtain it from the \ntwo\ case by adding a
neutral chiral superfield $\Phi$ to the theory and coupling it to the 
other fields
via the superpotential $W = \sqrt 2\Phi Q \tilde Q$.  This will
destabilize the \ntwo\ fixed point and cause it to flow to a new one,
as in \refiggg{NNtoNNNN}.

\begin{figure}[th]
\begin{center}
 \centerline{\psfig{figure=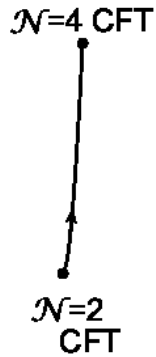,width=2cm,clip=}}
 \caption{\small There is a flow linking the two theories. }
\figgg{NNtoNNNN}
\end{center}
\end{figure}

\EX{ Check that the operator $\Phi Q\tilde Q$ is a relevant operator both at
the free \ntwo\ fixed point and at the infrared \ntwo\ conformal fixed point.}

Since the XYZ model is the same as \ntwo\ SQED in the infrared, we may
obtain the \nfour\ theory another way.  Let us go to the far infrared
of the XYZ model.  We just studied what happens when we add a
single field $\Phi$ and couple it to $M=Q\tilde Q$ in the
superpotential.  But we can simply change variables from SQED to XYZ;
from this dual point of view, what we did was couple $\Phi$ to $X$.
The low-energy physics of a model with $W= hXYZ + \Phi X$
should be the same as that of \nfour\ SQED.  But $\Phi X$ is just a
mass term which removes $\Phi$ and $X$ from the theory, leaving $Y$
and $Z$, with {\it no} superpotential.  {\it Thus the dual description
of the \nfour\ SQED fixed point is a free theory!}

In short, the low-energy limit of \nfour\ SQED is a conformal fixed
point which can be rewritten as a free theory --- a theory whose
massless particles are the vortices of SQED.

From this astonishing observation, a huge number of additional duality
transformations of other abelian gauge theories can be obtained.  In
this sense, it plays a role similar to ``bosonization'' (boson-fermion
duality) in two dimensions, which can be used to study and solve many
field theories.  These three-dimensions ``mirror'' duality
transformations, first uncovered by Intriligator and Seiberg and much
studied by many other authors, are a simple yet classic example of
dualities, and I strongly encourage you to study them.  A summary of
previous work and a number of new results on this subject appear in
work I did with Kapustin (1999).

An important aside: it is essential to realize that we have here an
example of a nontrivial {\it exact} duality, which is not merely
an infrared
duality.  We noted that the flow from the weakly coupled XYZ to the
\ntwo\ fixed point is different from the flow from weakly coupled SQED
to the \ntwo\ fixed point.  However, at the \ntwo\ fixed point the two
flows reach the same theory, and the operators $Q\tilde Q$ and $X$ are
identical there.  The relevant perturbations $\Phi Q\tilde Q$ and
$\Phi X$ may be added with arbitrarily tiny couplings; in this case
the two different flows approach and nearly reach the \ntwo\ fixed
point, stay there for a long range of energy, and then flow out,
together, along the same direction, heading for the \nfour\ fixed
point.  This is shown schematically in \refiggg{exactduality}.  In the limit
where the \ntwo\ fixed point is reached at arbitrarily high energies,
the flow to the \nfour\ fixed point is described exactly by two
different descriptions, one using the XYZ variables, the other using
those of SQED.  One will sometimes read in the string theory
literature that ``field theory has infrared dualities, but duality in
string theory is exact.''  Clearly this is not true; as we have seen
in this example, infrared dualities always imply the existence of
exact dualities.  You can look at my work with Kapustin (1999) for
some very explicit examples.

\begin{figure}[th]
\begin{center}
 \centerline{\psfig{figure=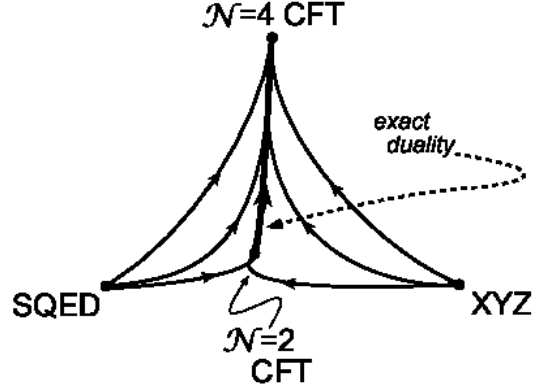,width=7cm,clip=}}
 \caption{\small By adjusting couplings and scales we may obtain two exactly
equivalent descriptions of the flow in \refiggg{NNtoNNNN}. }
\figgg{exactduality}
\end{center}
\end{figure}

\section{Non-Abelian Four-Dimensional Gauge Theory}

We now turn to nonabelian gauge theories in four dimensions.  This is  a
huge subject and we shall just scratch the surface, but hopefully this lecture
will give you some sense of the immensity of this field and teach you a
few of the key ideas you need to read the already existing review articles.

\subsection{The classical theory}

Let us first consider the classical pure gauge theory.  The only
difference from the abelian case (aside from some complications
in the
superfield formalism) is that the kinetic
terms reflect the fact that the pure vector multiplet is
self-interacting.  The gauge group, a Lie group such as $SU(N)$, is
generated by a Lie algebra with generators $T^A$, $A$ an index running
from 1 to the dimension of the group.\footnote{For example, for
$SU(N)$, $A$ runs from 1 to $N^2-1$.  The generators may themselves
appear as $N^2-1$ matrices in any representation of the group.  If we
take $T^A$ in the fundamental representation, then each $T^A$ is an
$N\times N$ matrix $(T^A)_i^{\bar j}$, where $i$ and $\bar j$ are
indices in the fundamental and antifundamental represention of
$SU(N)$.  The matrices are normalized by the condition $\tr(T^AT^B) =
\delta^{AB}$.  In the adjoint representation, $(T^A)_B^C$ is the
matrix $f^{ABC}$, the structure constants of the group.  In {\it any}
representation, $[T^A,T^B] = if^{ABC}T^C$.}  Gauge bosons $A_\mu =
A_{\mu}^AT^A$, gauginos $\lambda_a = \lambda_a^A T^A$, and auxiliary
fields $D = D^AT^A$ are all in the adjoint representation, and the
kinetic terms are the minimal ones
\begin{eqnarray}
S_{gauge} &=&
{i\tau\over 4\pi}\int\ d^4x\ 
\tr\left[-{1\over 4}(F^2+iF\tilde F) + i\bar\lambda\Dslash\lambda + \half D^2
\right]&  \cr
& \ & \ \ \ \ \ \  +  \  {\rm hermitean
\ conjugate.} 
\end{eqnarray}
where $\tau$ is again defined in \Eref{taudef} (with $e\to g$) and
\begin{eqnarray}
F_{\mu\nu} = F_{\mu\nu}^A T^A &=& 
\partial_\mu A_\nu-\partial_\nu A_\mu + i [A_\mu,A_\nu]
\cr \cr
&=& \left(\partial_\mu A_\nu^A-\partial_\nu 
A_\mu^A + f^{ABC} A_\mu^B A_\nu^C\right) T^A\ , \cr
\end{eqnarray} 
$D_\mu\lambda_a = \partial_\mu \lambda_a + i [A_\mu,\lambda_a]$,
and $D^2\equiv \sum_A |D^A|^2 $.  We will often choose to represent
fields in the adjoint representation using matrices $T^A$ that are in
the fundamental representation; this is usually the easiest
representation to work with.  In $SU(N)$, doing so allows us to write
$A_\mu,\lambda,D$ as $N\times N$ hermitean traceless matrices.

The addition of charged chiral fields involves a fairly minimal change
in the kinetic terms from the abelian case.  If we add chiral fields
$Q_r$ and $\tilde Q^s$, $r,s=1,\dots,N_f$, in the fundamental and
antifundamental representation of $SU(N)$, we obtain
\begin{eqnarray}
S_{kin} &= \sum_r \iddx  \Big[D_\mu q_r^\dagger D^\mu q_r
+i\bar\psi_r \Dslash \psi_r
 + F_r^\dagger F_r + \nonumber \cr
&\lambda\bar \psi_r q_r +
 \bar \lambda \psi_r q_r^\dagger + q_r^\dagger D q_r \Big]
\cr  & \cr
&+\sum_s \iddx 
 \Big[
D_\mu \tilde q^{s\dagger} D^\mu \tilde q^s
+i\overline{\tilde \psi^s} \Dslash \tilde \psi^s 
+ \tilde F^{s\dagger} \tilde F^s + \nonumber \cr
&\lambda\overline{ \tilde \psi}^s \tilde q^s 
+ \bar \lambda \tilde\psi^s \tilde q^{s\dagger} 
+\tilde q^{s\dagger}
D \tilde q^s
\Big] \cr  
\end{eqnarray}
where the contraction of gauge indices is in 
each case unique: for example in the term
$q_r^\dagger D q_r $ the indices are contracted as 
$$
(q_r^\dagger)_{\bar j} D^A (T^A)_i^{\bar j} (q_r)^i
$$ 
If we add a chiral superfield $\Phi$ in the adjoint representation,
the kinetic terms take the same form as above, but we should interpret
$\phi^\dagger D \phi$ as
$$
\tr\ \phi^\dagger[D, \phi] = -\tr\ D[\phi^\dagger, \phi] 
= -if^{ABC} D^A \phi^{B\dagger}\phi^C
$$
and similarly for the scalar-fermion-fermion terms.  (Be careful not
to confuse the derivative $D_\mu$ and the auxiliary field $D^A$!)

As in the abelian case, 
\begin{itemize}
\item We may obtain \none\ gauge theories by adding arbitrary charged
(and neutral) matter to the theory with arbitrary gauge-invariant
holomorphic gauge kinetic and superpotential functions and an abitrary
gauge-invariant K\"ahler potential.
\item We may obtain a pure \ntwo\ gauge theory by writing an \none\
gauge theory with a single chiral multiplet $\Phi$ in the adjoint
representation, a gauge kinetic term and K\"ahler potential term for
$\Phi$ which must be related, and zero superpotential.
\item We may add matter to the \ntwo\ gauge theory in the form of a
hypermultiplet (two chiral multiplets $Q$ and $\tilde Q$ in conjugate
representations) coupled in the superpotential $W=\sqrt{2}\tilde Q
\Phi Q$, with gauge indices contracted in the unique way.  We may also
add mass terms for the hypermultiplets, obtaining $W=\sqrt{2}\tilde Q
\Phi Q + mQ\tilde Q$.
\item Finally, if we have a massless hypermultiplet in the {\it
adjoint} representation, so that the theory has a total of {\it three}
chiral multiplets $\Phi_1=\Phi, \Phi_2=Q, \Phi_3=\tilde Q$ in the
adjoint, with superpotential $W = \sqrt{2}\ \tr\
\Phi_1[\Phi_2,\Phi_3]$, then the theory has \nfour\ supersymmetry.
\end{itemize}

Now the condition for a supersymmetric vacuum requires that $D^A = 0$,
$F^i_r=0$, $F^s_{\bar j}=0$.  If the superpotential is zero, then the
constraints all come from
\begin{eqnarray}\label{nabDterms}
0=D^A \propto
&(q_r^\dagger)_{\bar j} (T^A)_i^{\bar j} (q_r)^i 
(\tilde q^{s^\dagger})^i (T^A)_i^{\bar j} (\tilde q_r)_{\bar j} 
=   (T^A)_i^{\bar j} \left[(q_r^\dagger)(q_r)
-\tilde q^{s\dagger} \tilde q^s \right]_{\bar j}^i \cr \cr
& = \tr\ T^A \left[(q_r^\dagger)(q_r)-\tilde q^{s\dagger} \tilde q^s \right] 
\cr
\end{eqnarray}
(I have written a proportional sign since the precise relation depends
on the K\"ahler potential and gauge kinetic term, while the
proportionality relation does not!)

These equations are beautifully solved in the case of $SU(N)$ with
fields $N_f$ $Q$ in the ${\bf N}$ representation and $\tilde Q$ in the
${\bf \bar N}$ representation.\footnote{In this case there is an
$SU(N_f)_L$ and an $SU(N_f)_R$ symmetry acting on the $Q$ and $\tilde
Q$ fields respectively; there is also a baryon number under which $Q$
and $\tilde Q$ have charges $1$ and $-1$, and an {\it anomalous} axial
symmetry (present classically but explicitly violated quantum
mechanically) under which $Q$ and $\tilde Q$ both have charge 1.}  The
hermitean matrix $ \left[(q_r^\dagger)(q_r)-\tilde q^{s\dagger} \tilde
q^s\right]$ can be uniquely expanded as
$$
 \left[(q_r^\dagger)(q_r)-\tilde q^{s\dagger} \tilde q^s\right]_{\bar j}^i 
= c_0 \delta_{\bar j}^i 
+ \sum_B c_B (T^B)_{\bar j}^i 
$$
Then, using the fact that 
$$
\tr\ T^A T^B = \half\delta^{AB} \ ; \ \tr\ T^B = 0
$$ we find that the conditions \eref{nabDterms} become simply
\bel{sunDterms} 
\left[(q_r^\dagger)(q_r)-\tilde q^{s\dagger} \tilde
q^s\right]_{\bar j}^i = c_0 \delta_{\bar j}^i 
\ee 
for {\it any} $c_0$.
Before writing any solutions to these equations, we note the
following: given {\it any} expectation values of $q$ and $\tilde q$
which are a solution to these equations, a continuously infinite class
of solutions is generated by multiplying all of the $q$ and $\tilde q$
fields by a complex constant.  Thus there will generally be, as in the
abelian case, noncompact, continuous moduli spaces of vacua.  As
before, these vacua are not related by any symmetry, in that they have
fields with different masses.\footnote{More precisely, they are
related by scale invariance (since the vevs are the only scales in the
classical $d=4$ gauge theory) but since scale invariance will be
broken quantum mechanically, while the D-term conditions generally will
not be altered, we will see that the vacua really are physically very
different.}

\EX{ The scalars $q_r^i$ and $q^s_{\bar j}$ are $N_f\times N$ and
$N\times N_f$ matrices respectively.  Being careful with the indices,
show that the only solutions for $N_f<N$ are gauge and global symmetry
transformations of the particular solution $q_r^i = v\delta_r^i$,
$\tilde q^s_{\bar j} = v\delta^s_{\bar j}$ for $i,\bar j\leq N_f$.
Then show that the only solutions for $N_f\geq N$ are gauge and global
symmetry transformations of the particular solution $q_r^i =
v\delta_r^i$, $\tilde q^s_{\bar j} = \tilde v\delta^s_{\bar j}$ for
$r,s\leq N$.  Note that for $N_f\geq N$, $v$ and $\tilde v$ are
in general different and the constant
$c_0$ in \eref{sunDterms} is nonzero, while for $N_f<N$ 
$c_0$ must
be zero.}

As another example, consider $SU(2)$ with fields $\Phi_n$
($n=1,2,\dots,N_a)$ in the adjoint representation (the {\bf 3}).
Representing $(\Phi_n)_i^{\tilde j}$ as a traceless $2\times 2$
complex matrix, the D-term conditions are the matrix equation
$$
\sum_n [\Phi_n^\dagger, \Phi_n]_i^{\tilde j} = 0 \ .
$$

In the case of $N_a=1$,  appropriate to pure \ntwo\ gauge theory,
the solution is clearly that $\Phi$ is diagonal
$$
\vev\Phi= \left[\begin{matrix}a&0\cr 0 & -a\end{matrix}\right]
$$
This breaks the $SU(2)$ gauge group to $U(1)$.
In terms of the one independent gauge invariant
operator which can be built from $\Phi$
$$
\vev u \equiv \vev{\tr\ \Phi^2} = 2a^2
$$
the moduli space of the theory is the complex $u$ plane, classically.  The
K\"ahler potential for $u$ has a singularity
at $u=0$, where the gauge group is unbroken.  Let's check the counting:
$\Phi$ is a triplet, and two of its components are eaten when $SU(2)$ breaks
to $U(1)$, leaving one --- the chiral multiplet $u$.

Now suppose there are three fields $\Phi_n$, $n=1,2,3$, in the {\bf 3}
of $SU(2)$, and also a superpotential $W=\tr\ \Phi_1[\Phi_2,\Phi_3]$.
This is the \nfour\ $SU(2)$ gauge theory.  The easiest way to
establish the solutions to the D-term and F-term constraints is to
rewrite the fields as
$$\Phi_1 = X_4+iX_7 \ , \ \Phi_2=X_5+iX_8, \ , \ \Phi_3= X_6+iX_9 $$
in terms of which the potential $V(X_m)= \sum_A (D^A)^2 + \sum_n F_n^2$ 
can be rewritten as
$$
V(X_p) \propto \sum_{p,q=4}^9 \tr\left([X_p,X_q]\right)^2
$$ 
This formulation has the advantage that the $SO(6)$ symmetry
rotating the $X_p$ is manifest; the superpotential only exhibits a
$U(3)$ subgroup of that symmetry.  The condition $V=0$ implies all of
the $X_p$ are simutaneously diagonalizable.  In short, the solutions
are
$$
X_p= \left[\begin{matrix}c_p&0\cr 0 & -c_p\end{matrix}\right]
$$
with $(c_4,\cdots,c_9)$ forming a real six-vector living in a flat
six-real-dimensional moduli space.  Again, with the exception of the
point $c_p=0$, all of the vacua have $SU(2)$ broken to
$U(1)$.\footnote{This generalizes: for $SU(N)$, we have $N-1$ such
real six-vectors, corresponding to the $N-1$ sextuplets of eigenvalues
of the traceless matrices $X_p$.}

Let's add a mass $m\ \tr\ \Phi_2\Phi_3$ (which breaks \nfour\ but
preserves \ntwo) and integrate out the massive fields.   Their equations
of motion, which at low momenta reduce simply to
$$
\dbyd{W}{\Phi_2} = \sqrt 2[\Phi_1,\Phi_3] + m\Phi_3 = 0 ; 
\dbyd{W}{\Phi_3} = -\sqrt 2[\Phi_1,\Phi_2] + m\Phi_2 = 0 ; 
$$
only have a solution  $\Phi_2=\Phi_3=0$.  When substituted
back into the superpotential, this solution gives a low-energy theory with one
adjoint $\Phi_1$ and $W=0$ --- the pure $SU(2)$ \ntwo\ theory.  Thus it is
easy to flow from the \nfour\ theory to the pure \ntwo\ theory, and
this was studied in the $SU(2)$ case by Seiberg and Witten (1994).

By contrast, consider adding a mass $\half m\ \tr\ \Phi_3^2$.  The
situation here is much like the XYZ model with a mass for $X$; the
low-energy theory there had $W=Y^2Z^2$.  Here the equation of motion
for $\Phi_3$ reduces to
$$
\sqrt  2[\Phi_1,\Phi_2]= m\Phi_3
$$
leaving a superpotential
\bel{twoadjW}
W_L = p ([\Phi_1,\Phi_2])^2
\ee
where $p\propto 1/m$ is a coupling with classical dimension $-1$.  Like
$(YZ)^2$, it is an irrelevant operator in four dimensions, scaling to
zero in the infrared.

\subsection{Beta functions and fixed points}

Now we turn to the quantum mechanics of these theories.  We have
already studied the abelian case in great detail at the one-loop
level, and we know that the main difference in the nonabelian case
will be that gauge boson loops will give a negative contribution to
the beta function.  The contributions to the beta
function of various particles are shown in \refiggg{gaugebeta},
\begin{figure}[th]
\begin{center}
\centerline{\psfig{figure=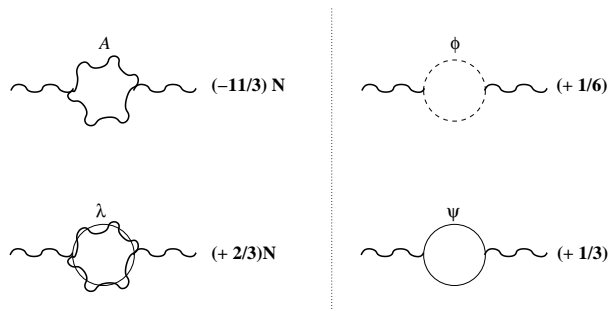,width=8cm,clip=}}
\caption{\small The contribution of various particles to the gauge beta
function. }
\figgg{gaugebeta}
\end{center}
\end{figure}
from which one can see that 
a vector multiplet contributes a factor of $-3N$ to the
one-loop beta function, while a chiral multiplet in the fundamental or
antifundamental representation contributes $1/2$, and one in the
adjoint representation gives a factor $N$.  Thus for supersymmetric
QCD with $N_f$ quarks/squarks and $N_f$ antiquarks/antisquarks, the
beta function at one-loop is (see the abelian case)
\bel{betagholo}
\beta_g = -{g^3\over 8\pi^2} (3N-N_f)  \Rightarrow
{1\over g^2(\mu) } =   {3N-N_f\over 8\pi^2} \ln\left({\Lambda\over\mu}\right)
\ee
where (defining $b_0=3N-N_f$)
\bel{LambdadefH}
\Lambda^{b_0} = \mu^{b_0} e^{{-8\pi^2/ g^2} + i\theta}  
=\mu^{b_0} e^{2\pi i \tau}
\ee
For $N_a$ adjoint fields the one-loop beta function has
$b_0=(3-N_a)N$; note it vanishes for \nfour\ Yang-Mills.
Just as in the abelian case, the entry of the theta angle
into $\Lambda$ implies there can be no perturbative
corrections to these formulas.

As an aside, let me stress that 
the existence of the quantum mechanical holomorphic parameter
$\Lambda$ is very important.  Although the effective superpotential is
still constrained by the perturbative nonrenormalization theorem, as
in non-gauge theories, the presence of $\Lambda$ permits, in many
cases, a nonperturbative renormalization of the superpotential.  There
is a lot of literature on this subject; the classic paper is that of
Affleck, Dine and Seiberg (1984), although a number of others obtained
similar results using somewhat less reliable techniques. There are
good reviews on this subject by other authors, including ones by
Intriligator and Seiberg  and one by Argyres.  For lack of time, in
these lectures we will {\it only} discuss cases where there is {\it no}
nonperturbative correction to the effective superpotential (or
at least no qualitative change in its structure.)  Large classes of
interesting theories have this property, but we are leaving out
other large classes; see the Appendix for some examples.

As before, the formula for the holomorphic coupling $g(\mu)$, and the
holomorphic renormalization scale $\Lambda$, can't represent the
physical properties of the theory.  This is obvious from the fact that
$g^2(\mu)$ blows up at small $\mu$ and can't make sense below
$\Lambda$.  Higher loop effects, and possibly nonperturbative effects,
appearing in the nonholomorphic parts of the theory can change this
formula significantly.  How can we define a physical gauge coupling?
A natural approach is to find a more physical definition of $\Lambda$,
so let the holomorphic object in \Eref{LambdadefH} be renamed $\hat
\Lambda$, and let us attempt to define a $\Lambda$ independent of
field redefinitions.  (The presentation here is related to recent work
of Arkani-Hamed and Rattazzi, although the resulting formula is due to
Novikov, Shifman, Vainshtein and Zakharov from the early 1980s.)

Recall how we defined a physical version of $y$, the coupling in the
Wess-Zumino model.  We noted that if we sent $\Phi\to a\Phi$, $a$ a
constant, this would affect both $\hat y$ in the superpotential and
$Z$ in the K\"ahler potential.  Let's do the same here for the charged
fields $Q$ and $\tilde Q$.  Suppose we multiply them all of them by
$a$, where $a$ is a {\it phase} $e^{i\alpha}$.  This is equivalent to a
transformation by an anomalous ``axial'' $U(1)$ global symmetry, under
which quarks and antiquarks have the same charge.  As happens in QCD,
this kind of transformation is an anomalous symmetry, and shifts the
$\theta$ angle; it therefore rotates $\Lambda^{b_0}$ by a phase.  The
phase by which $\Lambda^{b_0}$ rotates is $a^{2N_f}$.  But since $Q$
and $\Lambda$ are holomorphic, it must still be true that $\Lambda$
changes by $a^{2N_f}$ even if $a$ is {\it not} a phase but has
$|a|\neq 1$!  Therefore, since $Z\to |a|^{-2}Z$ under this
transformation, only
$$
(\hat\Lambda^{b_0})^\dagger
\left[\prod_{r=1}^{N_f} Z_r\right]
\left[\prod_{s=1}^{N_f} \tilde Z_s\right] \hat\Lambda^{b_0}  \ ,
$$
where $Z_r$ and $\tilde Z_s$ are the wave function factors for 
$Q_r$ and $\tilde Q_s$,
can be invariant under these field redefinitions.

We can go further by considering R-symmetry transformations.  Suppose
we rotate the gluino fields $\lambda$ by $e^{i\alpha}$ and the fields
$q,\tilde q$ by $e^{i\alpha}$, so that the quarks $\psi_q$ and
antiquarks $\tilde\psi_q$ do not rotate at all (recall there is a
$\lambda \psi_r q_r^\dagger$ 
interaction which fixes the charge of one field in
terms of the other two.)  Then the anomaly in this transformation
involves only the gluinos, and the $\theta$ angle shifts by
$e^{2Ni\alpha}$.  (The factor $2N$ is the group theory ``index'' of
the adjoint representation in $SU(N)$; it determines the size of the
anomaly.)  Again, we can generalize this by taking $\alpha$ to be
complex, so that $|e^{i\alpha}|\neq 1$.  An invariant which is
unchanged by both this and the previous transformation is
$$
\mu^{2b_0}e^{-16\pi^2/g^2(\mu)} = |\Lambda^2(\mu)| 
= (\hat\Lambda^{b_0})^\dagger
 Z_\lambda^{2N} \left[\prod_r Z_r\right]\left[ \prod_s \tilde Z_s\right] 
\hat\Lambda^{b_0}
$$

Taking a derivative with respect to $\mu$ of both sides, we obtain
$$
\beta_{8\pi^2\over g^2}=  b_0 + \half \sum_r \gamma_r 
+ \half \sum_s\gamma_s + N\gamma_\lambda
$$
But from the kinetic terms of the gauginos it is evident that
$Z_\lambda = {1/g^2(\mu)}$ itself!  Therefore 
$$
\gamma_\lambda = -\dbyd{\ln Z_\lambda}{\ln\mu} 
= {\beta_{8\pi^2/g^2}\over{8\pi^2/g^2}}
$$
from which we obtain the {\it exact} NSVZ beta function
$$
\beta_{8\pi^2\over g^2}=  {b_0 + \half \sum_r \gamma_r 
+ \half \sum_s\gamma_s \over
1 -  g^2N/8\pi^2} = -{16\pi^2\over g^3}\beta_g
$$
(Remember to keep track of the difference in sign between $\beta_g$
and $\beta_{8\pi^2/g^2}$.)  In supersymmetric QCD, where in the
absence of a superpotential all charged fields are related by
symmetry, and therefore have the same anomalous dimension $\gamma_0$,
we may write 
$$
\beta_{8\pi^2\over g^2}=  {3N - N_f[1-\gamma_0] \over
1 -  g^2N/8\pi^2}
$$ 
In a general theory with charged fields $\phi_i$ in representations
$R_i$ with $\tr\ T^AT^B = T_{R_i}\delta^{AB}$, and with anomalous
dimensions $\gamma_i$, we have
$$
\beta_{8\pi^2\over g^2}=  {3N - \sum_i T_{R_i}[1-\gamma_i] \over
1 -  g^2N/8\pi^2}=  {b_0+\sum_i T_{R_i}\gamma_i\over
1 -  g^2N/8\pi^2}
$$
These formulas continue to hold even when there are other gauge and
matter couplings in the theory.  This formula, which has a Taylor
expansion in $g^2$, summarizes all higher-loop corrections.  If there
is matter in the theory, then we do not know $\gamma_i$ exactly; but
in the absence of matter (the pure \none\ gauge theory) the formula is
a definite function of $g$.\footnote{The pole in the denominator is
still not fully understood, even after 20 years; notice that it
becomes dominant when $g^2 N \sim 8\pi^2$, an important issue for
those studying large-$N$ gauge theories at large 't Hooft coupling!}

We will now use this exact beta function to prove a few things.
Before doing so, let's consider the one-loop contributions to the
anomalous dimensions of charged fields.  As we saw, trilinear terms in
the superpotential give one-loop graphs as in \refiggg{ZinW}
\begin{figure}[th]
\begin{center}
\centerline{\psfig{figure=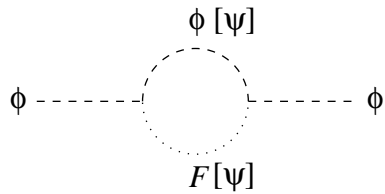,width=5cm,clip=}}
\caption{\small Typical contributions to $Z_\Phi$ from the superpotential.}
\figgg{ZinW}
\end{center}
\end{figure}
which {\it must} give a positive contribution, if there are no
gauge couplings around.  To leading order in the gauge coupling, this
must still be true; the one-loop graph from a superpotential
term must be positive.   However, there is no such constraint
for the one-loop diagram in \refiggg{Zingauge} involving the gauge 
interactions
\begin{figure}[th]
\begin{center}
\centerline{\psfig{figure=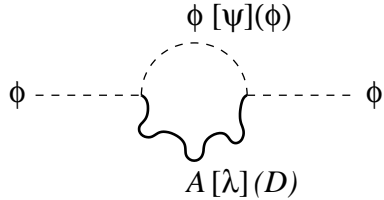,width=5cm,clip=}}
\caption{\small Contributions to $Z_\Phi$ from the gauge interactions.}
\figgg{Zingauge}
\end{center}
\end{figure}
and it is a determining, crucial feature of supersymmetric
gauge theories that the coefficient of this diagram has the opposite
sign.  (I don't know of an argument which explains this fact in
physical terms.)  Consequently the sign of $\gamma$ will flip as the
coupling constants are varied.  For a gauge theory with no
superpotential, the charged fields have negative anomalous dimensions.

First, let us prove that for large $N_f$, slightly less than $3N$, SQCD has
nontrivial conformal fixed points (and does not for $N_f\geq 3N$).  These
are sometimes called ``Banks-Zaks'' fixed points although they were
discussed earlier; they exist also in the nonsupersymmetric case!

\EX{ By examining the one- and two-loop beta functions (given in
Ellis's lectures at this school) verify that ordinary QCD has
conformal fixed points when $N_f$ and $N_c$ are large and the theory
is just barely asymptotically free.  Remember to check that all
higher-loop terms in the beta-function can be neglected!}

To see this, note that the loop expansion is really an expansion in $g^2 N$,
so if we can trap $g^2$ in a region where it is of order $1/N^2$, then
the loop expansion can be terminated at leading nonvanishing order.
The anomalous dimension $\gamma_0$ of the superfields $Q_r,\tilde Q_s$
will be of the form, on general grounds,
$$
- {\hat c} {g^2 N\over 8\pi^2} + {\rm order }\left[(g^2 N)^2\right]
$$
where ${\hat c}>0$.  For $N_f = 3N - k$, where $k$ is order 1, 
the beta function takes the form
$$ 
\beta_{8\pi^2\over g^2}= {k +{N_f\over 8\pi^2}\left(-{\hat c}{g^2 N}+ {\rm
order }\left[(g^2 N)^2\right]\right) \over 1 - g^2N/8\pi^2} \approx k -3N{\hat c}{g^2
N\over 8\pi^2} 
$$ 
where in the last expression we have dropped terms of order $g^2Nk$
and $N(g^2N)^2$.
For $k\leq 0$, that is, $N_f\geq 3N$, the beta function $\beta_g$
is positive ({\it i.e.} $\beta_{8\pi^2/g^2}<0$)
for small $g$, so the gauge coupling $g$ flows back to zero in the {\it
infrared} and is not asymptotically free in the ultraviolet.  However,
if $k>0$, and thus for $N_f<3N$, the beta function $\beta_g$
is negative at small $g$ but has a zero at
$$
g^2_* = {8\pi^2 k\over 3{\hat c} N^2}
$$
and therefore the coupling $g^2$ never gets larger than of order $1/N^2$.  
The above formula is therefore self-consistent in predicting that
the gauge coupling flows from zero to the above fixed point value.
(Note that the holomorphic coupling has no such fixed point!  thus the
physical properties and holomorphic properties of the theory are
vastly different!)  The flow is shown in \refiggg{gflow}.

\begin{figure}[th]
\begin{center}
 \centerline{\psfig{figure=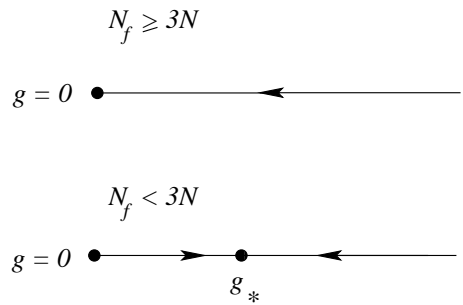,width=6cm,clip=}}
 \caption{\small The gauge coupling  is  marginally relevant
for $N_f<3N$, with a nearby fixed point at $g_*$. }
\figgg{gflow}
\end{center}
\end{figure}

Next, let's prove that in SQCD for $N_f\leq \frac32 N$ there can be no
such fixed points --- specifically, ones in which no fields
are free, and which are located at the origin of moduli
space, where no fields have expectation values.
In SQCD a fixed point requires
$$
\beta_{8\pi^2\over g^2} \propto b_0 + N_f\gamma_0 = 0 \ 
\Rightarrow \gamma_0 = {3N - N_f\over N_f} \ .
$$
(This in turn implies that the R-charge of $Q$ and 
$\tilde Q$, using the earlier formula
that $\dim Q = 1+\half \gamma_0 = {3\over 2}R_Q$, must be
\bel{RQ}
R_Q = 1 - {N\over N_f} \ ;
\ee
one may check that this particular R-symmetry is the unique
nonanomalous chiral symmetry of the theory!)  However, if
$\gamma_0\leq-1$, then the gauge-invariant operator $Q_r\tilde Q^s$
would have dimension $2(1+\half\gamma_0)\leq1$.  This is not allowed
at a nontrivial fixed point, so to have such a fixed point
(at least one in the simple class we have
been discussing) it must be that
$$
\gamma_0 >-1 \ \ \Rightarrow\  \ b_0 < N_f\  \ \Rightarrow\  
\ N_f > {\frac32} N
$$

What happens for adjoint fields?  If the number of fields is $N_a$, then
we have
$$
\beta_{8\pi^2\over g^2} \propto (3-N_a)N + N_a\gamma_0 
= 0 \ \Rightarrow \gamma_0 = 1 - {3 \over N_a} \ .
$$ 
This is too negative a $\gamma_0$ for $N_a=1$, so there is no
ordinary $SU(N)$ fixed point in the pure \ntwo\ theory.\footnote{There
are however some much more subtle fixed points discovered by Argyres
and Douglas in 1995.}  For $N_a=2$, the anomalous dimension at any
fixed point must be $-\half$; that is, the dimension of $\Phi_1$ and
$\Phi_2$ must be $\frac34$.  For $N_a=3$ the anomalous dimension at
any fixed point must be zero.  If $\Phi_n$ were gauge-invariant, then
by the earlier theorem this could only happen if $\Phi$ were free; but
outside of the abelian case, $\Phi_n$ is not gauge-invariant and this
is not a requirement.

\EX{ In an \ntwo\ gauge theory, the one-loop formula
$\beta_{8\pi^2/g^2}= b_0$ is exact.  Using the facts that (a) the
anomalous dimension of the adjoint field $\Phi$ is related to that of
the gauge bosons by \ntwo\ supersymmetry, and (b) \ntwo\ forbids
hypermultiplets to have anomalous dimensions, prove that the NSVZ beta
function is consistent with this statement.}

Now, Seiberg has suggested that there are fixed points in \none\ SQCD
for $3N>N_f>{\frac32}N$.  That is, he conjectured in 1994 that in this
range there is some value $g_*$ of the gauge coupling for which
$\gamma_0(g_*) = {3N - N_f\over N_f} \ .$ (Again, we know this is true
for $3N-N_f\ll N$; Seiberg's conjecture is that this continues
down to much lower $N_f$.) If he is right, as most experts think
that he is, then some remarkable and exciting phenomena immediately
follow.  These can be found by combining our minimal knowledge
concerning the properties of these putative fixed points with the
approach to the renormalization group outlined in the previous
lectures.

For example, consider adding the superpotential
\bel{QQQQ} W = \hat p \sum_{r,s=1}^{N_f} (Q_r\tilde Q^s) (Q_s\tilde
Q^r) \ee
with gauge indices contracted inside the parentheses. Very
importantly, this superpotential preserves a diagonal $SU(N_f)$ global
symmetry and charge conjugation; this is enough symmetry to ensure
that all of the fields share the same anomalous dimension
$\gamma_0(p,\tau)$, as was true for $p=0$.  As always we should study
the dimensionless coupling constant $\pi\equiv p\mu$ and ask how it
scales.  Classically it scales like $\mu$ (and thus has
$\beta_\pi = \pi$) but quantum mechanically, ---

Wait a minute.  This theory, whose potential
contains $({\rm scalar})^6$
terms, is nonrenormalizable.  Can we even discuss it?

Well, nonrenormalizable simply means that the operator in the
superpotential is irrelevant, so in the ultraviolet regime the
effective coupling is blowing up and perturbative diagrams in the
theory don't make sense. But we're interested in
the infrared anyway.  We'll deal with the ultraviolet later; for now
will think of $1/\hat p$ as setting a cutoff on the
theory.\footnote{Note that we did essentially the same thing with SQED
in four dimensions, which is {\it perturbatively renormalizable} but
{\it nonperturbatively nonrenormalizable}, since we cannot take the cutoff
on the theory to infinity without the gauge coupling diverging in the
ultraviolet.}  Perturbation theory may not converge, but we are asking
perfectly valid nonperturbative infrared questions which do not depend
on the details of the ultraviolet cutoff.

In particular, we know that we need to define a physical coupling
$\pi$, of the form $|\pi^2| = |\hat p \mu|^2 Z_0^{-4}$.  We see it has
a beta function
$$
\beta_\pi = \pi[1+2\gamma_0]
$$
 Now, this means $\pi$ is irrelevant if $\gamma_0>-\half$ and is
relevant if $\gamma_0<-\half$.  The formula for the gauge coupling is
unchanged
$$
\beta_{8\pi^2\over g^2} \propto 3N-N_f + N_f\gamma_0  \ .
$$ Now, remember that $\gamma_0$ is a function of $\tau$ and $\pi$
with the following properties: (1) if $g=0$, $\pi\neq 0$ then
$\gamma_0>0$; (2) if $\pi=0$, $0\neq g\ll 1$ then $\gamma_0<0$; and
(3) there is at least one nontrivial fixed point at $g=g_*$, $\pi=0$
with $\gamma_0 = {3N - N_f\over N_f} \ .$ Notice that at this fixed
point $\pi$ is irrelevant (as it is classically) if $N>2N_f$, marginal
if $N=2N_f$, and {\it relevant} if $N<2N_f$.  

From this we can guess
the qualitative features of the renormalization group flow.  For
$N>2N_f$, the qualitative picture is given in \refiggg{piflowNF}.
\begin{figure}[th]
\begin{center}
 \centerline{\psfig{figure=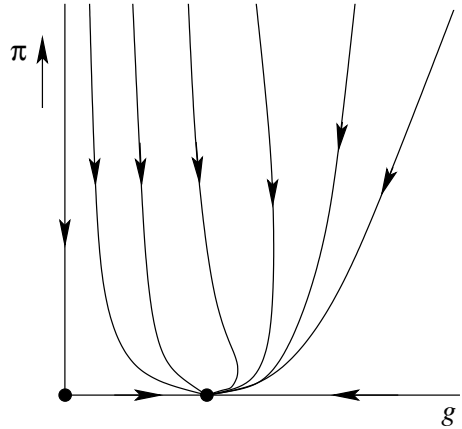,width=6cm,clip=}}
 \caption{\small For $N_f>2N$ the coupling $\pi$ is irrelevant both 
at $g=0$ and at $g=g_*$.}
\figgg{piflowNF}
\end{center}
\end{figure}
Even if $\pi\neq 0$, we still end up at the Seiberg fixed point.  For
$N<2N_f$, however, there is a very different picture, as in
\refiggg{piflownf}.  Notice that if we start at weak gauge coupling
initially, $\pi$ is irrelevant and flows toward zero as we would
expect classically; but as we flow toward the infrared, the gauge
coupling grows, $\gamma_0$ becomes more negative, and eventually the
coupling $\pi$ turns around and becomes {\it relevant}.  Although at
first it seems as though it will be negligible in the infrared, it in
fact {\it dominates.}  This is called a ``dangerous irrelevant''
operator, since although it is initially irrelevant it is dangerous to
forget about it!  In the infrared it becomes large, and we must be
more precise about what happens when it gets there.\footnote{What
happens is fascinating --- the theory flows to a {\it different}
Seiberg fixed point, that given by $SU(N_f-N)$ SQCD with $N_f$
flavors.  See Leigh and Strassler (1995) for an understanding of how
to treat the limit where the coupling is large.}

\begin{figure}[th]
\begin{center}
 \centerline{\psfig{figure=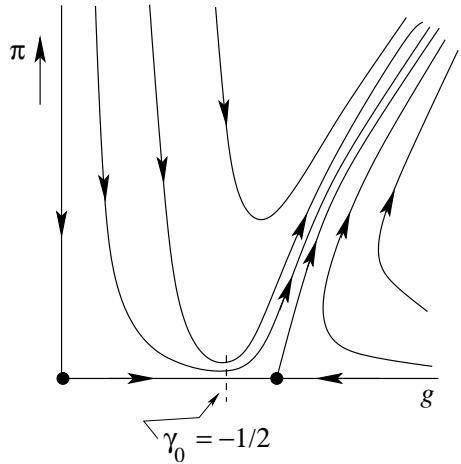,width=6cm,clip=}}
 \caption{\small For $N_f<2N$ the coupling $\pi$ is {\it relevant}
at $g_*$; its initial decrease
is reversed once $g$ is sufficiently large. }
\figgg{piflownf}
\end{center}
\end{figure}

What about $N_f=2N$?  You'll do this as an exercise, after I've done a
bit more.

\subsection{Using \none\ language to understand \nfour}

As another application of these ideas, let's argue that \nfour\
Yang-Mills is finite.  Consider an \none\ gauge theory with three
chiral superfields and a superpotential $W = h\ \tr\
\Phi_1[\Phi_2,\Phi_3]$.  I will use canonical normalization here 
for the $\Phi_n$, so
$h=g$ is the \nfour\ supersymmetric theory.  But let's not assume that
$h=g$.  For any $g,h$, the symmetry relating the three fields ensures
they all have the same anomalous dimension $\gamma_0$, which is a
single function of two couplings.  The beta functions for the
couplings are
$$
\beta_h = {\frac32}h\gamma_0 \ ; \ \beta_{g^2}= 
{-g^4\over 16\pi^2}{\gamma_0\over 1 -  g^2N/8\pi^2}
$$ 
These are proportional to one another, so the conditions for a
fixed point ($\beta_h=0$ and $\beta_g=0$) reduce to a single
equation, $\gamma_0(g,h)=0$.  But this is one
equation on two variables, {\it so if a solution exists, it will be a
part of a one-dimensional space of such solutions}.  

Now, does a solution exist?  We know that $\gamma(g,h=0)<0$ and that
$\gamma_0(g=0,h)>0$; so yes, by continuity, there must be a curve,
passing through $g=h=0$, along which $\gamma_0=0$ and
$\beta_h=\beta_g=0$ (and thus perturbation theory has {\it no} infinities
along this line.)  The renormalization group flow must look like the
graph in \refiggg{finitetheory}.
\begin{figure}[th]
\begin{center}
 \centerline{\psfig{figure=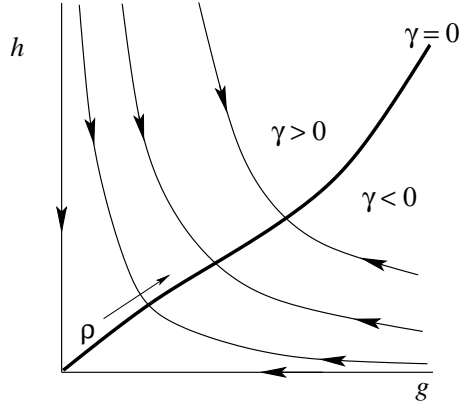,width=6cm,clip=}}
 \caption{\small In some \none\ theories one can argue for
a line of fixed points indexed
by an exactly marginal coupling $\rho$; perturbation theory
has no divergences on this line. Only in \nfour\ is the
equation for this line  $g=h$. }
\figgg{finitetheory}
\end{center}
\end{figure}
Both the theory with $h=0$ and the theory with $g=0$ are infrared
free; yet a set of nontrivial field theories lies between.  Notice
that we do not know, however, the precise position of the curve
$\gamma_0=0$.  In particular, we have not shown that $g=h$ gives
$\gamma_0=0$.  However, the existence of a finite theory (which is
renormalization-group stable in the infrared) requires only arguments
using \none\ symmetry.  Of course, since the theory at $g=h$ has more
symmetry (namely \nfour) it is natural to expect $g=h$ to be the
solution to $\gamma_0(g,h)=0$.

The motivation for introducing this \none-based reasoning is there are
many \none\ field theories which are also finite, as one can show
using similar arguments.  (For example, replace the \nfour\
superpotential with $W=h\ \tr\ \Phi_1\{\Phi_2,\Phi_3\}$; the discussion
is almost unchanged, except that $g=h$ is not the solution to
$\gamma_0=0$.)  The existence of these theories was discovered in the
1980s; the slick proof presented above is in Leigh and Strassler
(1995).

As mentioned in Sec.~\ref{subsec:QXYZ}, 
the coupling which parametrizes the line
of conformal fixed points (which is actually a complex line, since the
couplings are complex) is called an``exactly marginal coupling.''
Let's call this complex coupling $\rho$.  (In the \nfour\ case we can
identify $\rho$ as equal to the gauge coupling $i/\tau$, but in a more
general \none\ finite theory these will not be simply related.)
Unlike $\lambda$ in $\lambda\phi^4$, which is marginal at $\lambda=0$
but irrelevant at $\lambda\neq 0$, $\rho$ is marginal at $\rho=0$, and
remains marginal for any value of $\rho$.  Thus $\rho$ is a truly
dimensionless coupling, indexing a continuous class of scale-invariant
theories.  It is very common for such classes of theories to be acted
upon by duality transformations.  In fact, for \nfour\,
electric-magnetic duality (S-duality) acts on this coupling $\rho$ as
in \refiggg{Sduality}, identifying those theories at large $\rho$ with
those at small $\rho$.

\begin{figure}[th]
\begin{center}
 \centerline{\psfig{figure=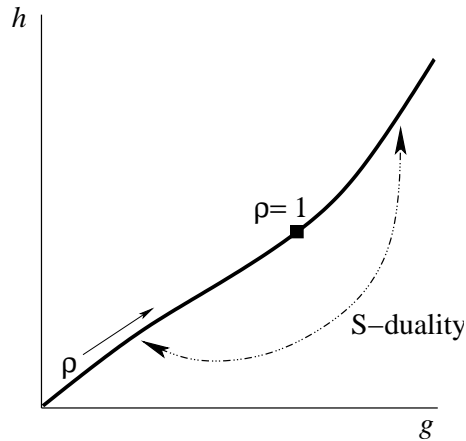,width=6cm,clip=}}
 \caption{\small The action of S-duality on the line of fixed points. }
\figgg{Sduality}
\end{center}
\end{figure}

\subsection{The two-adjoint model}
We conclude with a discussion of a theory with two adjoint chiral
multiplets, obtained from the \nfour\ gauge theory by adding a mass for
the third adjoint $\Phi_3$.  It has a superpotential \eref{twoadjW}:
$$
W_L = p ([\Phi_1,\Phi_2])^2 \ .
$$ 
This quartic superpotential is nonrenormalizable, but low-energy
effective theories often are.  We are interested (for the moment) in
the infrared behavior, so the fact that $\pi = p\mu\propto \nu^{-1}$
(where $\nu={m\over \mu}$) blows up in the ultraviolet is not our
immediate concern.

What are the beta functions for $\pi$ and for $g$?  We have
$$
\beta_\pi = \pi[1+2\gamma_0] \ ; \ 
\beta_{8\pi^2/g^2}\propto 3N - 2N(1-\gamma_0) = N + 2N\gamma_0
$$ 
and thus $\beta_\pi\propto \beta_g$.  This means that, as before,
the conditions for a fixed point to exist, namely
$\beta_\pi=0=\beta_g$, reduce to a single condition (except at
$g=\pi=0$):
$$
1+2\gamma_0 = 0 \ \Rightarrow \ \gamma_0=-\half \ .
$$
Again, this is one condition on two couplings, {\it so any solution
will be part of a one-dimensional space of solutions.}  Following
Seiberg, we might well expect that the theory with two adjoints and
$W=0$ has a fixed point at some $g_*$ where $\gamma_0(g_*) = -\half$.
If this is true, then the renormalization group flow of the theory
will look like \refiggg{pimarginal}.
\begin{figure}[th]
\begin{center}
 \centerline{\psfig{figure=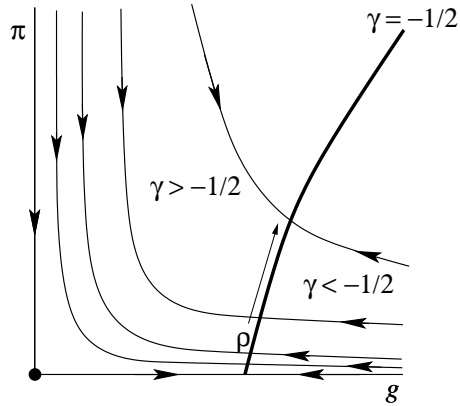,width=6cm,clip=}}
 \caption{\small In the two-adjoint theory, $\pi$ is marginal at
$g=g_*$ and there is a line of fixed points emerging from
the fixed point at $(g,\pi)=(g_*,0)$.}
\figgg{pimarginal}
\end{center}
\end{figure}

So here our irrelevant operator has been converted into
an exactly marginal one!  The coupling $\rho$ which parametrizes the
line of fixed points, and on which duality symmetries might act, now has
nothing to do with the gauge coupling.   In fact $\rho=0$ 
corresponds to $g=g_*$.  Nowhere are these conformal
field theories near $g=\pi=0$, so we have no hope of
seeing them in any perturbation expansion.

Now, since the theory is nonrenormalizable, we probably should at
least say something about the ultraviolet.  But we know a perfectly
good ultraviolet theory into which we can embed this theory, namely
the \nfour\ gauge theory with $\Phi_3$ massive.  Classically, we know
what this flow would look like.  Just as in the XYZ model with $X$
massive, shown in \refiggg{RGinXYZmass}, the theory would start from a
conformal field theory indexed by $\tau$ and flow into a classical
fixed point, with a nonzero gauge coupling and $W=0$, along the
irrelevant operator $([\Phi_1,\Phi_2])^2$.  But quantum mechanically
the endpoint of the theory is not $W=0$; instead, it is one of the
conformal field theories we found above in the two-adjoint theory.  In
fact, we can expect that each \nfour\ field theory flows to a unique
two-adjoint theory, along a flow which looks schematically like
\refiggg{thewholeshebang}.
\begin{figure}[th]
\begin{center}
 \centerline{\psfig{figure=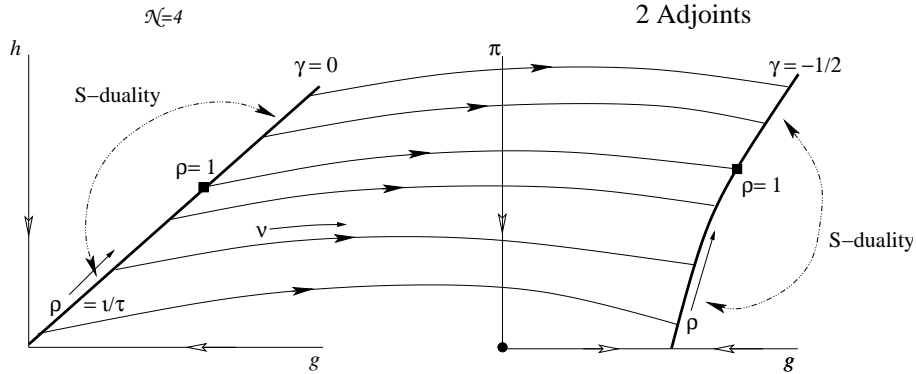,width=12cm,clip=}}
 \caption{\small The two-adjoint theory inherits S-duality
from the \nfour\ theory through the flow which takes
one to the other. }
\figgg{thewholeshebang}
\end{center}
\end{figure}
It is natural therefore to identify $\rho$ with $i/\tau$, as
we did in the \nfour\ case, but this ``$\tau$'' is not the gauge
coupling of the two-adjoint theory.  Rather, we have
defined here a physical mechanism for using the label $\tau$ of the
\nfour\ fixed points as a label for the two-adjoint
fixed points.  
Since S-duality acts on
$\tau$ in \nfour\, we are essentially guaranteed that duality will
also act on $\rho$ in the two-adjoint theory.

\EX{  Examine the beta functions and 
sketch the renormalization group flow for SQCD with $N_f=2N$
and the quartic superpotential \eref{QQQQ}.}
 
\section{Regrets}

What didn't I talk about in these lectures?  The list seems to be
infinite.  There was nothing on the phases of gauge theories; nothing
on confinement; nothing on nonperturbative renormalizations of
superpotentials; nothing on Seiberg duality; nothing on two, five or
six dimensions; nothing on spontaneous dynamical supersymmetry
breaking; nothing on exact methods for studying renormalization in
theories with broken supersymmetry; nothing on D-brane or other
stringy constructions of these theories; and above all, nothing on
applications of this material to real-world physics!  But there are
good reviews on almost all of these subjects.  By contrast, there are
no reviews on the material I have discussed here, which is necessary
for an understanding of duality, plays a very important role in the
AdS/CFT correspondance, and (as Ann Nelson and I have suggested) may
even be responsible for the pattern of quark and lepton masses and for
the low rate of proton decay.  So I hope that this somewhat unorthodox
introduction to this subject will serve you well; and I hope I have
convinced you that this is a profound and fascinating subject, where
much is to be learned and much remains to be understood.

\newpage

\section{Appendix: Comments on \none\ $SU(2)$ SQCD}

We will focus our attention on the case of $SU(2)$.
This case is a bit special because the ${\bf 2}$ and ${\bf \bar2}$
representation are identical [you already know this from quantum mechanics:
there is no conjugate-spin-1/2 representation of $SU(2)$] so we
actually should combine the $Q_r$ and $\tilde Q^s$ into $2N_f$ fields
$Q_u$, with an $SU(2N_f)$ global symmetry, and a D-term condition
\bel{sutwoDterms}
\left[\sum_{u=1}^{2N_f}(q_u^\dagger)(q_u)\right]_{\bar j}^i = c_0
\delta_{\bar j}^i 
\ee 
One solution to this condition is $q_1 = (\sqrt{c_0/2},0)$,
$q_2=(0,\sqrt{c_0/2})$, with all others zero.  As in the abelian case, it
is most convenient to express this result using gauge invariant
combinations of the chiral superfields.  The $2N_f\times 2N_f$
antisymmetric matrix of gauge-invariants $M_{uv}=Q_u^iQ_v^j\epsilon_{ij}$ has
$M_{12}=-M_{21}= c_0/2$, with all other components zero.

In fact, all solutions to the condition \eref{sutwoDterms} can be
written as $SU(2N_f)$-flavor and $SU(2)$-gauge rotations of the above
particular solution.  The gauge rotations leave $M_{uv}$ invariant,
and the flavor rotations leave invariant the fact that it has rank at
most two, with either zero or two equal non-vanishing eigenvalues.

Note that unless $M_{uv}=0$, the gauge group is completely broken.
Let's check this is the case for $N_f=1$. There are two chiral fields
$Q_1$ and $Q_2$, each in the ${\bf 2}$ of $SU(2)$, for a total of four
complex fields.  Three of these must be eaten by the three gauge
bosons if $SU(2)$ is completely broken.  Consequently, there should be
one remaining.  Indeed, there is only one (unconstrained) field
$M_{12}$.  Let's check it for $N_f=2$: in this case there are six
fields $M_{uv}$ ($u,v=1,2,3,4$) but also a single constraint that the
rank must be 2, not 4, which is the condition Pf$(M)=0$.  (The
Pfaffian is just the square root of the determinant, and is defined as
$\epsilon_{uvwx}M^{uv}M^{wx}$ in this case.)  This leaves five
unconstrained fields.  Initially there are four doublets
$Q_1,Q_2,Q_3,Q_4$ for a total of eight fields, with three being eaten when
the gauge group is broken; this too leaves five.

\EX{ For $SU(3)$ the $Q_r$ and $\tilde Q^s$ are in distinct
representations.  The allowed operators are $M_{r}^s = Q_r\tilde Q^s$,
$B=QQQ$ and $\tilde B = \tilde Q\tilde Q\tilde Q$ (indices
suppressed.)  Show that the conditions we have just obtained from the
$SU(N)$ D-terms imply that for $N_f<N$ the rank of $M$ is $N_f$ or
less; for $N_f=N$ $\det M = B\tilde B$; and for $N_f>N$ the rank of
$M$ is $N$ or less.  Show also that for $N_f\geq N$ there are branches
with $B\neq 0$ but $M=0$, and that $B\tilde B$ always equals a
subdeterminant of rank $N$ of $M$.}

What then happens quantum mechanically for $SU(2)$?  Let's note that
for $N_f=1,2,3,4,5$, the nonanomalous R-charge of the $Q_r$ is
$-1,0,\third,\half,{\frac35}$; for $N_f\geq 6$ the theory is no longer
asymptotically free.  For $N_f=4,5$ we might well have a conformal
fixed point in the infrared, but not for $N_f<4$.  We should check for
renormalizations.  The classical superpotential is zero; the one-loop
holomorphic gauge coupling is renormalized; but neither can get any
further perturbative renormalization for the reasons we discussed
earlier.  All of the higher-loop effects in the NSVZ beta function
come through the K\"ahler potential.  However, we did not check that
nonperturbative effects were absent.  In particular, while the
{\it perturbative} superpotential cannot depend on the theta angle, this is
not true {\it nonperturbatively}.  We should therefore look for a
superpotential of the form
$$
W_{nonpert}(M_{uv},\hat\Lambda)
$$ 
which is invariant under all of the global symmetries.  The only
globally-symmetric holomorphic object which we can build from $M$ is
its Pfaffian $\Pf M$, which has dimension $2N_f$ and has R-charge
$2(N_f-2)$, and its powers.  The superpotential has dimension 3 and
R-charge 2, so its form is very highly constrained; in fact
\bel{allsups}
W_{nonpert}=c\left({\Pf M\over \hat \Lambda_{N_f}^{6-N_f}}\right)^{1/(N_f-2)}
\ee
where $c$ is a constant, is the {\it only} possibility.  (This was
pointed out by Affleck, Dine and Seiberg in 1984.)  You should check
that this formula is also consistent with the {\it anomalous} $U(1)$
symmetries which we used to write the physical version of $\Lambda$.  For
this reason, the above formula even holds for $N_f=0$, where there is no
anomaly-free R-symmetry.

Now let us examine whether the coefficient $c$ can ever be nonzero.
Affleck, Dine and Seiberg pointed out that $c$ is in fact nonzero in
the case $N_f=1$; they showed that an instanton effect does indeed
give a mass to the fermion in the multiplet $M\equiv M_{12}$, and calculated
it, showing that
$$
W_{N_f=1} = {\Lambda_1^5\over M}
$$
This is rather strange; the potential 
$$
V(M)\sim {1\over |M|^2}
$$ 
blows up at small $M$ (though the K\"ahler potential cannot be
calculated there) and runs gradually to zero as $M\to\infty$ (where
the gauge theory is broken at a very high scale, and thus at weak
coupling, where the K\"ahler potential is easy to calculate.)  In
short, this theory has no supersymmetric vacuum except at $M=\infty$;
it has a runaway instability!

However, if we add a mass for the two doublets
$$
W_{classical} = mQ_1Q_2
$$
then the effective superpotential becomes
$$
W_{full} =  {\Lambda_1^5\over M} + m M
$$
which has supersymmetic minima
\bel{Msquared}
M^2 = m\Lambda^5
\ee
or in other words two vacua, $M = \pm \sqrt{m\Lambda^5}$.  Notice that
the superpotential in \Eref{allsups}, for $N_f=0$, gives $W=\pm
c\sqrt{\Lambda_0^6}$, which, for $c=2$ and the matching condition
$\Lambda_0^6 = m\Lambda_1^5$, is consistent with \eref{Msquared}.  The
interpretation of this result, originally due to Witten (1980), is
that the pure \none\ $SU(2)$ gauge theory has a fermion bilinear
condensate
$$
\vev{\lambda\lambda} \propto \sqrt{\Lambda_0^6}
$$
which breaks a discrete chiral symmetry, somewhat analogous to QCD's
breaking of chiral symmetries, and generates a nonzero superpotential
$W\propto \lambda\lambda$.

What about $N_f=2$?  In this case the theory has six mesons
$M_{uv}$ subject to the constraint $\Pf M =0$.  There can be no
nontrivial superpotential here built just from $M$, but Seiberg (1994)
pointed out that it was useful to implement this constraint using a
Lagrange multiplier field $X$, of R-charge 2 and dimension -1, in the
tree-level superpotential:
$$
W_{classical} = X(\Pf M)
$$
Then $\dbyd{W}{X} = \Pf M = 0$ defines the classical moduli space.  
However, quantum mechanically we are allowed
by the symmetries to add
$$
W_{nonpert} = cX\Lambda_2^4
$$
which means
$$\dbyd{W}{X} = \Pf M +c\Lambda_2^4 = 0$$
so the classical moduli space is modified quantum mechanically.  
In particular, the symmetric point $\Pf M=0$ is removed!  This means that
the chiral $SU(4)$ symmetry is nowhere restored on the moduli space ---
there is quantum breaking of a chiral symmetry in this theory!

\EX{ By adding mass terms for the fields $Q_3,Q_4$ and comparing
with the $N_f=1$ case, show that $c$ cannot be zero.}

And for $N_f=3$?  Here the proposed quantum superpotential
$$
W={\Pf M\over \Lambda_3^3}
$$ 
is exactly right for imposing the classical constraint that $M$ have rank
2.  The interpretation Seiberg gave is that the $M$ fields are mesons
built from confined quarks, and they have an $XYZ$-like superpotential
quantum mechanically, one which is marginally irrelevant.  In
the infrared, the $M$ fields are free, and at the origin, the $SU(6)$
symmetry is {\it unbroken}.  This is the first example known of
confinement {\it without} chiral symmetry breaking.

For $N_f=4,5$, the proposed superpotential is singular at $M=0$, and
cannot be valid there.  Seiberg (1994) therefore suggested that there
are nontrivial infrared fixed points at $M=0$ for $N_f=4,5$.  The
evidence in favor of this suggestion is now very strong, although it
is still not really proven beyond a shadow of doubt.  Personally I
don't doubt it, but I would love to see a conclusive proof someday.

\section{Suggested Reading}

There are many great papers, and many excellent reviews, for you to
look at in your further explorations of this subject.  I learned
supersymmetry, the Wess-Zumino model, non-renormalization theorems,
and so forth from Wess and Bagger and from West; both books have
advantages and problems.  Philip Argyres has a set of very useful
lectures; they can be accessed from his website.  Renormalization you
must learn from many places; no one book does it all well.  The two
classic papers of Seiberg and Witten (1994) on duality in \ntwo\ and
the various papers of Seiberg on holomorphy and on duality (1993-1994)
in \none\ are must-reads for everyone.  There are pedagogical reviews
(try Bilal (1995) for \ntwo, Intriligator and Seiberg (1995) for
\none) that unpack these papers somewhat.  Three-dimensional
supersymmetric abelian gauge theories were first studied in papers by
Seiberg and Witten (1996) and by Intriligator and Seiberg (1996); see
also de Boer et al. (1996, 1997), and Aharony et al. (1997).  Vortex
solutions appear in Nielsen and Olesen, and earlier in work of
Abrikosov in the context of superconductivity. Duality is best
understood by first studying the classic work on the Ising model, and
by reading a lovely paper on bosonization by Burgess and Quevedo
(1995).  The work of the author and Kapustin (1998) follows in this
spirit and points in new directions.  You can also get a quick tour of
duality (though not as quick as in these lectures) and vortices in my
Trieste 2001 lectures.  The papers of Shifman and Vainshtein, many
cowritten with Novikov and Zakharov (1980-1988), painstakingly
explored and finally drained the swamp surrounding the distinction
between the holomorphic and physical gauge couplings.  The work of
Leigh and the author (1995) on exactly marginal couplings builds on
their results, as well as on related results in two dimensions (see
for example Martinec (1989) and Lerche, Vafa and Warner (1989).)
A summary and list of references concerning recent refinements 
in the study of beta functions can be found
in an appendix of a paper by Nelson and the author (2002).

  This is as sketchy a bibliography as can be imagined; there are
literally hundreds of interesting papers which are relevant to these
lecture notes.  Well, such is the fate of most papers that we write;
we may love them dearly, but it is wise to remember that the next
generation of students will never read them.

\end{document}